\documentclass[%
    preprint, 
 amsmath,amssymb,
aps,prfluids,showkeys,10pt]{revtex4-2}
\usepackage{graphicx}
\usepackage{dcolumn}
\usepackage{bm}

\usepackage{flushend}
\usepackage{natbib}
\usepackage{fancyhdr}
\usepackage{amsmath}
\usepackage{mathrsfs}
\usepackage{physics}
\usepackage{amssymb}
\usepackage{subcaption}
\usepackage{titlesec}
\usepackage{graphicx}
\usepackage{xcolor}
\usepackage{tikz}

\begin{document}

\title{Improved uncertainty representation for reducing artificial energy production in structured input–output stability analysis}

\author{Ofek Frank-Shapir}
\email{Contact author: ofekfr@campus.technion.ac.il}
 \altaffiliation[Also at ]{The Stephen B. Klein Faculty of Aerospace Engineering, Technion – Israel Institute of Technology.}
\author{Igal Gluzman}%
\affiliation{%
  The Stephen B. Klein Faculty of Aerospace Engineering, Technion – Israel Institute of Technology, Haifa, 3200003, Israel.
}%

\date{\today}

\begin{abstract}

This work employs a new form of fixed structured uncertainty within the structured small-gain theorem approach proposed by Frank-Shapir~\&~Gluzman (J. Fluid Mech., vol. 1030, 2026, pp A8) for the stability analysis of incompressible shear flows subject to finite-magnitude disturbances. Within this framework, the nonlinear advection term in the Navier-Stokes equations is replaced by a structured feedback uncertainty interconnection with the linearized dynamics to account for the impact of nonlinear feedback. Herein, a new uncertainty representation is derived via linear transformations of the input and output channels, transforming the feedback loop such that the resulting structured uncertainty has a repeated-diagonal structure. This structure aims to preserve the component-wise pathways of the nonlinear advection term while keeping the structured singular value computation tractable.  We apply the method to two canonical base flows: Couette and plane Poiseuille flows. The resulting thresholds on disturbance magnitude to preserve stability are less conservative and more accurate. We compare the novel methodology presented here with previously proposed repeated and non-repeated block approximations of the uncertainty structure, where our stability threshold provided the closest agreement with previous numerical and experimental studies. We show that repeated and non-repeated block structures that were proposed in past studies result in an artificial energy-production term arising from using constant structured uncertainty in the structured input-output formulation, violating the divergence-free assumption. This energy-production term is smallest when using the methodology presented in this work, providing a more faithful representation of the impact of nonlinear feedback interconnection with the linearized dynamics of the Navier-Stokes system.

\end{abstract}


\maketitle

\section{INTRODUCTION}

Understanding the stability of shear flows in the presence of finite-magnitude disturbances remains a central challenge in transition research. Classical linear stability theory (LST) predicts the onset of instability under infinitesimal perturbations and has been highly successful in identifying modal instability mechanisms via eigenvalue analysis, such as Tollmien--Schlichting (TS) waves. However, transitional flows are often subject to finite-amplitude disturbances, and subcritical transition may arise through mechanisms not captured by purely linear modal analysis. In boundary layers, for example, oblique-wave interactions and streak-related structures play a major role in bypass transition and may dominate the route to instability well below the critical Reynolds number predicted by LST \cite{asai1995boundary,elofsson2000experimental,cherubini2011minimal,vavaliaris2020optimal}. 
Within the LST framework, there exist several flow stability criteria. 
For example, the well-known inviscid Rayleigh's inflection point criterion and Fj{\o}rtoft's criterion both provide sufficient conditions for stability (or, equivalently, necessary conditions for instability) in inviscid flows \citep{schmid2002stability}. In the recent work of \cite{peet2025inflectional}, the authors show that Rayleigh's inflection point criterion and Fj{\o}rtoft's criterion are valid only under the normal-mode assumption, and for general perturbations do not provide a sufficient condition for stability. The authors suggest a Lyapunov-based approach to analyzing inviscid flows without assuming a normal-mode decomposition, producing a sufficient condition for flow stability. However, these criteria cannot predict instability associated with viscous mechanisms that may occur in non-inflectional velocity profiles. For such cases, the eigenvalue analysis for given initial conditions can provide a necessary condition for flow stability. This means that in scenarios where flow systems represented by the Navier-Stokes equations (NSE) are stable (i.e., remain laminar), LST will predict stable eigenvalues. Equivalently, violation of the LST condition for stability (by crossing the critical Reynolds number) will result only in a sufficient condition for instability of the flow. Thus, while crossing the critical Reynolds number predicted by LST leads to flow instability, transition to turbulence can occur at much lower Reynolds numbers than the theory predicts due to transient growth or bypass transition. This phenomenon is well known and applies to various base flows \cite{asai1995boundary,elofsson2000experimental,cherubini2011minimal,vavaliaris2020optimal}. 
One possible pitfall of a conservative sufficient stability criterion is that it may prematurely expect the onset of instability after the criterion has been violated. 
Thus, a crucial goal for achieving accurate predictions of the onset of instability using a sufficient stability condition is to reduce conservatism as much as possible. In this study, we will rely on the input-output analysis paradigm to achieve this goal, as we discuss next.

Within non-modal flow analysis approaches,  input-output analysis has emerged as a useful framework for studying disturbance amplification in shear flows, characterizing how forcing is amplified by the linearized Navier-Stokes (LNS) dynamics. Examples include studies of the externally forced LNS equations that model different actuation geometries \citep{GG21,frank2025choosing} and examine dynamic processes, structural features, and energy pathways across a variety of base flows \citep{farrell1993stochastic,bamieh2001energy,jovanovic2004unstable,jovanovic2005componentwise,HC10,mckeon2010critical,mckeon2013experimental,mckeon2017engine,madhusudanan2019coherent,liu2020input,symon2021energy}. 
 Recently, in \cite{bovzic2026oblique}, the authors integrated the weakly nonlinear theory (expanding the perturbation dynamics to higher orders) into the input-output analysis framework to study the energy transfer, evolution, and interaction of oblique and streak flow structures in initial and later stages of transition. This framework was also applied to study the dynamics of large-scale structures in turbulent boundary layers by analyzing the flow response to external periodic perturbations \citep{liu2022spatial}. Yet, classifying the conditions under which the transition process begins remains a relevant question. These conditions often involve finite-amplitude disturbances that may trigger flow instability and lead to bypass transition. 
For such cases, we recently proposed a sufficient stability criterion utilizing the small-gain theorem coupled with structured input-output analysis \cite{frank2026stability}. This approach enables the derivation of stability bounds on the finite disturbance amplitude in shear flows. To determine these bounds, we use structured input-output methodologies to account for nonlinear interactions via a structured feedback interconnection.

The structured input-output analysis framework proposed by \cite{liu2021} is based on the concept of structured uncertainty \citep{packard1993,Zhou1995-dl} from the field of robust control theory. Using this approach, the nonlinear advection term in the NSE is replaced by a structured feedback uncertainty interconnection with the linearized dynamics, allowing for the approximation of the impact of nonlinear feedback instead of considering a worst-case (unstructured) feedback. Previous works suggested different fixed-structure uncertainties for recovery of flow features and dominant modes of transitional flows in Couette, plane Poiseuille, Couette-Poiseuille, and stratified Couette flows \citep{liu2021,liu2022structured,shuai2023structured,bhattacharjee2023structured,mushtaq2023,mushtaq2024structured}, via the computation of structured singular values (SSVs). Structured input-output analysis was also used for identification of control effects on the growth of different types of disturbances in boundary layers \cite{rath2024structured}, and compliant-wall turbulent channel flows at high Reynolds numbers \citep{song2026structured}. Furthermore, structured input-output works have been successful in identifying optimal perturbation shape in the wall-normal direction. This can be done either by analyzing the disturbance field corresponding to the worst-case structured uncertainty itself \citep{luiACC23} or by scaling the resolvent operator in a unique way tied to the results of the structured input-output analysis, and obtaining the most amplified modes via singular value decomposition (SVD) \citep{shuai2023structured,song2026structured}.

A key difficulty within structured input-output analysis is faithfully representing the nonlinear advection term using an uncertainty structure. In structured input–output analysis, the nonlinear term in the NSE is modeled as a fixed-structure uncertainty that is interconnected to the linear frequency-response operator obtained from the LNS equations. Previous structured input-output studies have employed two uncertainty modeling approaches: non-repeated blocks \citep{liu2021} and repeated blocks \citep{mushtaq2023}, both used to obtain tractable SSV computations \citep{liu2021,mushtaq2023,shuai2023structured,mushtaq2024structured,frank2026stability}. While such approximations provide valuable insights, they do not preserve the exact structure of the nonlinear advection term in the NSE and may introduce additional feedback pathways that are not physically associated with the advection term's effect on the flow. Therefore, these two approaches lead to different, and sometimes conflicting, predictions about the most amplified flow structures and flow-stability bounds. 

Both the repeated and non-repeated block structures impose stricter bounds on the disturbance threshold to maintain the stability of the interconnected system than the uncertainty structure associated with the nonlinear advection term \citep{frank2026stability}. While repeated blocks yield less conservative stability bounds, this structure leads to higher computational demand than non-repeated blocks because of the additional constraint of repeated uncertainty entries. These structures result in different predictions of dominant flow structures, and whether enforcing the repeated-entry constraint for block uncertainty is beneficial remains an open question. In \cite{luiACC23}, a transformation was applied to the non-repeated block uncertainty structure, allowing analysis of the velocity correlation fields corresponding to each individual velocity component, rather than each block corresponding to the entire velocity field. This method produced similar SSV values and dominant flow structures as previous works using non-repeated block uncertainties. 

Herein, we propose a structured input-output formulation of the NSE to represent the uncertainty matrix structure that more faithfully approximates the impact of nonlinear feedback associated with the advection term in the full Navier-Stokes system while remaining computationally feasible. We achieve this by applying linear transformations to the transfer function representing linear flow interactions and to the structured uncertainty used to model nonlinear flow interactions. The transformed uncertainty is diagonal, allowing a tractable SSV computation. With this new structure, we avoid adding uncertainty terms over the exact structure that models the nonlinear advection term, leading to reduced conservatism in structured input-output analysis.

Using the proposed new uncertainty structure, we examine how disturbance thresholds (for assessing flow stability) vary with the Reynolds number and wavenumber pair for incompressible Couette (purely shear-driven) and plane Poiseuille (pressure-driven) flows. 
These two canonical flows provide a solid testing basis for our approach, showing that our new structure yields consistently less restrictive bounds than previous structured input-output approaches. Additionally, we study the effect of repeated uncertainty entries by comparing predictions using non-repeated and repeated block uncertainty structures, and by applying the new methodology with and without the repeated-entry constraint. We show that modeling the nonlinear advection term within a structured input-output framework leads to artificial energy production which is not found in the original NSE. This energy production is shown to be caused by the structured uncertainty not being divergence-free. Analyzing this artificial energy production reveals that for a group of wavenumber pairs, applying the constraint of repeated uncertainty entries significantly reduces it, while leading to less conservative and more accurate stability bound predictions. The methodology that yields the lowest artificial energy production, computed under the integration process presented in this work, is the proposed new uncertainty structure. This reduced conservatism allows us to provide more accurate stability thresholds via our structured small-gain theorem stability analysis \cite{frank2026stability} compared to results from the literature that utilized high-fidelity nonlinear analysis, experiments, and direct numerical simulations.

In~\S\ref{sec:math} we provide a mathematical framework for formulating our stability analysis methodology using structured input-output analysis. Details on the uncertainty structures used in this work, along with the transformation we apply to the SSV problem to achieve a less conservative stability analysis using the uncertainty structures presented here, are in~\S\ref{sec:uncertainty}. In~\S\ref{sec:production}, we suggest a methodology to analyze the artificial energy production term that arises when modeling the nonlinear advection term using constant structured uncertainty. Results of the stability analysis performed using the methodology in this work are shown in~\S\ref{sec:stability_results} for Couette and plane Poiseuille flows. A comparison between these results and high-fidelity simulation results from the literature is also provided in this section. We analyze the artificial energy production associated with each uncertainty structure in~\S\ref{sec:repeated}, showing that lifting the constraint of repeated uncertainty entries increases artificial energy production and reduces stability margins for a particular group of oblique flow structures. Finally, conclusions are discussed in~\S\ref{sec:conclusions}.

\section{MATHEMATICAL MODEL}
\label{sec:math}

We consider incompressible shear flows, for which the LNS equations for perturbations in fluid velocity and pressure $(\mathbf{u}, p)$ about the base flow $(\boldsymbol{\overline{u}}, \overline{p})$ are as follows: 
\begin{equation}
\label{eq:LNS}
\begin{matrix}
 \frac{\partial \mathbf{u}}{\partial t} = -\mathbf{\bar{u}} \cdot \nabla \mathbf{u} - \mathbf{u} \cdot \nabla \mathbf{\bar{u}} - \nabla p + \frac{1}{Re} \Delta \mathbf{u} + \boldsymbol{d}, \\
\nabla \cdot \mathbf{u} = 0.
\end{matrix}
\end{equation}
Here, $\mathbf{u}=\begin{bmatrix} u & v & w\end{bmatrix}^T$ is the perturbation velocity vector, corresponding to the streamwise $x$, wall-normal $y$, and spanwise $z$ directions, respectively. Throughout this work, we consider base velocity profiles of the form $\mathbf{\bar{u}} = \begin{bmatrix} U(y) & 0 & 0 \end{bmatrix}^T$. Additionally, $\boldsymbol{d}=\begin{bmatrix} d_x & d_y & d_z\end{bmatrix}^T$ is the term representing body forcing. $\nabla$ is the del operator,  and $\Delta\equiv\nabla^2$ is the Laplacian operator.
 
In addition, we assume spatial invariance of the parallel flow field in the horizontal directions to allow Fourier transforms in the $x$ and $z$ directions, providing a system of two partial differential equations depending on spatial wavenumbers $k_x$, $k_z$, and $y$, the position in the wall-normal direction \citep{kim1987turbulence}. This assumption holds for Couette and plane Poiseuille flows.
This process allows us to write the LNS equations in a state-space form, which simplifies performing input-output analysis, via a formulation we adopt from \cite{jovanovic2005componentwise}:
 
\begin{equation}
\label{eq:state}
\begin{matrix}
\frac{\partial \boldsymbol{\psi}}{\partial t}(k_x,y,k_z,t) = [\mathscr{A}(k_x,k_z)\boldsymbol{\psi}(k_x,y,k_z,t)](y) + [\mathscr{B}(k_x,k_z)\boldsymbol{d}(k_x,k_z,t)](y), \\
\boldsymbol{\phi}(k_x,y,k_z,t) = \mathscr{C}(k_x,k_z)\boldsymbol{\psi}(k_x,k_z,t)](y),
\end{matrix}
\end{equation}
where $\boldsymbol{\psi} \equiv \begin{bmatrix} v & \omega_y \end{bmatrix} ^T$ is the state vector, comprised of the wall-normal velocity perturbation $v$, and the wall-normal vorticity perturbation $\omega_y = {\partial u}/{\partial z} - {\partial w}/{\partial x}$. The output vector $\boldsymbol{\phi}$ in our analysis is set to be $\boldsymbol{\phi} \equiv \mathbf{u} = \begin{bmatrix} u & v & w\end{bmatrix}^T$.
Lastly, $\boldsymbol{d}$ is the body forcing term applied to the flow system. 
The operators denoted as $\mathscr{A}$, $\mathscr{B}$ and $\mathscr{C}$ are defined below \citep{jovanovic2005componentwise}:

\begin{equation}
\label{eq:A}
\begin{aligned}
\mathscr{A} \equiv 
\begin{bmatrix}
 \mathscr{A}_{11} & 0 \\
 \mathscr{A}_{21} & \mathscr{A}_{22}
\end{bmatrix}
\equiv
\begin{bmatrix}
 -ik_x\Delta^{-1}U\Delta+ik_x\Delta^{-1}U''+\frac{1}{Re}\Delta^{-1}\Delta^2 & 0 \\
 -ik_zU' & -ik_xU+\frac{1}{Re}\Delta
\end{bmatrix},
\end{aligned}
\end{equation}

\begin{equation}
\label{eq:B}
\begin{aligned}
\mathscr{B} \equiv 
\begin{bmatrix}
 \mathscr{B}_{x} & \mathscr{B}_{y} & \mathscr{B}_{z}
\end{bmatrix}
\equiv
\begin{bmatrix}
 \Delta^{-1} & 0 \\
 0 & I
\end{bmatrix}
\begin{bmatrix}
 -ik_x\frac{\partial}{\partial y} & -(k_x^2 + k_z^2) & -ik_z\frac{\partial}{\partial y} \\
 ik_z & 0 & -ik_x
\end{bmatrix},
\end{aligned}
\end{equation}

\begin{equation}
\label{eq:C}
\begin{aligned}
\mathscr{C} \equiv 
\begin{bmatrix}
 \mathscr{C}_{u} \\ 
 \mathscr{C}_{v} \\ 
 \mathscr{C}_{w}
\end{bmatrix}
\equiv
\frac{1}{k_x^2+k_z^2}
\begin{bmatrix}
 ik_x\frac{\partial}{\partial y} & -ik_z \\
 k_x^2+k_z^2 & 0 \\
 ik_z\frac{\partial}{\partial y} & ik_x
\end{bmatrix}.
\end{aligned}
\end{equation}
Here, $U'=\frac{dU(y)}{dy}$ and $U''=\frac{d^{2}U(y)}{dy^{2}}$. The Laplacian operator is defined as $\Delta=\frac{\partial^2}{\partial y^2}-(k_x^{2}+k_z^{2})$ because of the Fourier transforms done in directions $x$ and $z$. $\mathscr{A}$ is the operator representing the dynamics of the system and is comprised of the Orr-Sommerfeld operator ($\mathscr{A}_{11}$), the coupling operator  ($\mathscr{A}_{21}$), and the Squire operator ($\mathscr{A}_{22}$). $\mathscr{B}$ is the forcing operator, representing the effect of forcing input on the flow, and $\mathscr{C}$ is the operator relating the state variables---wall-normal velocity and vorticity perturbations---to the three velocity perturbations ($u,v,w$). We note that the operator $\mathscr{C}$ can be modified to acquire the desired flow variables of interest as output of the system. Each of the operators $\mathscr{A}$, $\mathscr{B}$ and $\mathscr{C}$ implicitly depends on the wall-normal coordinate $y$.

Herein, the boundary conditions on $v$ and $\omega_y$ depend on the specific base flow. For Couette and plane Poiseuille flows, the boundary conditions are:

\begin{equation}
\label{eq:BC}
 v(y= \pm 1) = \frac{\partial v}{\partial y}(y= \pm 1) = \omega_{y}(y= \pm 1) =0, \forall k_x,k_z \in \mathbb{R}, t \geq 0.
\end{equation}
Lastly, the Reynolds number is defined as $Re = \mathcal{U}L / \nu$, where $\mathcal{U}$ is the reference velocity, $L$ is the reference height, and $\nu$ is the kinematic viscosity of the fluid. These parameters also render the variables in this work dimensionless. In particular, for Couette flow, $\mathcal{U}$ is half the difference between the base-flow velocities at the lower and upper walls; for plane Poiseuille flow, it is the centerline velocity. $L$ is the channel half-height for both Couette and plane Poiseuille flows.

\begin{figure}
 \centering
 \includegraphics[width = \textwidth]{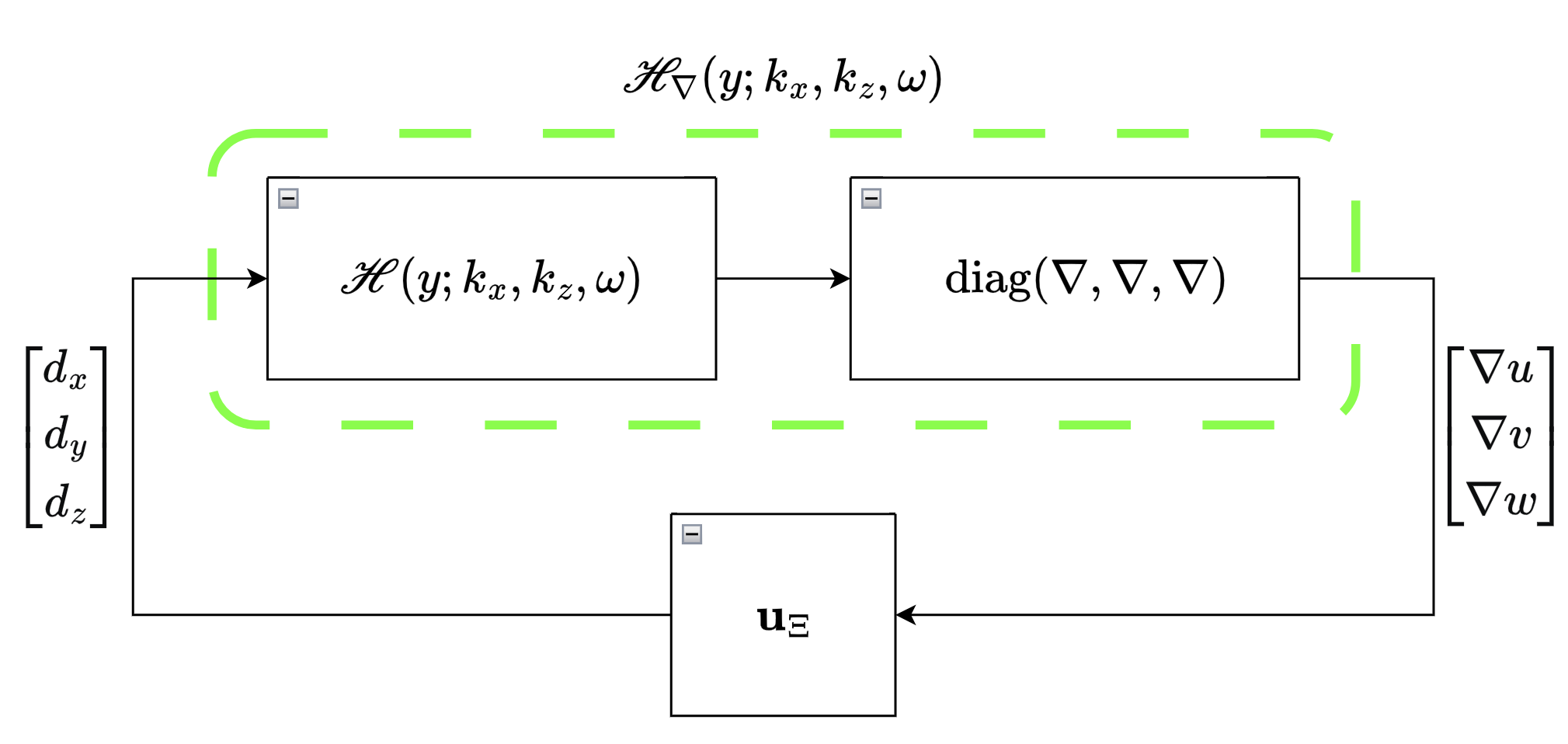}
 \caption{Interconnection loop block diagram for the nonlinear interactions. Adapted from \cite[][Fig. 3]{liu2021}.}
 \label{fig:1}
\end{figure}

Here, we consider a forcing term $\boldsymbol{d}$, which is harmonic in time and in the $x$ and $z$ directions, i.e., 
$\boldsymbol{d}(t,x,y,z) = \overline{\boldsymbol{d}}(y)\exp[i(k_x x + k_z z + \omega t)]$,
where $\omega \in \mathbb{R}$ is the temporal frequency and $k_x,k_z\in\mathbb{R}$ are wavenumbers of the streamwise and spanwise flow directions, respectively, and $\overline{\boldsymbol{d}}(y)$ is some function of $y$, following the formulation in \cite{jovanovic2005componentwise}. In our study, we do not consider any specific form of $\overline{\boldsymbol{d}}(y)$; instead, the forcing component is viewed as the feedback in the interconnected system in Fig.~\ref{fig:1} to account for the effect of the nonlinear advection term of the full Navier-Stokes (NS) system.

The frequency response operator that maps the harmonic forcing term applied to the above state-space system in Eq.~\eqref{eq:state} to the resulting perturbation velocity field is given by \citep{Zhou1995-dl}:
\begin{equation}
\label{eq:2.8}
\begin{aligned}
\mathscr{H}(y;k_x,k_z,\omega)=[\mathscr{C}(k_x,k_z)[i\omega I - \mathscr{A}(k_x,k_z)]^{-1}\mathscr{B}(k_x,k_z)](y).
\end{aligned} 
\end{equation}
As shown in equations \eqref{eq:B} and \eqref{eq:C}, the operators $\mathscr{B}$ and $\mathscr{C}$ each have 3 components, corresponding to different inputs and outputs, respectively. Thus, $\mathscr{H}$ can be decomposed as follows, as suggested by \cite{jovanovic2005componentwise}: 
\begin{equation}
\label{eq:H1x3}
\begin{matrix}
\mathscr{H}(y;k_x,k_z,\omega)=
\begin{bmatrix}
  \mathscr{C}_{u} \\ 
  \mathscr{C}_{v} \\ 
  \mathscr{C}_{w}
\end{bmatrix}
(i\omega I - \mathscr{A}(k_x,k_z))^{-1}\mathscr{B} =
\begin{bmatrix}
  \mathscr{H}_{u}(y;k_x,k_z,\omega)\\
  \mathscr{H}_{v}(y;k_x,k_z,\omega)\\ 
  \mathscr{H}_{w}(y;k_x,k_z,\omega)  
\end{bmatrix}.
\end{matrix} 
\end{equation}
This decomposition allows us to examine the dynamics of each velocity component separately.

The frequency response operator is widely used in input-output analysis \cite{jovanovic2005componentwise}. Herein, we define an alternative operator, as proposed by \cite{liu2021} and used in several structured input-output works \citep[such as][]{mushtaq2023,shuai2023structured}:
\begin{equation}
\label{eq:Hnabla}
\mathscr{H}_{\nabla}(y;k_x,k_z,\omega) \equiv
\text{diag} (\nabla,\nabla,\nabla)  \mathscr{H}(y;k_x,k_z,\omega).
\end{equation}
This frequency response operator of the system $\mathscr{H}_{\nabla}$ relates the forcing term to the resulting gradients of velocity perturbations. 

We consider the full NS equations for velocity and pressure perturbations to incorporate nonlinear behavior in our analysis. The NSE with no forcing term is presented below:

\begin{equation}
\label{eq:NS}
\begin{matrix}
 \frac{\partial \mathbf{u}}{\partial t} = -\mathbf{\bar{u}} \cdot \nabla \mathbf{u} - \mathbf{u} \cdot \nabla \mathbf{\bar{u}} - \nabla p + \frac{1}{Re} \Delta \mathbf{u} - \mathbf{u} \cdot \nabla \mathbf{u}, \\
\nabla \cdot \mathbf{u} = 0.
\end{matrix}
\end{equation}
Eq. \eqref{eq:NS} is very similar to Eq. \eqref{eq:LNS}, with the only difference being that the forcing term is replaced by the nonlinear advection term {$\mathbf{u}\cdot \nabla \mathbf{u}$}. 
In the structured input-output approach, we model the nonlinear interactions arising from the advection term, using the block diagram representation in Fig.~\ref{fig:1}.
Here, $\mathbf{u}_{\Xi}$ is the gain applied to the vectorized velocity gradient $\begin{bmatrix} \nabla^T u & \nabla^T v & \nabla^T w \end{bmatrix}^T$ that retains the component-wise structure of the advection term $\mathbf{u} \cdot \nabla \mathbf{u}$ \citep{liu2021}. To interpret the feedback interconnection in Fig.~\ref{fig:1} physically, we view the nonlinear advection term as an internal forcing mechanism acting on the LNS dynamics. In a standard 'unstructured' analysis, the feedback block ${\mathbf{u}}_\Xi$ is treated as a generic operator bounded only by its magnitude. Thus, unstructured analysis can be thought of as a worst-case model of nonlinearity, since there are no constraints on the individual terms in ${\mathbf{u}}_\Xi$. By contrast, the 'structured' feedback connection explicitly imposes constraints on ${\mathbf{u}}_\Xi$, such that some terms in it are fixed to be zero. Structure constraints are imposed to limit feedback pathways for the goal of replicating the structure of the nonlinearity in the NSE \citep{liu2021}.

The exact expression for ${\mathbf{u}}_\Xi$ replicates the nonlinear advection term, meaning that ${\mathbf{u}}_\Xi \begin{bmatrix} \nabla^T u & \nabla^T v & \nabla^T w \end{bmatrix}^T = \mathbf{u} \cdot \nabla \mathbf{u}$, is (see \cite{liu2021}, equation 2.9):
\begin{equation}
\label{eq:u_Xi}
 {{\mathbf{u}}}_\Xi \equiv \begin{bmatrix}
    -{\mathbf{u}}^T & \mathbf{0} & \mathbf{0} \\
    \mathbf{0} & -{\mathbf{u}}^T & \mathbf{0} \\
    \mathbf{0} & \mathbf{0} & -{\mathbf{u}}^T
\end{bmatrix}.
\end{equation}
Discretization in the wall-normal direction results in a finite-dimensional matrix representation. For the rest of this work, we consider a constant-valued approximation of the discretized version of the operator $\mathbf{u}_\Xi$, which we denote as $\mathbf{U_\Xi}$. 
To incorporate this nonlinearity, we compute the system's SSVs. SSVs generalize singular values for a transfer function or frequency-response operator that is feedback-interconnected with an uncertain matrix with a predetermined structure \citep{Zhou1995-dl}.
In detail, we compute SSVs for the frequency response operator $\mathscr{H}_{\nabla}$ by solving the following minimization problem \citep{packard1993,Zhou1995-dl}:

\begin{equation}
\label{eq:SSV}
\mu_{\mathbf{\Delta}}[\mathscr{H}_{\nabla}(k_x,k_z,\omega)] = \frac{1}{\text{min}\{ \overline{\sigma} ({\mathbf{U}}_\Xi) : {\mathbf{U}}_\Xi \in \mathbf{\Delta}, \text{det}[I - \mathscr{H}_{\nabla}(k_x,k_z,\omega) {\mathbf{U}}_\Xi] = 0 \}}.
\end{equation}
In Eq.~\eqref{eq:SSV}, $\overline{\sigma}(\cdot)$ represents the largest singular value, $\text{det}[\cdot]$ is the determinant of a matrix, and $\boldsymbol{\Delta}$ denotes the set of matrices that have a certain structure. Solving the SSV problem under the same (or similar) structure as ${{\mathbf{U}}}_\Xi$ allows modeling the "worst-case" nonlinear response to external velocity perturbations. In this framework, the matrix ${{\mathbf{U}}}_\Xi$ is not known a priori. Rather, it is computed by solving Eq.~\eqref{eq:SSV} such that it is the smallest structured uncertainty (in terms of the largest singular value) that causes instability of the system. Although ${{\mathbf{U}}}_\Xi$ models nonlinear interactions, we treat it within the SSV framework as a modeling uncertainty. Thus, we refer to ${{\mathbf{U}}}_\Xi$ for the rest of this work as the \emph{structured uncertainty}. A detailed discussion about uncertainty structures is provided in~{\S}\ref{sec:uncertainty}.

We quantify amplification associated with nonlinear interactions modeled by a structured uncertainty ${\mathbf{U}}_\Xi \in \mathbf{\Delta}$ using the supremum over all temporal frequencies:

\begin{equation}
\label{eq:H_ssv}
\norm{\mathscr{H}_{\nabla}}_{\mu_{\mathbf{\Delta}}}(k_x,k_z) \equiv \sup_{\omega \in \mathbb{R}}{\mu_{\mathbf{\Delta}}[ \mathscr{H}_{\nabla}(k_x,k_z,\omega)]}, 
\end{equation}
this is the "worst-case" amplification of perturbation gradients under the uncertainty structure $\mathbf{\Delta}$. This amplification is not an amplification of the energy of the whole flow field, but rather the gain applied to the most amplified mode. Additionally, we use the notation $\norm{\cdot}_{\mu}$ for consistency with previous works \citep{packard1993,liu2021}, despite this operator not being a proper norm since it does not satisfy the triangle inequality. 

Using the {small-gain} theorem, the following condition for system stability is proposed in \cite{frank2026stability}:

\begin{equation}
    \label{eq:stability_structured}
    \max_{y\in\mathcal{Y}}\norm{\mathbf{u}(y)}_2  < \norm{\mathscr{H}_{\nabla}}_{\mu_{{\Delta}}}^{-1}(k_x,k_z).
\end{equation}
Here, $\max_{y\in\mathcal{Y}}\norm{\mathbf{u}(y)}_2$ represents the largest velocity perturbation found in the wall-normal domain $\mathscr{Y}$. 
Eq.~\eqref{eq:stability_structured} provides a bound of value $\norm{\mathscr{H}_{\nabla}}_{\mu_{{\Delta}}}^{-1}$ on the magnitude of perturbations that are allowed in the flow to maintain stability.

\section{UNCERTAINTY STRUCTURES}
\label{sec:uncertainty}

The uncertainty structure that represents ${\mathbf{u}}_\Xi$, for reproducing the structure of the advection term in the NSE, within the interconnected system shown in Fig.~\ref{fig:structure1}, is
\begin{equation}
\label{eq:structexact}
\boldsymbol{\Delta}_{\mathbf{u}} = \left\{\begin{bmatrix}
    \Delta_u & \mathbf{0} & \mathbf{0} \\
    \mathbf{0} & \Delta_u & \mathbf{0} \\
    \mathbf{0} & \mathbf{0} & \Delta_u 
\end{bmatrix} , \Delta_u = [\text{diag}(u_{\xi}), \text{diag}(v_{\xi}), \text{diag}(w_{\xi})]\right\}.
\end{equation}
Here, $u_{\xi},v_{\xi},w_{\xi} \in \mathbb{C}^{N_y \times 1}$ and $N_y$ is the number of points that are used in the discretization of the wall-normal direction. Any additional terms that are not included in the set $\boldsymbol{\Delta}_{\mathbf{u}}$ may be regarded as fictitious amplification pathways that are not associated with the structure of the non-linear advection term of the NSE (within a structured input-output framework) and, thus, do not represent physical flow behaviors. The solution for uncertainties in $\boldsymbol{\Delta}_{\mathbf{u}}$ requires solving the optimization problem in Eq.~\eqref{eq:SSV} with a very complex set of constraints that are required for preserving the structure in Eq.~\eqref{eq:structexact}. To our knowledge, no such solution exists. 

Previous works in the field of structured input-output analysis employ two different simplified approximations for the structure of $\mathbf{U}_\Xi$:

\begin{enumerate}
\item $\mathbf{\Delta}_{RB}$ (repeated blocks), the set of block-diagonal matrices with three repeating full blocks with compatible sizes, as was used in the works of \cite{mushtaq2023,mushtaq2024structured}, 
\begin{equation}
\label{eq:D_RB}
\mathbf{\Delta}_{RB} = \left\{ \begin{bmatrix}
    \Delta_r & \mathbf{0} & \mathbf{0} \\
    \mathbf{0} & \Delta_r & \mathbf{0} \\
    \mathbf{0} & \mathbf{0} & \Delta_r 
\end{bmatrix} : \Delta_r \in \mathbb{C}^{N_y \times 3N_y}\right\}.
\end{equation}
Here, $\Delta_r$ can be any matrix of compatible dimensions ($N_y \times 3N_y$).

\item $\mathbf{\Delta}_{NRB}$ (non-repeated blocks), the set of block-diagonal matrices with three full, independent \textbf{non-repeating} blocks with compatible sizes, {as was used in the works of \cite{liu2021,shuai2023structured}}: 

\begin{equation}
\label{eq:D_NRB}
\mathbf{\Delta}_{NRB} = \left\{\begin{bmatrix}
    \Delta_1 & \mathbf{0} & \mathbf{0} \\
    \mathbf{0} & \Delta_2 & \mathbf{0} \\
    \mathbf{0} & \mathbf{0} & \Delta_3 
\end{bmatrix} : \Delta_1,\Delta_2,\Delta_3 \in \mathbb{C}^{N_y \times 3N_y}\right\}.
\end{equation}
The difference here between $\mathbf{\Delta}_{NRB}$ and $\mathbf{\Delta}_{RB}$ lies in the fact that $\mathbf{\Delta}_{NRB}$ does not include the constraint of equal blocks, as $\Delta_1$, $\Delta_2$, and $\Delta_3$ can be different matrices.

\end{enumerate}

These relations between the above two sets are reflected in the SSV values for each set. Based on the monotonicity of the SSV with respect to set inclusion  \cite[see][section 4.10]{scherer2001theory}, solving the minimization problem (Eq.~\eqref{eq:SSV}) over the sets of matrices that satisfy the relations in Eq.~\eqref{eq:delta_subset1} yields:
\begin{equation}
\label{eq:hierarchy}
\norm{\mathscr{H}_{\nabla}}_{\mu_{\mathbf{\Delta}_{u}}}(k_x,k_z) \leq \norm{\mathscr{H}_{\nabla}}_{\mu_{\mathbf{\Delta}_{RB}}}(k_x,k_z) \leq \norm{\mathscr{H}_{\nabla}}_{\mu_{\mathbf{\Delta}_{NRB}}}(k_x,k_z).
\end{equation}

In this work, we propose two new uncertainty structures:

\begin{enumerate}
 \setcounter{enumi}{2}
    \item  $\boldsymbol{\Delta}_{RD}$ (repeated diagonals),  which is defined as follows:
\begin{equation}
    \label{eq:RD}
    \boldsymbol{\Delta}_{RD} = \{\text{diag}(\delta_1 \boldsymbol{I}_3,\dots,\delta_{3N_y}\boldsymbol{I}_3):\,\delta_i \in \mathbb{C}\}.
\end{equation}
Here, $\delta_i\in\mathbb{C}$ are scalar quantities, {where $i = 1,2,...,3N_y$,} and $\boldsymbol{I}_3$ is a $3\times3$ identity matrix.

\item $\boldsymbol{\Delta}_{NRD}$ (non-repeated diagonals),  which is defined as follows:
\begin{equation}
    \label{eq:NRD}
    \boldsymbol{\Delta}_{NRD} = \{\text{diag}(\delta_1,\dots,\delta_{9N_y}):\,\delta_i \in \mathbb{C}\}.
\end{equation}
Here, $\delta_i\in\mathbb{C}$ are scalar quantities. This structure is obtained from lifting the repeated-entry constraint from the structure $\boldsymbol{\Delta}_{RD}$.

\end{enumerate}

The uncertainties in Eq.~\eqref{eq:structexact}, Eq.~\eqref{eq:D_RB}, Eq.~\eqref{eq:NRD} and \eqref{eq:D_NRB} satisfy,
\begin{subequations}
     \label{eq:D_hyr}
     \begin{align}
        \label{eq:delta_subset1}\mathbf{\Delta}_{\mathbf{u}} \subset \mathbf{\Delta}_{RB} \subset \mathbf{\Delta}_{NRB}\subset \mathbb{C}^{3N_y \times 9N_y},
     \end{align}
     \begin{align}
     \label{eq:delta_subset2}
\mathbf{\Delta}_{RD} \subset \mathbf{\Delta}_{NRD} \subset \mathbb{C}^{9N_y\times9N_y}.
     \end{align}
\end{subequations}
Before we can establish the relation between the SSV values for the four structures,  a transformation of the SSV problem in Eq.~\eqref{eq:SSV} is required to utilize the uncertainty structures,  $\boldsymbol{\Delta}_{RD}$ and $\boldsymbol{\Delta}_{NRD}$, as will be discussed next in~\S\ref{sec:stability}.

\subsection{Procedure to utilize $\boldsymbol{\Delta}_{RD}$ and $\boldsymbol{\Delta}_{NRD}$ within structured input-output analysis}
\label{sec:stability}
\begin{figure}[ht!]
    \centering
    
    \begin{subfigure}[b]{0.48\textwidth} 
        \centering
        \includegraphics[width=\textwidth]{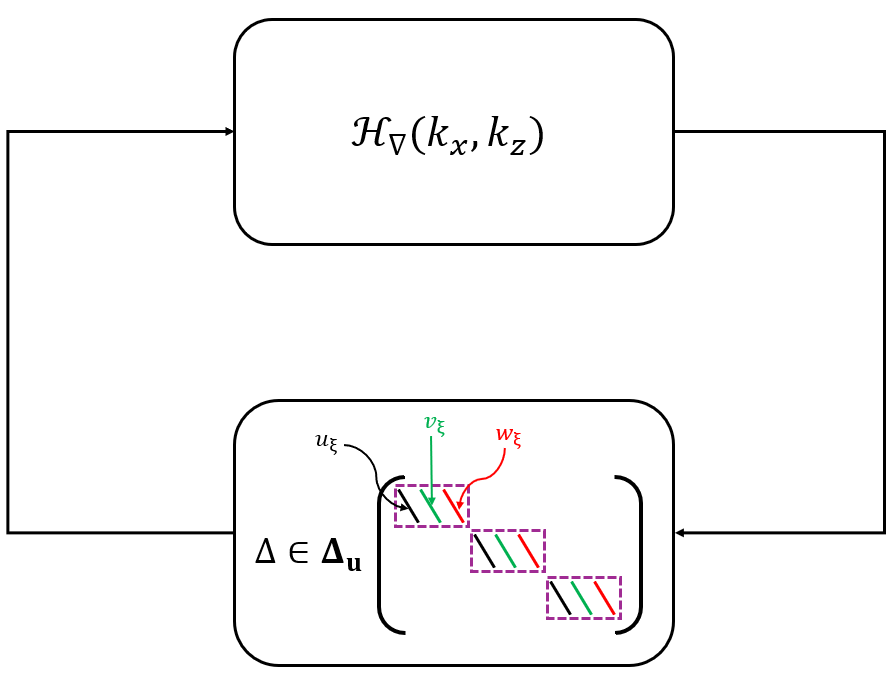}
        \caption{ }
        \label{fig:structure1}
    \end{subfigure}\hfill
    \begin{subfigure}[b]{0.48\textwidth} 
        \centering
        \includegraphics[width=\textwidth]{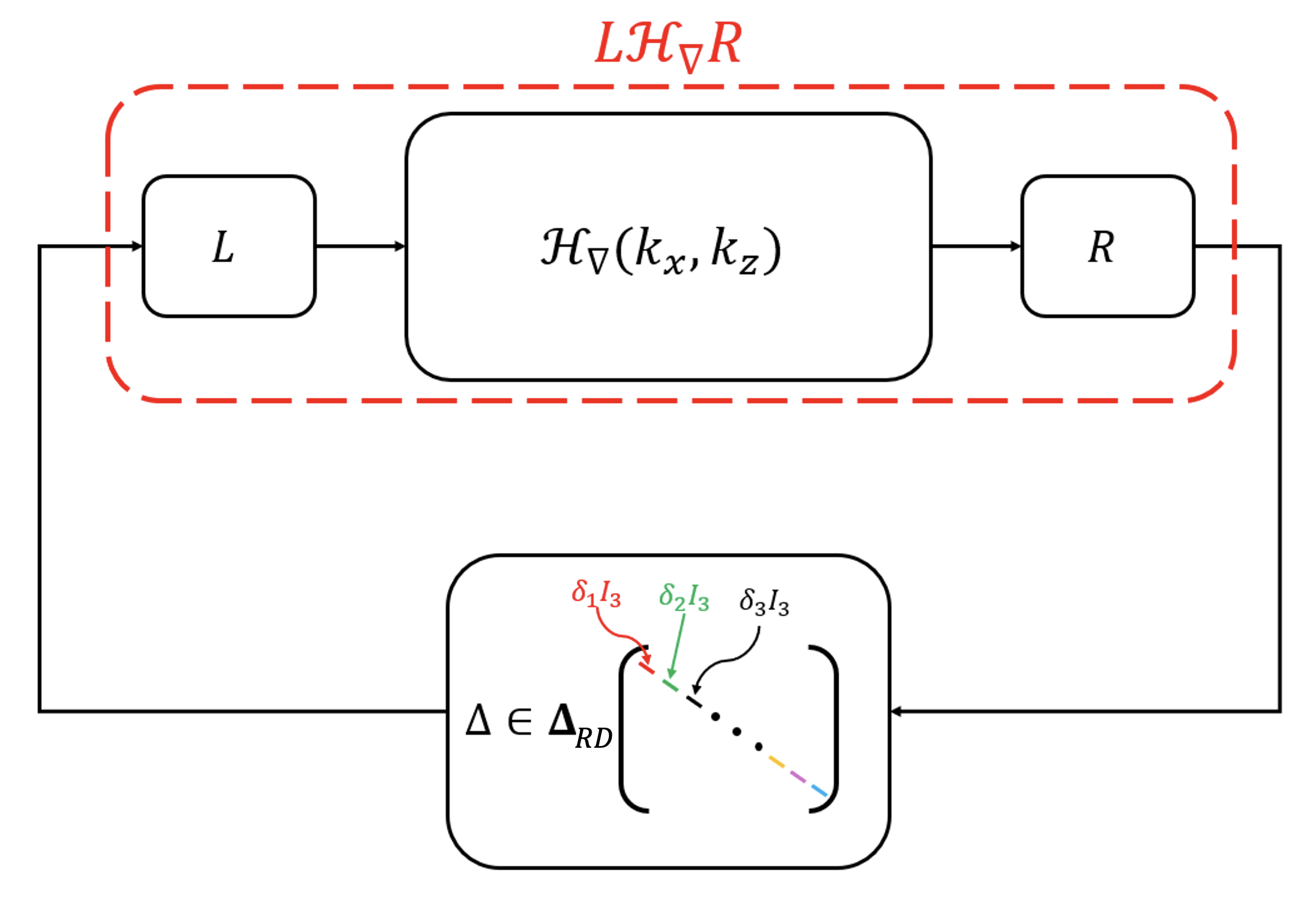}
        \caption{ }
        \label{fig:structure3}
    \end{subfigure}
    
    \caption{Representations of the NSE via: (a)~Interconnection loop block diagram with uncertainty structure $\boldsymbol{\Delta}_{\mathbf{u}}$. (b)~Our modified block diagram with uncertainty structure $\boldsymbol{\Delta}_{RD}$.}
    \label{fig:structure}
\end{figure}

We model the nonlinear advection term introduced to the linearized system as an external forcing with the following form: $\boldsymbol{f}={\mathbf{u}}_\Xi \begin{bmatrix} \nabla^T u & \nabla^T v & \nabla^T w \end{bmatrix}^T$, where $\mathbf{u}_\Xi$ is a fixed matrix feedback gain that has a certain structure in the interconnected system shown in Fig.~\ref{fig:structure1}. Structure constraints limit feedback pathways to replicate the nonlinearity structure in the NSE, which the SSV framework treats as uncertainty.
In our study, we propose a novel representation of the interconnected system, which is used to model the NSE in Fig.~\ref{fig:structure1}, via modifying the transfer function $\mathscr{H}_\nabla$ to efficiently compute SSVs while preserving the amplification pathways that are associated with the structure in Eq.~\eqref{eq:structexact}. This process is described in Fig.~\ref{fig:structure3},  where we apply the transformation represented by the matrices $L$ and $R$ to the frequency operator $\mathscr{H}_\nabla$, such that $\textbf{U}_\Xi \in \boldsymbol{\Delta}_{\mathbf{u}}$ and $\textbf{U}_\Xi = L \textbf{V}_\Xi R$, where $\textbf{V}_\Xi$ is a new uncertainty matrix satisfying $\textbf{V}_\Xi \in \boldsymbol{\Delta}_{RD}$ and the details on the two matrices $L$ and $R$ are provided in Appendix~\ref{sec:A}. This process is somewhat similar to the transformation used in \cite{luiACC23}. In this previous work, a matrix transformation is used to convert the uncertainty structure from $\mathbf{\Delta}_{NRB}$ into a more detailed, 9-block structure. This transformation was applied to allow for an analysis of the block-structured uncertainty as three separate velocity components, rather than each block representing the entire velocity field. In the present work, we apply a more complex transformation that achieves a similar ability to analyze separate velocity components using the structured uncertainty. However, the main contribution of the process we propose is the ability to tractably compute SSVs without imposing additional uncertainty terms on top of those found in $\boldsymbol{\Delta}_{\mathbf{u}}$, keeping only the terms that correspond to the structure of the nonlinear advection term in the NSE system.
In detail: we start with the transfer function $\mathscr{H}_\nabla=\mathscr{C}_\nabla(i\omega I - \mathscr{A})^{-1}\mathscr{B}$, where $\mathscr{C}_\nabla=\text{diag} (\nabla,\nabla,\nabla)\mathscr{C}$. Next, we apply the transformation to get $R\mathscr{H}_\nabla L=R\mathscr{C}_\nabla(i\omega I - \mathscr{A})^{-1}\mathscr{B}L$. By defining $\widetilde{\mathscr{H}}_\nabla = R\mathscr{H}_\nabla L$, $\widetilde{\mathscr{C}}_\nabla = R\mathscr{C}_\nabla$, and $\widetilde{\mathscr{B}} = \mathscr{B} L$, we obtain the following form: $\widetilde{\mathscr{H}}_\nabla=\widetilde{\mathscr{C}}_\nabla(i\omega I - \mathscr{A})^{-1}\widetilde{\mathscr{B}}$. In this process, only the input and output maps are affected, while the resolvent $(i\omega I - \mathscr{A})^{-1}$ remains unaffected by the transformation. Thus, the transformed transfer function receives the input transformed by matrix  $L$ and returns the output transformed by matrix $R$. The physical interactions of the LNS system, captured by the resolvent, remain unchanged under this transformation.
The resulting interconnected system using this transformation is shown in Fig.~\ref{fig:structure3}. 
The transformation yields a simplified uncertainty structure, $\mathbf{\Delta}_{RD}$, defined in \eqref{eq:RD}. By using this transformation, we obtain the following relation between $\norm{\mathscr{H}_{\nabla}}_{\mu_{\boldsymbol{\Delta}}}$ and  $\norm{R\mathscr{H}_{\nabla}L}_{\mu_{\boldsymbol{\Delta}_{RD}}}$   (for details see Appendix~\ref{sec:A}),  
\begin{equation}
\label{eq:size_bound}
    \norm{R\mathscr{H}_{\nabla}L}_{\mu_{\boldsymbol{\Delta}_{RD}}}^{-1}\leq\norm{\mathscr{H}_{\nabla}}_{\mu_{\boldsymbol{\Delta_{\mathbf{u}}}}}^{-1} \leq\sqrt{3}\norm{R\mathscr{H}_{\nabla}L}_{\mu_{\boldsymbol{\Delta}_{RD}}}^{-1}, 
\end{equation}
where $\norm{R\mathscr{H}_{\nabla}L}_{\mu_{\boldsymbol{\Delta_{RD}}}}$ is computationally feasible to obtain. 
We use this result to determine the stability threshold on velocity perturbation magnitude, based on Eq.~\eqref{eq:stability_structured}
\begin{equation}
\label{eq:vel_bounds}
    \max_{y\in\mathcal{Y}}\norm{\mathbf{u}(y)}_2  < \sqrt{3}\norm{R\mathscr{H}_{\nabla}L}_{\mu_{\boldsymbol{\Delta_{RD}}}}^{-1}. 
\end{equation}
The relation in Eq.~\eqref{eq:vel_bounds} is suggested to be used for stability analysis within the structured {small-gain} theorem framework in  \cite{frank2026stability}.  Crucially, combining Eq.~\eqref{eq:hierarchy} and Eq.~\eqref{eq:size_bound} leads to the following hierarchal relationship between stability bounds:
\begin{equation}
    \label{eq:hirarchy2}
    \norm{\mathscr{H}_{\nabla}}_{\mu_{\mathbf{\Delta}_{NRB}}}^{-1} \leq \norm{\mathscr{H}_{\nabla}}_{\mu_{\mathbf{\Delta}_{RB}}}^{-1} \leq
\norm{\mathscr{H}_{\nabla}}_{\mu_{\boldsymbol{\Delta_{\mathbf{u}}}}}^{-1} \leq \sqrt{3}\norm{R\mathscr{H}_{\nabla}L}_{\mu_{\boldsymbol{\Delta}_{RD}}}^{-1}.
\end{equation}

Thus, both the repeated and non-repeated block uncertainty structures under-predict the stability threshold imposed by the structure associated with the nonlinear advection term. Using these uncertainty structures yields overly conservative, inaccurate stability bounds (a more detailed discussion of the stability thresholds produced by block uncertainties appears in \cite{frank2026stability}). In contrast, the bound we propose in this work, $\sqrt{3}\norm{R\mathscr{H}_{\nabla}L}_{\mu_{\boldsymbol{\Delta}_{RD}}}^{-1}$, is found on the other side of the inequality, meaning that it will never under-predict the stability threshold, eliminating all conservatism. As shown in \eqref{eq:size_bound}, which derivation is detailed in Appendix~\ref{sec:A}, the bound we suggest is tight up to a factor of $\sqrt{3}$, meaning that it will predict stability thresholds which are at most $\sqrt{3}$ times the threshold set by $\norm{\mathscr{H}_{\nabla}}_{\mu_{\boldsymbol{\Delta_{\mathbf{u}}}}}^{-1}$.

We note that the relations in Eq.~\eqref{eq:delta_subset2}, along with the monotonicity of the SSV with respect to set inclusion \citep{scherer2001theory}, lead to the result
\begin{equation}
\label{eq:vel_bounds2}
    \sqrt{3}\norm{R\mathscr{H}_{\nabla}L}_{\mu_{\boldsymbol{\Delta_{NRD}}}}^{-1} \leq \sqrt{3}\norm{R\mathscr{H}_{\nabla}L}_{\mu_{\boldsymbol{\Delta_{RD}}}}^{-1}.
\end{equation}

To assess the performance of the four considered uncertainty structures ($\mathbf{\Delta}_{RB},\mathbf{\Delta}_{NRB},\mathbf{\Delta}_{RD}$, and $\mathbf{\Delta}_{NRD}$), we propose a measure based on artificial energy production arising from violations of the divergence-free velocity-field assumption in the structured input-output formulation, as discussed next in \S~\ref{sec:production}.

\section{ARTIFICIAL ENERGY PRODUCTION}
\label{sec:production}
In the derivation of the Reynolds–Orr energy equation, the convective nonlinear term contribution drops, showing the role of this term as a redistributor of the kinetic energy among the modes, but contributing nothing to the net budget  \cite{schmid2002stability}. Thus, nonlinear evolution modifies the disturbance shape over time, but the linearized equations determine the instantaneous amplification mechanism \cite{Henningson1996}. Therefore, in the structured input-output formulation, the nonlinearity in the structured feedback should only redistribute energy rather than producing it. Yet, as we show next, modeling the NSE nonlinearity with constant structured uncertainty without enforcing a divergence-free constraint on the system leads to artificial energy production that is not part of the NS dynamics. This artificial energy production is represented by the term, as detailed in Appendix~\ref{sec:app_b}:
\begin{equation}
    \label{eq:P_xi_body}
    \mathcal{P}_{\Xi} =
    \frac{1}{2}
    \int
    \left[
        u^2 (\nabla\cdot {U_\xi}_1)
        +
        v^2 (\nabla\cdot {U_\xi}_2)
        +
        w^2 (\nabla\cdot {U_\xi}_3)
    \right]
    \,\mathrm{d}V,
\end{equation}
where ${U_\xi}_i$ are comprised of sub-blocks of the uncertainty matrix as follows,
\begin{equation}
    \label{eq:uncertainty_vec_main}
    \begin{cases}
        i=1:\:{U_\xi}_1 = \begin{bmatrix}
        {U_\xi}_{1,1} & {U_\xi}_{1,2} & {U_\xi}_{1,3}
    \end{bmatrix} \\
        i=2:\:{U_\xi}_2 = \begin{bmatrix}
        {U_\xi}_{2,1} & {U_\xi}_{2,2} & {U_\xi}_{2,3}
    \end{bmatrix} \\
        i=3:\:{U_\xi}_3 = \begin{bmatrix}
        {U_\xi}_{3,1} & {U_\xi}_{3,2} & {U_\xi}_{3,3}
    \end{bmatrix}
    \end{cases}.
\end{equation}

Applying the assumption of spatial invariance in the $x$ and $z$ directions as in~\S\ref{sec:math}, we can write
\begin{equation}
    \label{eq:div_Uxi}
    \nabla\cdot {U_{\xi}}_i = ik_x{U_{\xi}}_{i,1} + \frac{\partial}{\partial y}{U_{\xi}}_{i,2} + ik_z{U_{\xi}}_{i,3}.
\end{equation}

To quantify the relative contributions of the three velocity components, we 
compute their energy amplification using the linearized dynamics. As was shown by \cite{jovanovic2005componentwise}, $\mathcal{H}_2$-norm like quantities such as $\norm{\mathscr{H}_u}$, $\norm{\mathscr{H}_v}$, and $\norm{\mathscr{H}_w}$ can be computed using $\text{trace}(\mathscr{C}_u \mathscr{X}_u \mathscr{C}_u^*)$, $\text{trace}(\mathscr{C}_v \mathscr{X}_v \mathscr{C}_v^*)$, and $\text{trace}(\mathscr{C}_w \mathscr{X}_w \mathscr{C}_w^*)$, respectively. Here $\mathscr{X}_i$ are controllability Gramians \citep{Zhou1995-dl}. These $\mathcal{H}_2$-norm-like quantities represent the energy/variance amplification of different velocity components. The diagonal entries of $\mathscr{C}_u \mathscr{X}_u \mathscr{C}_u^*$, $\mathscr{C}_v \mathscr{X}_v \mathscr{C}_v^*$, and $\mathscr{C}_w \mathscr{X}_w \mathscr{C}_w^*$ represent the variance of velocity components at individual $y$ positions due to excitation by the input \citep{singh2006computing}. Thus, we define the amplification from the prescribed forcing to the squared magnitude of each velocity component at a given wall-normal location as:

\begin{subequations}
\label{eq:G}
\begin{align}
    G_{u^2} = \begin{bmatrix}
        (\mathscr{C}_u \mathscr{X}_u \mathscr{C}_u^*)_{1,1} &\dots &(\mathscr{C}_u \mathscr{X}_u \mathscr{C}_u^*)_{N_y,N_y}
    \end{bmatrix}^T,
\end{align} 
\begin{align}
    G_{v^2} = \begin{bmatrix}
        (\mathscr{C}_v \mathscr{X}_v \mathscr{C}_v^*)_{1,1} &\dots &(\mathscr{C}_v \mathscr{X}_v \mathscr{C}_v^*)_{N_y,N_y}
    \end{bmatrix}^T,
\end{align} 
\begin{align}
    G_{w^2} = \begin{bmatrix}
        (\mathscr{C}_w \mathscr{X}_w \mathscr{C}_w^*)_{1,1} &\dots &(\mathscr{C}_w \mathscr{X}_w \mathscr{C}_w^*)_{N_y,N_y}
    \end{bmatrix}^T.
\end{align}
\end{subequations}
The artificial energy production associated with the structured uncertainty matrix, $U_{\Delta}\in\mathbf{\Delta}$, is obtained by first, replacing the squared velocity components in Eq.~\eqref{eq:P_xi_body} with the corresponding $G_{u^{2}}$, $G_{v^{2}}$, and $G_{w^{2}}$ from Eq.~\eqref{eq:G}. Second, using Eq.~\eqref{eq:div_Uxi} for the terms with the divergence operator in Eq.~\eqref{eq:P_xi_body}. Applying these steps leads to the following expression:
\begin{subequations}    
\begin{align}
&\begin{aligned}
\label{eq:P_D}
\mathcal{P}_{\mathbf{\Delta}}
    = \int \mathcal{E}_{\mathbf{\Delta}}dy,
\end{aligned} \\
&\begin{aligned}   
\label{eq:E_delta}
\mathcal{E}_{\mathbf{\Delta}} =\frac{1}{2}
        [&G_{u^{2}}(ik_x{U_{\Delta}}_{1,1} + D_y{U_{\Delta}}_{1,2} + ik_z{U_{\Delta}}_{1,3})
        + \\ &G_{v^{2}}(ik_x{U_{\Delta}}_{2,1} + D_y{U_{\Delta}}_{2,2} + ik_z{U_{\Delta}}_{2,3})
        + \\ &G_{w^{2}}(ik_x{U_{\Delta}}_{3,1} + D_y{U_{\Delta}}_{3,2} + ik_z{U_{\Delta}}_{3,3})].
\end{aligned}
\end{align}
\end{subequations}
Here, $\mathcal{E}_{\mathbf{\Delta}}$ represents the energy production profile in the $y$ direction, while $\mathcal{P}_{\mathbf{\Delta}}$ is the integrated quantity that measures how much artificial energy was produced.
Our goal is to understand which uncertainty structure yields the largest artificial energy production for a given uncertainty size. Thus, we normalize $\mathcal{P}_{\mathbf{\Delta}}$ to obtain a measure of artificial energy production that is not based on uncertainty size, i.e., 
\begin{equation}    
\widetilde{\mathcal{P}}_{\mathbf{\Delta}} = \frac{{\mathcal{P}}_{\mathbf{\Delta}}}{\norm{\mathscr{H}}_2^2\norm{{U_{\Delta}}}_2}, \quad \widetilde{\mathcal{E}}_{\mathbf{\Delta}} = \frac{{\mathcal{E}}_{\mathbf{\Delta}}}{\norm{\mathscr{H}}_2^2\norm{{U_{\Delta}}}_2}.
\end{equation}

Here, $\norm{\mathscr{H}}_2$ is the $\mathcal{H}_2$ norm of the transfer function $\mathscr{H}$, defined in Eq.~\eqref{eq:H1x3}. The $\mathcal{H}_2$ norm is computed similarly to that in  \cite{jovanovic2005componentwise}.  $\norm{{U_{\Delta}}}_2$ is the 2-norm of the structured uncertainty matrix $U_\Delta$, defined as $\norm{{U_{\Delta}}}_2=\bar{\sigma}({U_{\Delta}})$.
For the uncertainty structures considered in this work ($\mathbf{\Delta}_{NRB}$, $\mathbf{\Delta}_{RB}$, $\mathbf{\Delta}_{NRD}$, and $\mathbf{\Delta}_{RD}$) we compute $\widetilde{\mathcal{E}}_{\mathbf{\Delta}}$ and use the notation $\widetilde{\mathcal{E}}_{\mathbf{\Delta}_{NRB}}$, $\widetilde{\mathcal{E}}_{\mathbf{\Delta}_{RB}}$, $\widetilde{\mathcal{E}}_{\mathbf{\Delta}_{NRD}}$, and $\widetilde{\mathcal{E}}_{\mathbf{\Delta}_{RD}}$ respectively. To compute these quantities, we use the structured uncertainties that are detailed in~\S\ref{sec:uncertainty}. This lets us compare the artificial energy production resulting from the worst-case structured uncertainties obtained by solving the optimization problem in Eq.~\eqref{eq:SSV}.

Next, we utilize the new structure $\mathbf{\Delta}_{RD}$ to obtain a threshold on disturbance magnitude in stability analysis via our structured small-gain theorem framework in \ref{sec:stability_results}. Then, we study the differences between different SSV values for the four different uncertainty structures ($\mathbf{\Delta}_{RB},\mathbf{\Delta}_{NRB},\mathbf{\Delta}_{RD}$ and $\mathbf{\Delta}_{NRD}$), and specifically between repeated and non-repeated uncertainty representations in~\S\ref{sec:repeated}.

\section{STABILITY ANALYSIS USING $\mathbf{\Delta}_{RD}$}
\label{sec:stability_results}
In this section, we utilize the suggested bound on the disturbance threshold that provides the least conservative estimate based on $\mathbf{\Delta}_{RD}$ structure, applying it to Couette and plane Poiseuille flows. 
We note that the results utilizing $\mathbf{\Delta}_{NRB}$ that were proposed by \cite{liu2021} and  $\mathbf{\Delta}_{RB}$ that were proposed by \cite{mushtaq2023} are provided in \cite{frank2026stability}, and thus omitted from the discussion here for brevity. However, these uncertainty structures will be used in the following section \S\ref{sec:repeated} focusing on the differences between them, and specifically between repeated and non-repeated uncertainty representations.

Here, we compute SSVs using the matrix set $\mathbf{\Delta}_{RD}$ via the MATLAB function \emph{mussv}. A discretization is applied in the wall-normal direction with $N_y=61$ Chebyshev discretization points.
 These settings were determined to be adequate for obtaining reliable results, as doubling the number of source points did not affect the trends and values of the curves, nor did it cause any noticeable changes in the contour plots presented in the rest of this section.
For the wavenumber domain, we used a $50\times90$ grid of logarithmically spaced values with the following boundaries: $(k_{x_{min}}=10^{-4},k_{x_{max}}=3.02)$ and $(k_{z_{min}}=10^{-2},k_{z_{max}}=15.84)$, which are the same as in \cite{jovanovic2005componentwise}.

First, we examine and analyze here the contour plots of $\sqrt{3}\norm{R\mathscr{H}_{\nabla}L}_{\mu_{\boldsymbol{\Delta_{RD}}}}^{-1}$ over the $k_x,k_z$ domain at specific Reynolds numbers. With this approach, we identify the modes of interest that determine the stability-threshold bound in our analysis. 
Then, we analyze the behavior of three pre-selected modes across a wide range of Reynolds numbers, which we identified as dominant modes based on the results of Fig.~\ref{fig:kxkz_C}. We focus on three specific modes, each represented by a unique wavenumber pair $(k_x, k_z)$.  Specifically, the considered modes are listed in Table~\ref{tbl:modes}. 

\begin{table}[ht!]
\centering
\begin{tabular}{|c|c|c|c|}
\hline
 & OW & TS & SPS \\
\hline
Couette & $(k_x,k_z)=(0.0024,1.0315)$  & $(k_x,k_z)=(0.0029,10^{-6})$ & $(k_x,k_z)=(10^{-4},1.0315)$\\
\hline
Poiseuille & $(k_x,k_z)=(0.0055,0.9496)$  & $(k_x,k_z)=(1.02,10^{-6})$ & $(k_x,k_z)=(10^{-4},1.0315)$\\
\hline
\end{tabular}
\caption{Selected modes: OW = oblique wave, TS = Tollmien--Schlichting, SPS = spanwise periodic streak.}
\label{tbl:modes}
\end{table}

\subsection{Results for Couette flow}
\label{sec:Couette}
\begin{figure}[ht!]
    \centering
    
    \begin{subfigure}[b]{0.48\textwidth} 
        \centering
        \includegraphics[width=\textwidth]{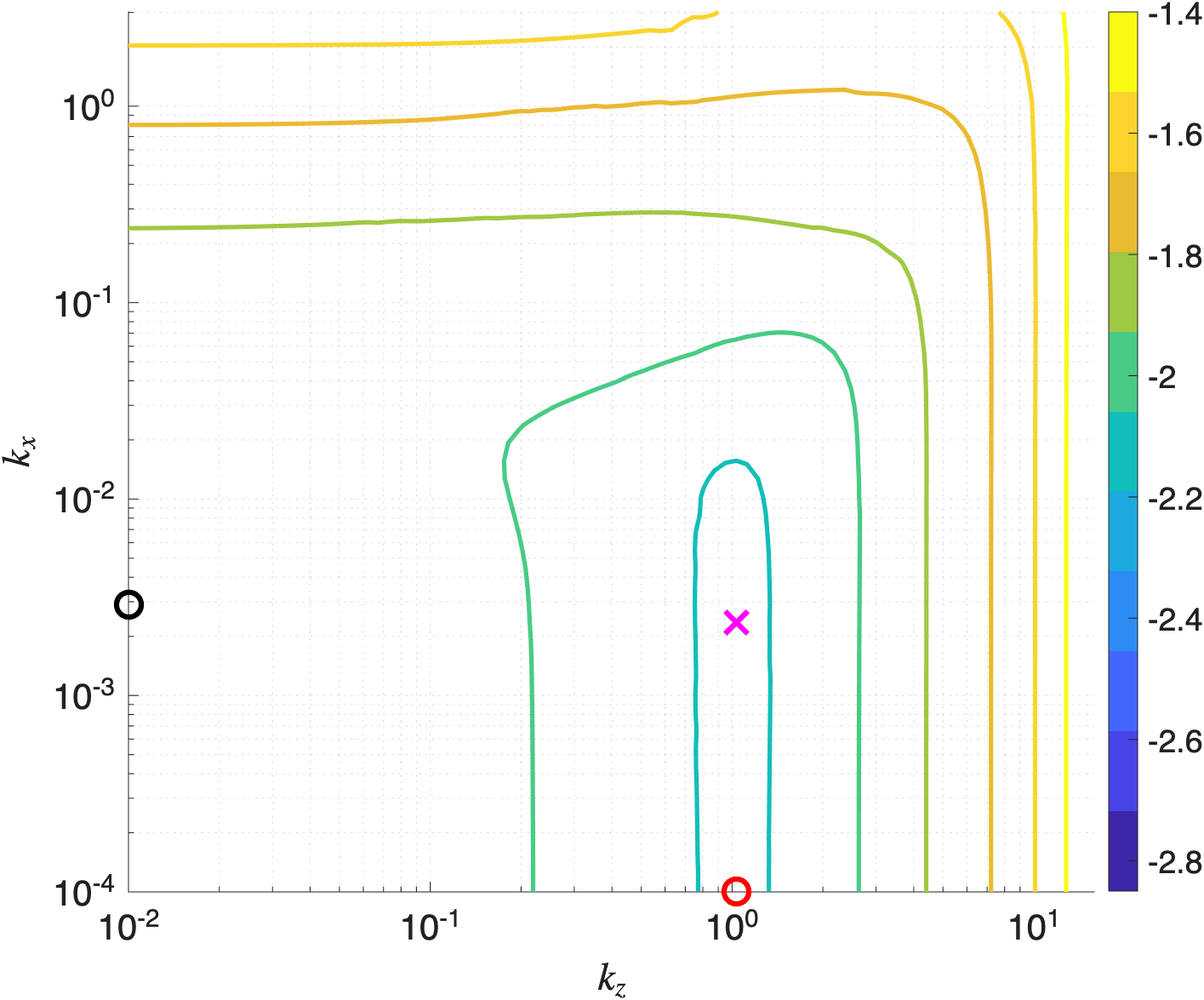}
        \caption{$Re=358$}
        \label{fig:Re358}
    \end{subfigure}\hfill 
    \begin{subfigure}[b]{0.48\textwidth} 
        \centering
        \includegraphics[width=\textwidth]{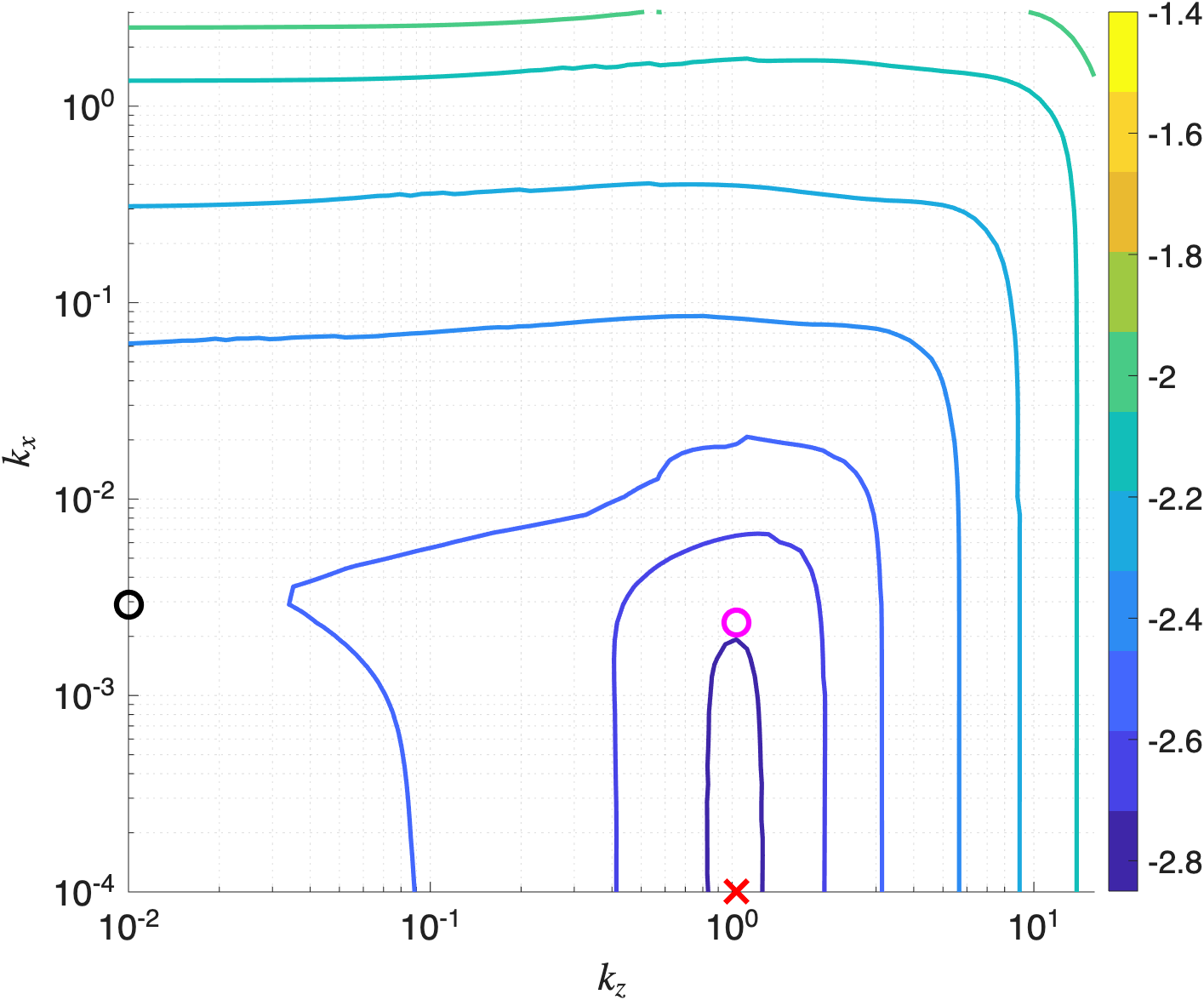}
        \caption{$Re=2000$}
        \label{fig:Re2000}
    \end{subfigure}
        
    \caption{Contour plots (in logarithmic scale) of $\sqrt{3}\norm{R\mathscr{H}_{\nabla}L}_{\mu_{\boldsymbol{\Delta_{RD}}}}^{-1}(k_x,k_z)$ for Couette flow. The X symbol denotes the most dominant mode, whereas the other modes from Table~\ref{tbl:modes} are marked by colored circles (magenta - OW mode, black - TS mode, red - SPS mode).
    }
    \label{fig:kxkz_C}
\end{figure}

In Fig.~\ref{fig:kxkz_C}, we show the  contour maps of $\sqrt{3}\norm{R\mathscr{H}_{\nabla}L}_{\mu_{\boldsymbol{\Delta_{RD}}}}^{-1}(k_x,k_z)$, for the Reynolds numbers $Re=358$ (Fig.~\ref{fig:Re358}) and $Re=2000$ (Fig.~\ref{fig:Re2000}) for Couette flow. 
We overlay the modes from Table~\ref{tbl:modes} over these contour maps. The most dominant mode (associated with the lowest stability threshold) is marked with an X. In Fig.~\ref{fig:Re358}, the smallest perturbation magnitude corresponding to one of the contours is $10^{-2.1}\approx7.9\times 10^{-3}$, while the smallest perturbation magnitude in Fig.~\ref{fig:Re2000} is $10^{-2.85}\approx1.4\times 10^{-3}$. This observation is consistent with the results shown in \cite{frank2026stability}, indicating that Couette flow can become unstable under smaller perturbations as the Reynolds number increases. 
Additionally, the most dominant flow structure---the mode that is associated with the lowest bound on perturbation magnitude---changes with the Reynolds number. For the value $Re=358$, the dominant flow structure is an oblique wave as shown in Fig.~\ref{fig:Re358}. In contrast, Fig.~\ref{fig:Re2000} shows that for higher Reynolds numbers the dominant flow structure changes and becomes a spanwise-periodic streak.

\begin{figure}[ht!]
    \centering

    \begin{tikzpicture}
        \node[anchor=south west,inner sep=0] (image) at (0,0)
            {\includegraphics[width=0.95\linewidth]{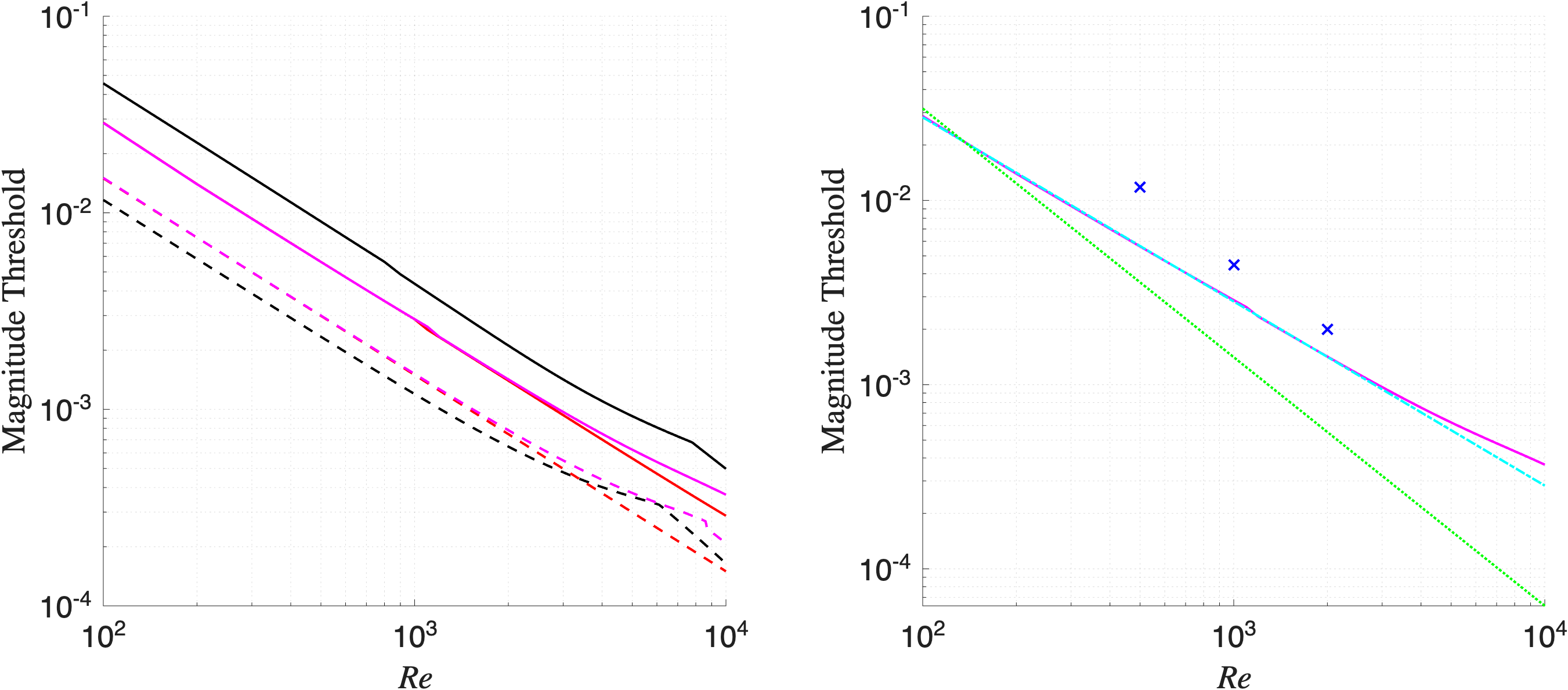}};
    
        \begin{scope}[x={(image.south east)},y={(image.north west)}]
            \node at (0,0.95) {\large\textbf{(a)}};
            \node at (0.53,0.95) {\large\textbf{(b)}};
        \end{scope}
    \end{tikzpicture}
    
    \caption{(a) Evolution curves of imposed perturbation magnitude threshold due to flow structures associated with pre-selected modes of interest denoted in Table~\ref{tbl:modes} (magenta - OW mode, black - TS mode, red - SPS mode) for Couette base flow as a function of the Reynolds number in terms of our new threshold $\sqrt{3}\norm{R\mathscr{H}_{\nabla}L}_{\mu_{\boldsymbol{\Delta_{RD}}}}^{-1}$ (solid curves) and threshold determined by previous approach using repeated block uncertainty $\norm{\mathscr{H}_{\nabla}}_{\mu_{\boldsymbol{\Delta_{RB}}}}^{-1}$  (dashed curves). (b) Comparison between $\sqrt{3}\norm{R\mathscr{H}_{\nabla}L}_{\mu_{\boldsymbol{\Delta_{RD}}}}^{-1}$ (OW mode, magenta solid curve) and results from literature for an oblique mode transition (blue x symbols - \cite{reddy1998stability}), 
    dashed-dotted cyan line - \cite{duguet2010towards}, dotted green line - \cite{duguet2013minimal}).}
    \label{fig:modes_Re_Couette}
\end{figure}

Next, in Fig.~\ref{fig:modes_Re_Couette}(a), we show the evolution of the imposed bounds $\sqrt{3}\norm{R\mathscr{H}_{\nabla}L}_{\mu_{\boldsymbol{\Delta_{RD}}}}^{-1}$ computed via our approach (denoted by solid curves) as a function of Reynolds number over the wide range, $Re \in [10^2,10^4]$ for the pre-selected modes that are listed in Table~\ref{tbl:modes}. We also compute and show on the same plot the bounds that we denote as $\norm{\mathscr{H}_{\nabla}}_{\mu_{\boldsymbol{\Delta_{r}}}}^{-1}$ (via dashed curves) using the repeated blocks uncertainty structure that was suggested by \cite{mushtaq2024structured}.  
This plot yields the following insights.
First, for each respective mode we observe that $\norm{\mathscr{H}_{\nabla}}_{\mu_{\boldsymbol{\Delta_{RB}}}}^{-1} < \sqrt{3}\norm{R\mathscr{H}_{\nabla}L}_{\mu_{\boldsymbol{\Delta_{RD}}}}^{-1}$. This result is to be expected, as we suggest a bound aimed to be closer to the  $\norm{\mathscr{H}_{\nabla}}_{\mu_{\boldsymbol{\Delta_{\mathbf{u}}}}}^{-1}$ bound compared to bounds obtained via the repeated-block approach.
Therefore, a less conservative uncertainty structure leads to less conservative stability estimates, as explained in \cite{frank2026stability}. 

Second, the lowest threshold is set by different modes, depending on the Reynolds number. For $Re \lesssim 1000$,
the thresholds associated with OW and SPS modes barely differ, with the OW mode, corresponding to an oblique flow structure, producing slightly lower stability thresholds. As the Reynolds number increases, a spanwise-periodic streaky structure (SPS mode) sets the lowest threshold. These observations are consistent with the results in Fig.~\ref{fig:kxkz_C}, and demonstrate that the instability mechanism is not governed by one flow structure.

Fig.~\ref{fig:modes_Re_Couette}(b) shows a comparison between the stability threshold obtained via $\sqrt{3}\norm{R\mathscr{H}_{\nabla}L}_{\mu_{\boldsymbol{\Delta_{RD}}}}^{-1}$ for the OW mode from Table~\ref{tbl:modes}, which imposes lowest threshold value, and simulation results from the works of \cite{reddy1998stability,duguet2010towards,duguet2013minimal}. These works compute energy thresholds for transition to turbulence using high-fidelity simulations and various optimization methods. These thresholds were transformed from representing disturbance energy to representing disturbance magnitude using the relation $A_c = \sqrt{2E_c}$ (i.e., $E_c=A_c^2/2$, as the factor of $1/2$ was applied in all of the works we compared our results), where $E_c$ represents disturbance energy and $A_c$ represents disturbance magnitude. 
The threshold predicted by our analysis falls within the range imposed by the considered simulation results. In particular, our threshold is very close to the result of \cite{duguet2010towards}. While both \cite{reddy1998stability} and \cite{duguet2010towards} consider specific oblique wave scenarios, \cite{duguet2013minimal} employs a nonlinear optimization method to consider a more general transition scenario, leading to lower critical energy predictions. This might explain the difference between our results and \cite{duguet2013minimal}, as in the current work we also focus on a specific oblique mode transition scenario, imposed by the OW mode in Table~\ref{tbl:modes}. Crucially, it is clear that the threshold imposed by  $\sqrt{3}\norm{R\mathscr{H}_{\nabla}L}_{\mu_{\boldsymbol{\Delta_{RD}}}}^{-1}$ matches simulation results far better than the one imposed by $\norm{\mathscr{H}_{\nabla}}_{\mu_{\boldsymbol{\Delta_{RB}}}}^{-1}$, strengthening the notion that eliminating the conservatism accompanying block-structure uncertainties results in more accurate stability bounds.

\subsection{Results for plane Poiseuille flow}
\label{sec:Poiseuille}

\begin{figure}[ht!]
    \centering
    
    \begin{subfigure}[b]{0.48\textwidth} 
        \centering
        \includegraphics[width=\textwidth]{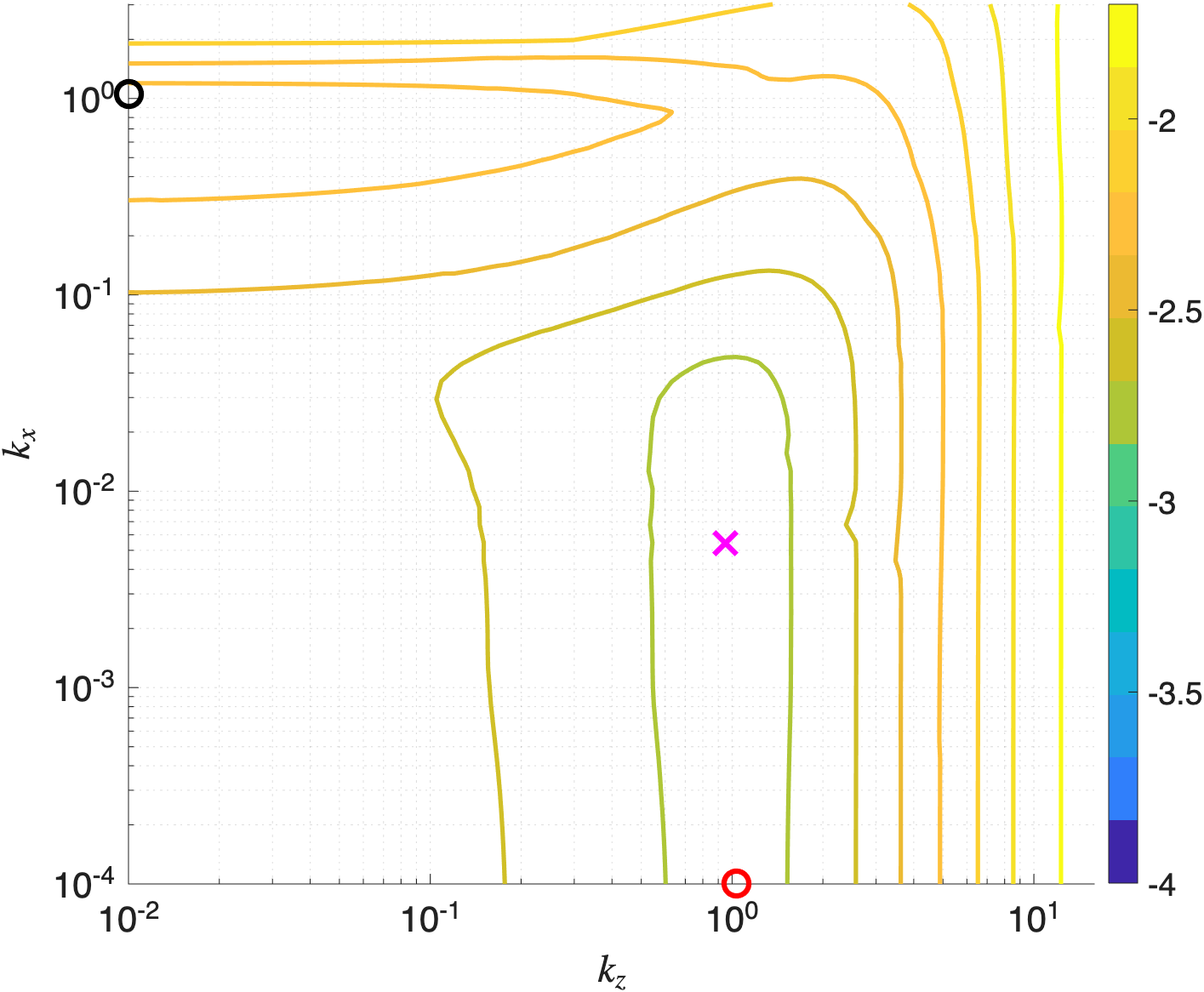}
        \caption{$Re=690$}
        \label{fig:Re690}
    \end{subfigure}\hfill
    \begin{subfigure}[b]{0.48\textwidth} 
        \centering
        \includegraphics[width=\textwidth]{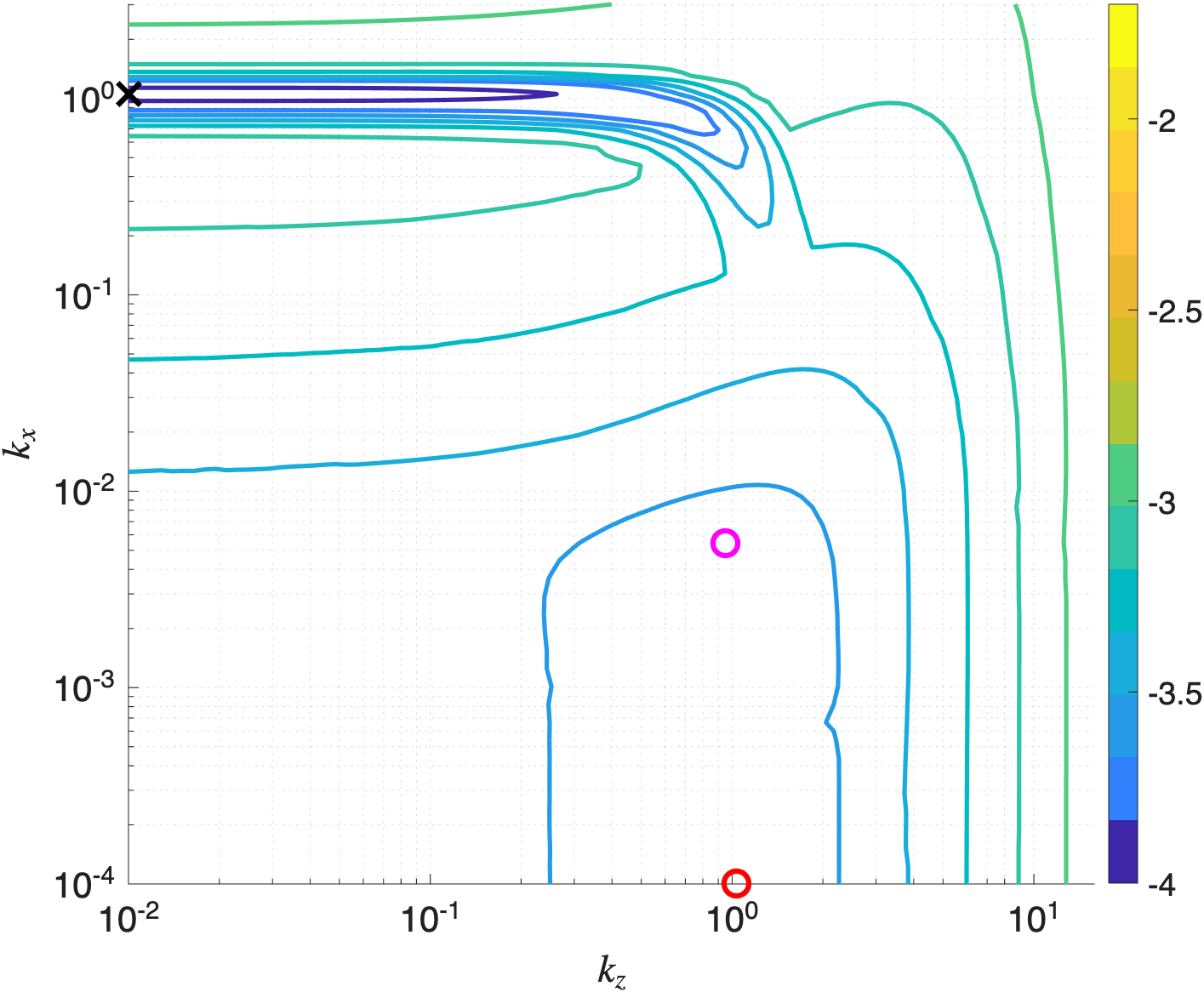}
        \caption{$Re=5700$}
        \label{fig:Re5700}
    \end{subfigure}
        
    \caption{Contour plots (in logarithmic scale) of $\sqrt{3}\norm{R\mathscr{H}_{\nabla}L}_{\mu_{\boldsymbol{\Delta_{RD}}}}^{-1}(k_x,k_z)$ for plane Poiseuille flow. The X symbol denotes the most dominant mode, whereas the other modes from Table~\ref{tbl:modes} are marked by colored circles (magenta - OW mode, black - TS mode, red - SPS mode).
    }
    \label{fig:kxkz_P}
\end{figure}

Fig.~\ref{fig:kxkz_P} shows the contour maps of $\sqrt{3}\norm{R\mathscr{H}_{\nabla}L}_{\mu_{\boldsymbol{\Delta_{RD}}}}^{-1}(k_x,k_z)$, for the Reynolds numbers $Re=690$ (Fig.~\ref{fig:Re690}) and $Re=5700$ (Fig.~\ref{fig:Re5700}) for plane Poiseuille flow. 
Over these contour maps, we overlay the modes from Table~\ref{tbl:modes}. The most dominant mode (associated with the lowest stability threshold) is marked with an X. In Fig.~\ref{fig:Re690}, the smallest perturbation magnitude corresponding to one of the contours is $10^{-2.35}\approx4.5\times 10^{-3}$, while the smallest perturbation magnitude in Fig.~\ref{fig:Re5700} is $1\times 10^{-4}$. This observation is consistent with the results shown in \cite{frank2026stability}, indicating that plane Poiseuille flow can become unstable under smaller perturbations as the Reynolds number increases.
We see once again that the most dominant flow structure---the mode that is associated with the lowest bound on perturbation magnitude---changes with the Reynolds number. For Reynolds numbers significantly lower than $Re_c=5772$, the critical value predicted by LST \citep{schmid2002stability}, the dominant flow structure is an oblique wave, as shown in Fig.~\ref{fig:Re690}. In contrast, Fig.~\ref{fig:Re5700} shows that closer to $Re_c=5772$, the dominant flow structure changes and becomes the TS mode. This result is consistent with observations in \cite{frank2026stability} for other uncertainty structures, showing that near the critical Reynolds number, the TS mode predicted by LST imposes the lowest stability threshold.

\begin{figure}[ht!]
    \centering

    \begin{tikzpicture}
        \node[anchor=south west,inner sep=0] (image) at (0,0)
            {\includegraphics[width=0.95\linewidth]{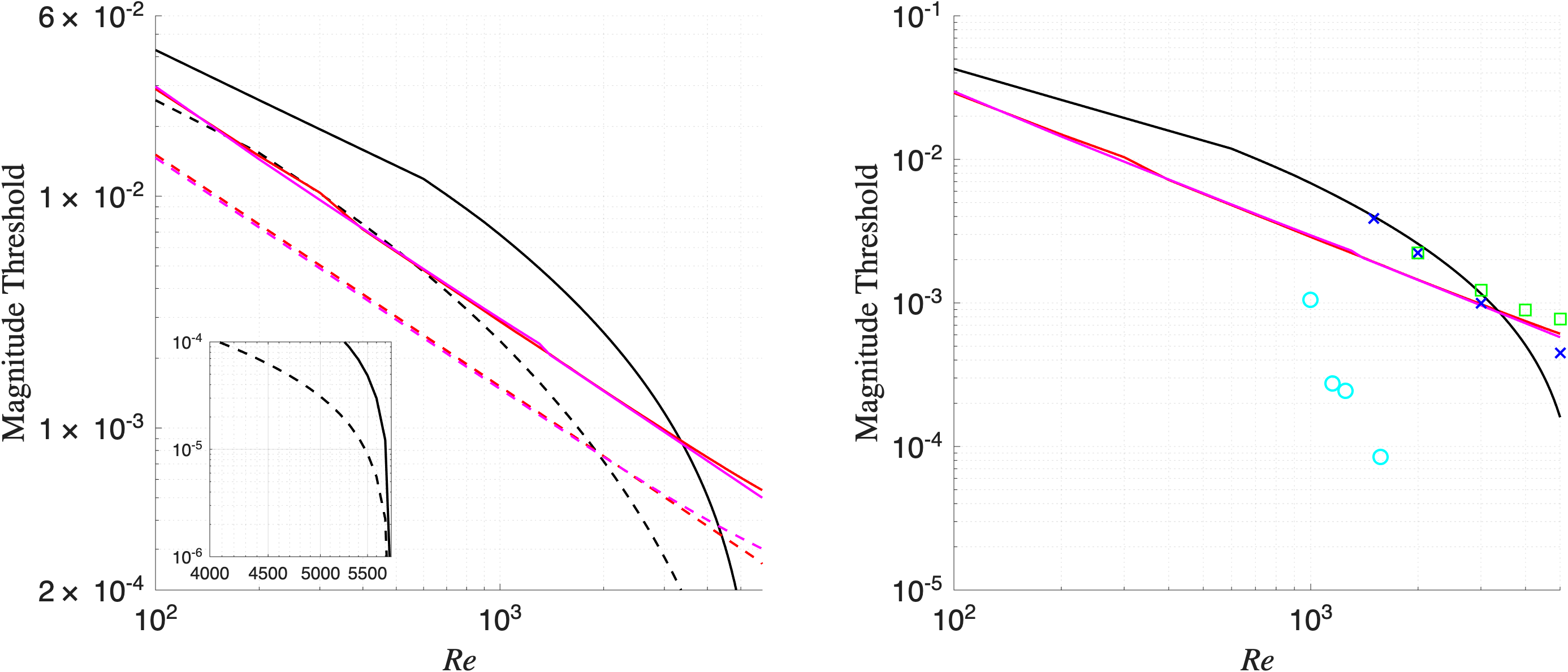}};
    
        \begin{scope}[x={(image.south east)},y={(image.north west)}]
            \node at (0,0.95) {\large\textbf{(a)}};
            \node at (0.53,0.95) {\large\textbf{(b)}};
        \end{scope}
    \end{tikzpicture}

    \caption{(a) Evolution curves of imposed perturbation magnitude threshold due to flow structures associated with pre-selected modes of interest denoted in Table~\ref{tbl:modes} (magenta - OW mode, black - TS mode, red - SPS mode) for Poiseuille base flow as a function of the Reynolds number in terms of our new threshold $\sqrt{3}\norm{R\mathscr{H}_{\nabla}L}_{\mu_{\boldsymbol{\Delta_{RD}}}}^{-1}(k_x,k_z)$  (solid curves) and threshold determined by previous approach using repeated block uncertainty $\norm{\mathscr{H}_{\nabla}}_{\mu_{\boldsymbol{\Delta_{RB}}}}^{-1}$  (dashed curves). (b) Comparison between $\sqrt{3}\norm{R\mathscr{H}_{\nabla}L}_{\mu_{\boldsymbol{\Delta_{RD}}}}^{-1}$ and results from literature for an oblique mode transition (blue x symbols - \cite{reddy1998stability} OW scenario), 
    cyan circles - \cite{parente2022minimal}, green squares - \cite{farano2015hairpin}).}
    \label{fig:modes_Re_Poiseuille}
\end{figure}

Next, in Fig.~\ref{fig:modes_Re_Poiseuille}(a), we show the evolution of the {imposed bounds $\sqrt{3}\norm{R\mathscr{H}_{\nabla}L}_{\mu_{\boldsymbol{\Delta_{RD}}}}^{-1}$ computed via our approach} (denoted by solid curves) as a function of Reynolds number over the wide range, $Re \in [100,5700]$ for the pre-selected modes that are listed in Table~\ref{tbl:modes}. We also compute and show on the same plot the bounds that we denote as $\norm{\mathscr{H}_{\nabla}}_{\mu_{\boldsymbol{\Delta_{r}}}}^{-1}$ (via dashed curves) using the repeated blocks uncertainty structure that was suggested by \cite{mushtaq2024structured}.

This plot yields the following insights.
Once again, the inequality $\norm{\mathscr{H}_{\nabla}}_{\mu_{\boldsymbol{\Delta_{RB}}}}^{-1} < \sqrt{3}\norm{R\mathscr{H}_{\nabla}L}_{\mu_{\boldsymbol{\Delta_{RD}}}}^{-1}$ hold true for each respective mode tested. This result further demonstrates the conservative nature of full-block uncertainty structures compared to our approach of constructing the structured uncertainty.

Second, the curve corresponding to the TS mode rapidly decreases from $Re\approx 1000$, approaching very low threshold values (around $10^{-6}$) for $Re=5772$. We treat this result as approaching an infinitesimally small stability threshold, up to numerical precision. Thus, for Reynolds numbers approaching $Re=5772$, the threshold~$\sqrt{3}\norm{R\mathscr{H}_{\nabla}L}_{\mu_{\boldsymbol{\Delta_{RD}}}}^{-1}$ corresponds to the flow being unstable even for asymptotically small perturbations, and thus results which match LST predictions (the least stable TS mode becomes unstable at $Re=5772$ \citep{schmid2002stability}) are to be expected. This result is consistent with our analysis in \cite{frank2026stability} that uses other uncertainty structures.

At low, subcritical Reynolds numbers, the SPS and OW modes govern the lowest threshold. These flow structures impose the lowest threshold for Reynolds numbers below $3500$ in our approach and below $2000$ in the repeated block approach. This means that below these Reynolds numbers, these modes require significantly smaller perturbations to lose stability than the TS mode.
This result is consistent with previous studies of plane Poiseuille flow showing that oblique-wave and streamwise-periodic-streak transition scenarios can require substantially lower disturbance energies than TS-wave transition \citep{reddy1998stability}. Nonlinear optimal-growth studies have also identified three-dimensional, hairpin-like structures as effective finite-amplitude perturbations for triggering transition in plane Poiseuille flow \citep{farano2015hairpin}. These observations are also consistent with minimal-seed studies based on nonlinear variational optimization, which determine the finite-energy perturbations required to trigger sustained turbulence in channel flow \citep{parente2022minimal}.

Fig.~\ref{fig:modes_Re_Poiseuille}(b) directly compares the stability thresholds imposed by $\sqrt{3}\norm{R\mathscr{H}_{\nabla}L}_{\mu_{\boldsymbol{\Delta_{RD}}}}^{-1}$ to the results available from literature of \cite{reddy1998stability,farano2015hairpin,parente2022minimal}. These works compute energy thresholds for transition to turbulence using high-fidelity simulations and various optimization methods. We transformed these energy thresholds from disturbance energy to disturbance magnitude using the previously denoted $A_c$. The thresholds predicted by our analysis fall within the range observed in the simulation results considered. In \cite{parente2022minimal}, the disturbance energy is computed by integrating over the entire computational domain and dividing by its volume. Thus, the computation of $A_c$ based on their results corresponds to the average disturbance magnitude in the flow, rather than the maximal one. This explains why the results of \cite{parente2022minimal} impose significantly lower thresholds than our analysis, since our analysis considers the largest disturbance magnitude in the flow domain, which is naturally higher than the average disturbance magnitude. This discrepancy did not appear in Fig.~\ref{fig:modes_Re_Couette}(b) for Couette flow, as the simulation results considered there all used an oblique transition scenario. In \cite{reddy1998stability}, the authors consider a specific oblique wave transition scenario, which may not be the optimal transition route, perhaps explaining why this work yields a slightly higher threshold than our approach. The work of \cite{farano2015hairpin} specifically computes short-time nonlinear optimal perturbation, suggesting that lower thresholds will emerge when allowing for longer transition time in the optimization process, which might explain why their thresholds are higher than those we predict.

\section{EFFECT OF DIFFERENT UNCERTAINTY STRUCTURES ON ARTIFICIAL ENERGY PRODUCTION}
\label{sec:repeated}
In this section, we compare the repeated and non-repeated uncertainty representations in the resulting flow structures that determine the smallest finite-size threshold in our stability analysis framework. We also examine the artificial energy production term for each uncertainty representation as a performance measure of the considered uncertainty structure to determine whether it is more faithful (closer to the divergence-free measure) within a structured input-output framework.  
In particular, we analyze the quantities $\norm{\mathscr{H}_{\nabla}}_{\mu_{\boldsymbol{\Delta_{NRB}}}}^{-1}$, $\norm{\mathscr{H}_{\nabla}}_{\mu_{\boldsymbol{\Delta_{RB}}}}^{-1}$, $\sqrt{3}\norm{R\mathscr{H}_{\nabla}L}_{\mu_{\boldsymbol{\Delta_{NRD}}}}^{-1}$, and $\sqrt{3}\norm{R\mathscr{H}_{\nabla}L}_{\mu_{\boldsymbol{\Delta_{RD}}}}^{-1}$ to observe the differences between the wave-number maps that show the bounds on velocity perturbation magnitude for Couette and Plane Poiseuille flows.
The quantities $\norm{\mathscr{H}_{\nabla}}_{\mu_{\boldsymbol{\Delta_{NRB}}}}^{-1}$, $\sqrt{3}\norm{R\mathscr{H}_{\nabla}L}_{\mu_{\boldsymbol{\Delta_{NRD}}}}^{-1}$, and $\sqrt{3}\norm{R\mathscr{H}_{\nabla}L}_{\mu_{\boldsymbol{\Delta_{RD}}}}^{-1}$ were computed via the MATLAB function \textit{mussv} while the quantity $\norm{\mathscr{H}_{\nabla}}_{\mu_{\boldsymbol{\Delta_{RB}}}}^{-1}$ was computed using the methodology presented in \cite{mushtaq2023}. The number of points $N_y$ and the wavenumber domain are the same as those used in~\S\ref{sec:stability_results}.

\subsection{Results for Couette flow: artificial energy production}
\label{sec:P_Couette}

\begin{figure}[ht!]
    \centering
    \begin{subfigure}[b]{0.48\textwidth}
        \centering
        \includegraphics[width=\textwidth]{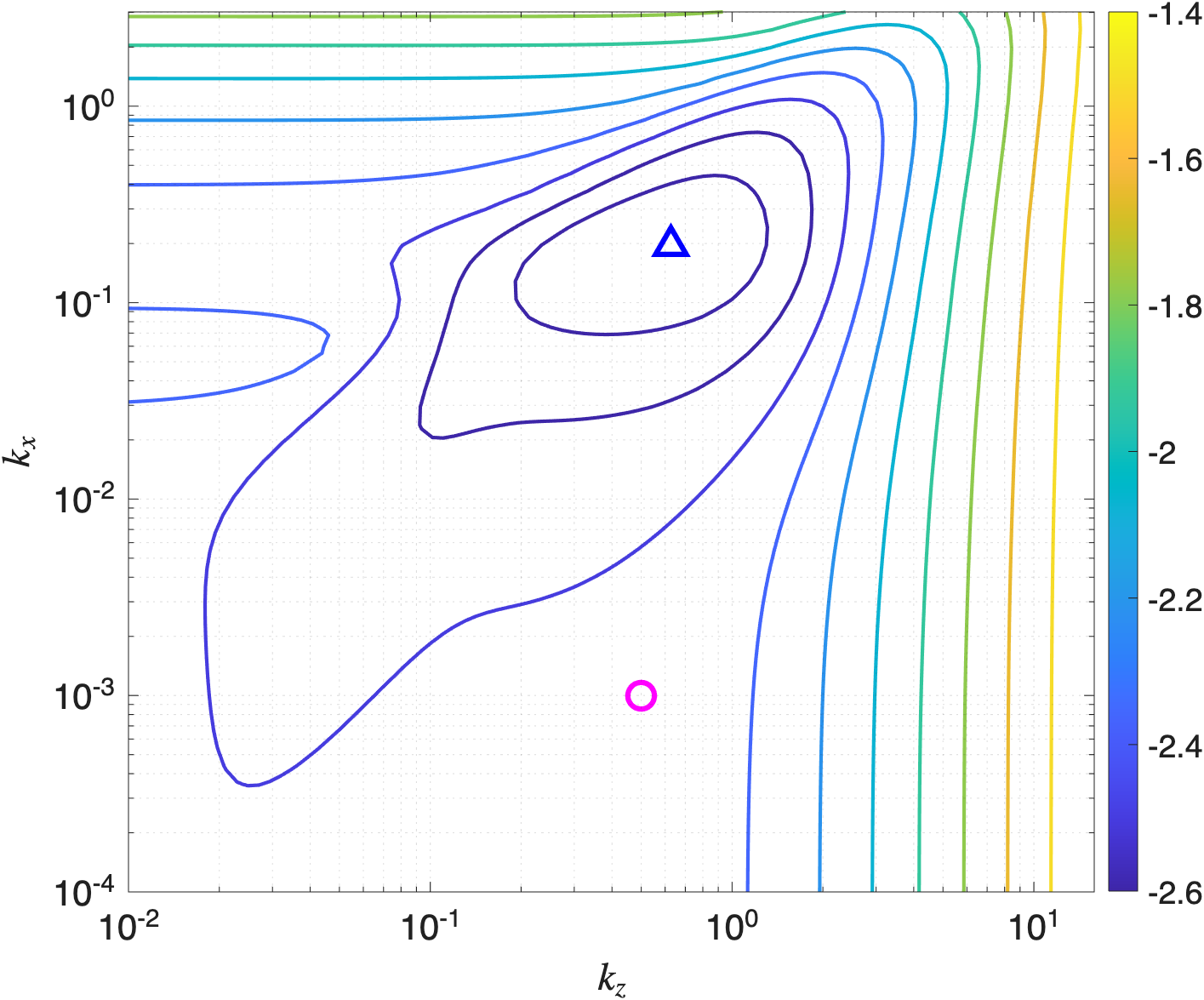}
        \caption{$\norm{\mathscr{H}_{\nabla}}_{\mu_{\boldsymbol{\Delta_{NRB}}}}^{-1}(k_x,k_z)$}
        \label{subfig:couette_NRB}
    \end{subfigure}
    \hfill
    \begin{subfigure}[b]{0.48\textwidth}
        \centering
        \includegraphics[width=\textwidth]{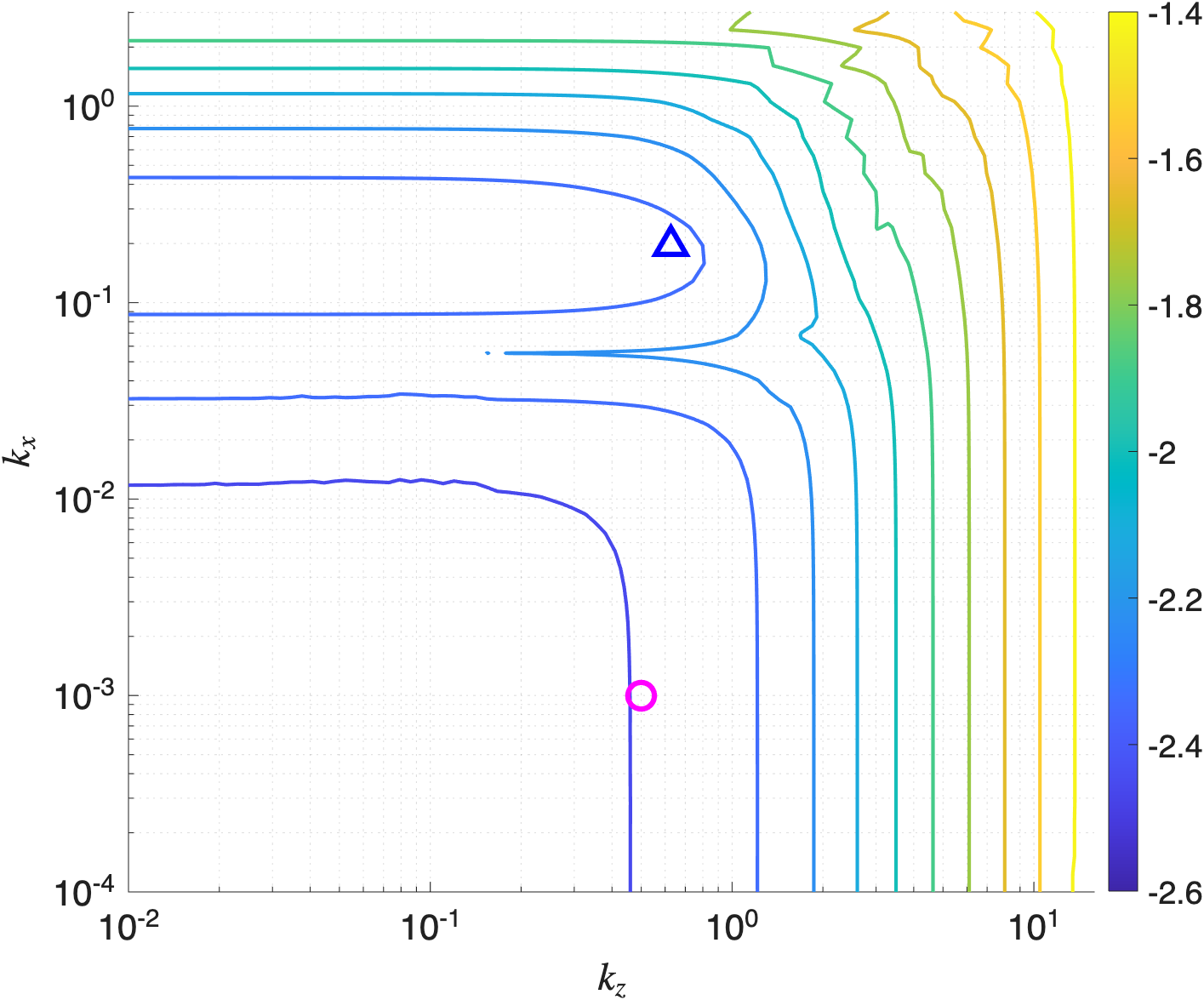}
        \caption{$\norm{\mathscr{H}_{\nabla}}_{\mu_{\boldsymbol{\Delta_{RB}}}}^{-1}(k_x,k_z)$}
        \label{subfig:couette_rb}
    \end{subfigure}
    
    \vspace{0.5cm}
    
    \begin{subfigure}[b]{0.48\textwidth}
        \centering
        \includegraphics[width=\textwidth]{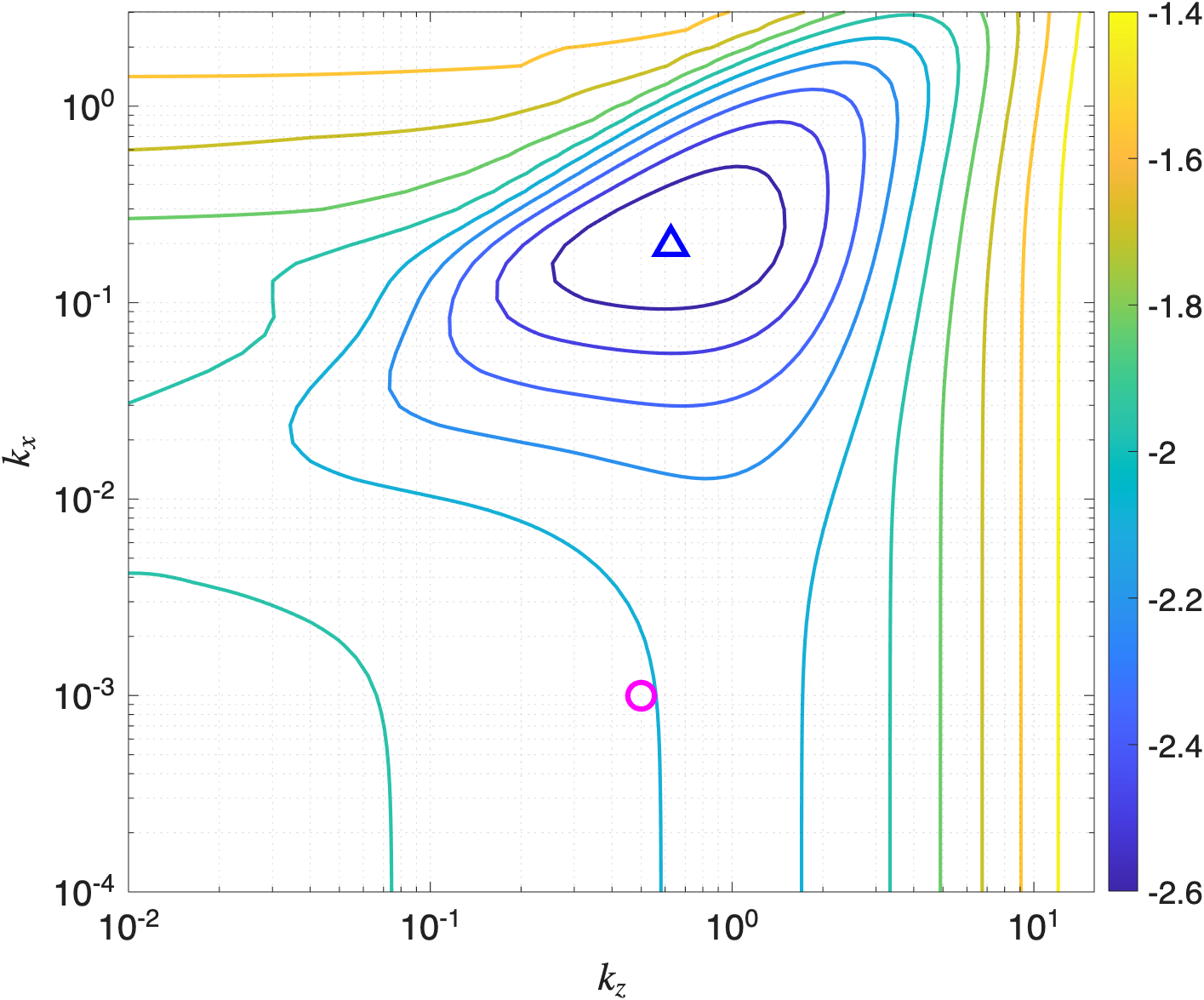}
        \caption{$\sqrt{3}\norm{R\mathscr{H}_{\nabla}L}_{\mu_{\boldsymbol{\Delta_{NRD}}}}^{-1}(k_x,k_z)$}
        \label{subfig:couette_deld}
    \end{subfigure}
    \hfill
    \begin{subfigure}[b]{0.48\textwidth}
        \centering
        \includegraphics[width=\textwidth]{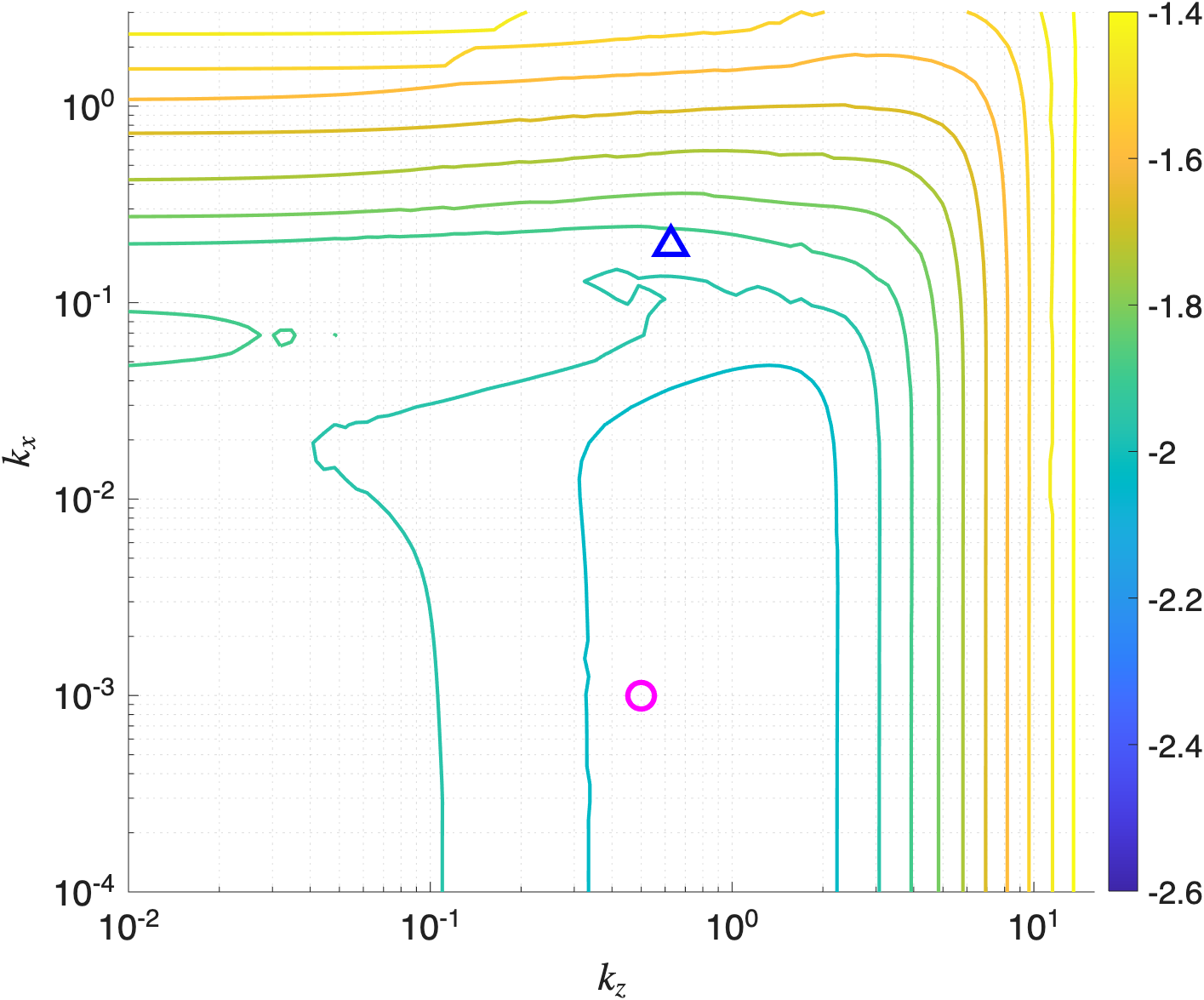}
        \caption{$\sqrt{3}\norm{R\mathscr{H}_{\nabla}L}_{\mu_{\boldsymbol{\Delta_{RD}}}}^{-1}(k_x,k_z)$}
        \label{subfig:couette_delu}
    \end{subfigure}
    
    \caption{Stability analysis of Couette flow at $Re=358$ using the uncertainty structures (a) $\mathbf{\Delta}_{NRB}$, (b) $\mathbf{\Delta}_{RB}$, (c) $\mathbf{\Delta}_{NRD}$, (d) $\mathbf{\Delta}_{RD}$. Blue triangle - $k_x=0.196,k_z=0.628$, magenta circle - $k_x = 10^{-3}, k_z = 0.5$.}
    \label{fig:prf_couette}
\end{figure}

Fig.~\ref{fig:prf_couette} shows contour plots of the four bounds on velocity perturbation magnitude for Couette flow at $Re=358$. 
This Reynolds number was chosen to be the same as in \cite{liu2021,mushtaq2023}, to validate our results of $\norm{\mathscr{H}_{\nabla}}_{\mu_{\boldsymbol{\Delta_{NRB}}}}^{-1}$ and $\norm{\mathscr{H}_{\nabla}}_{\mu_{\boldsymbol{\Delta_{RB}}}}^{-1}$. 
Over these contour plots, we highlight two wavenumber pairs - one where the repeated approaches produce very different results than the non-repeated ones (namely $k_x=0.196, k_z=0.628$), and one where they produce similar results (namely $k_x=10^{-3},k_z=0.5$).

We observe that $\sqrt{3}\norm{R\mathscr{H}_{\nabla}L}_{\mu_{\boldsymbol{\Delta_{RD}}}}^{-1}$ produces the highest and least conservative stability thresholds out of all of the methods that were tested.
Fig.~\ref{fig:prf_couette} shows a distinct region of oblique modes which have significantly lower stability thresholds for the non-repeated quantities ($\norm{\mathscr{H}_{\nabla}}_{\mu_{\boldsymbol{\Delta_{NRB}}}}^{-1}$ and $\sqrt{3}\norm{R\mathscr{H}_{\nabla}L}_{\mu_{\boldsymbol{\Delta_{NRD}}}}^{-1}$) when compared to the repeated quantities ($\norm{\mathscr{H}_{\nabla}}_{\mu_{\boldsymbol{\Delta_{RB}}}}^{-1}$ and $\sqrt{3}\norm{R\mathscr{H}_{\nabla}L}_{\mu_{\boldsymbol{\Delta_{RD}}}}^{-1}$ respectively). Apart from this specific region, the non-repeated quantities yield very similar stability thresholds to the repeated quantities.

The results in Fig.~\ref{fig:prf_couette} comply with the relationships in Eq.~\eqref{eq:hierarchy} and Eq.~\eqref{eq:vel_bounds2}. The region of oblique modes with lower stability thresholds corresponds to strict inequalities between repeated and non-repeated quantities (i.e., $\norm{\mathscr{H}_{\nabla}}_{\mu_{\mathbf{\Delta}_{NRB}}}^{-1} < \norm{\mathscr{H}_{\nabla}}_{\mu_{\mathbf{\Delta}_{RB}}}^{-1}$ and $\sqrt{3}\norm{R\mathscr{H}_{\nabla}L}_{\mu_{\boldsymbol{\Delta_{NRD}}}}^{-1} < \sqrt{3}\norm{R\mathscr{H}_{\nabla}L}_{\mu_{\boldsymbol{\Delta_{RD}}}}^{-1}$). In the rest of the wavenumber domain, we obtain equality between the repeated and corresponding non-repeated quantities (i.e., $\norm{\mathscr{H}_{\nabla}}_{\mu_{\mathbf{\Delta}_{NRB}}}^{-1} = \norm{\mathscr{H}_{\nabla}}_{\mu_{\mathbf{\Delta}_{RB}}}^{-1}$ and $\sqrt{3}\norm{R\mathscr{H}_{\nabla}L}_{\mu_{\boldsymbol{\Delta_{NRD}}}}^{-1} = \sqrt{3}\norm{R\mathscr{H}_{\nabla}L}_{\mu_{\boldsymbol{\Delta_{RD}}}}^{-1}$). As discussed in~\S\ref{sec:stability}, the non-repeated quantity $\sqrt{3}\norm{R\mathscr{H}_{\nabla}L}_{\mu_{\boldsymbol{\Delta_{NRD}}}}^{-1}$ can under-predict the stability threshold while the repeated quantity $\sqrt{3}\norm{R\mathscr{H}_{\nabla}L}_{\mu_{\boldsymbol{\Delta_{RD}}}}^{-1}$ is a strict upper bound. Similarly, based on Eq.~\eqref{eq:hierarchy}, the repeated quantity $\norm{\mathscr{H}_{\nabla}}_{\mu_{\boldsymbol{\Delta_{RB}}}}^{-1}$ is closer to $\norm{\mathscr{H}_{\nabla}}_{\mu_{\boldsymbol{\Delta_{\mathbf{u}}}}}^{-1}$ than $\norm{\mathscr{H}_{\nabla}}_{\mu_{\boldsymbol{\Delta_{NRB}}}}^{-1}$. Thus, we conclude that the region where the non-repeated quantities differ from the repeated quantities is caused by disregarding the constraint of repeated entries as appears in the structure $\boldsymbol{\Delta}_{\mathbf{u}}$. As we will show next, this result is consistent with the fact that non-repeated uncertainty structures cause an increase in artificial energy creation that is not part of the NSE in this region.

Next, we compute the artificial energy production using $N_y=121$ discretization points over the wall-normal direction, chosen to achieve convergence of the $\widetilde{\mathcal{E}}_{\mathbf{\Delta}}$ profiles. We focus on two wavenumber pairs - one where the repeated and non-repeated approaches yield very different stability thresholds, and one where they yield similar stability thresholds. These wavenumber pairs are $(k_x,k_z)=(0.196,0.628)$ and $(k_x,k_z)=(10^{-3},0.5)$, indicated in Fig.~\ref{fig:prf_couette} by the blue triangle and magenta circle, respectively.

\begin{figure}[ht!]
    \centering
    \begin{subfigure}[b]{0.32\textwidth}
        \centering
        \includegraphics[width=\textwidth]{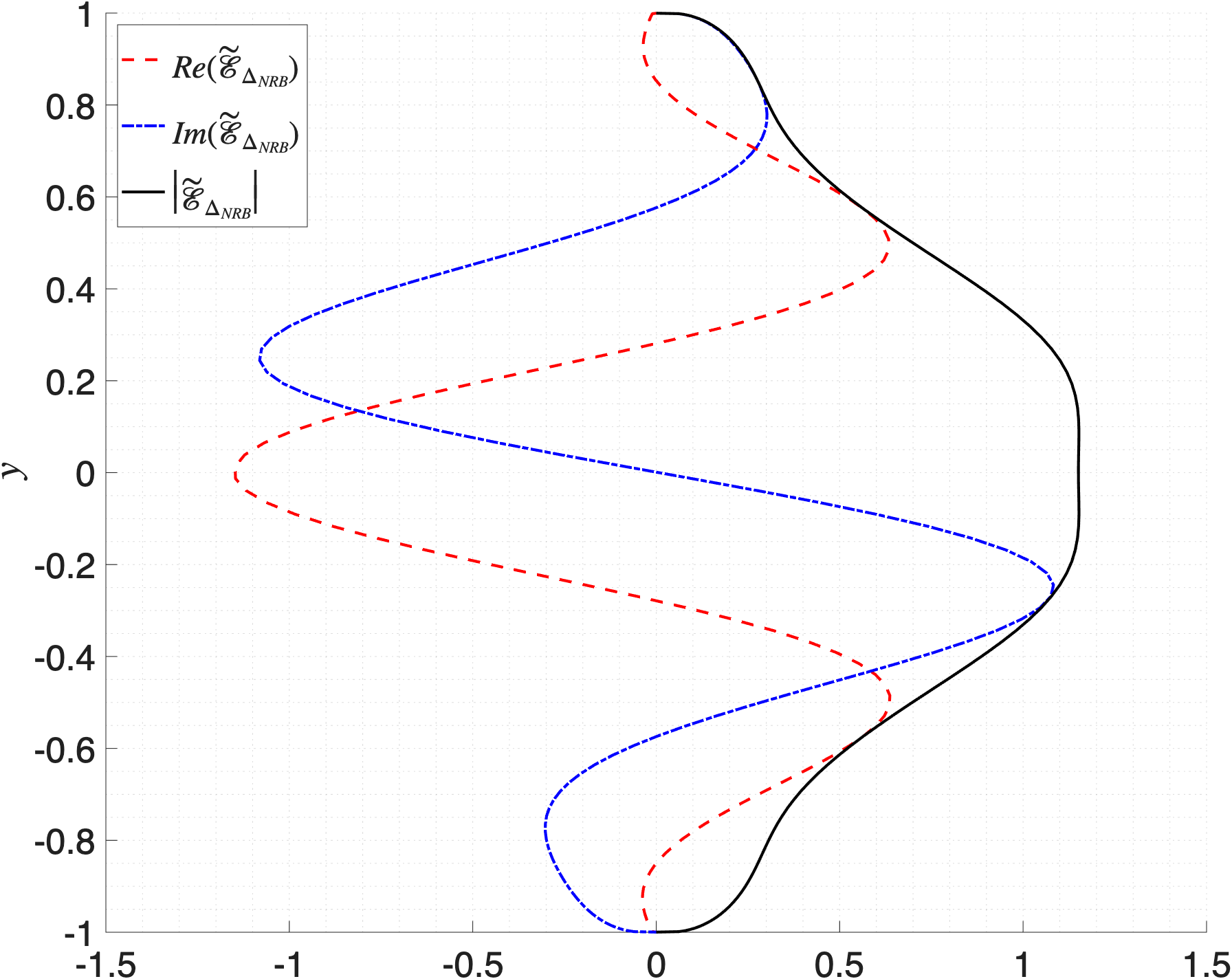}
        \caption{{$\mathcal{E}_{\mathbf{\Delta}_{NRB}}$}}
        \label{subfig:e_NRB_dif}
    \end{subfigure}
    \hfill
    \begin{subfigure}[b]{0.32\textwidth}
        \centering
        \includegraphics[width=\textwidth]{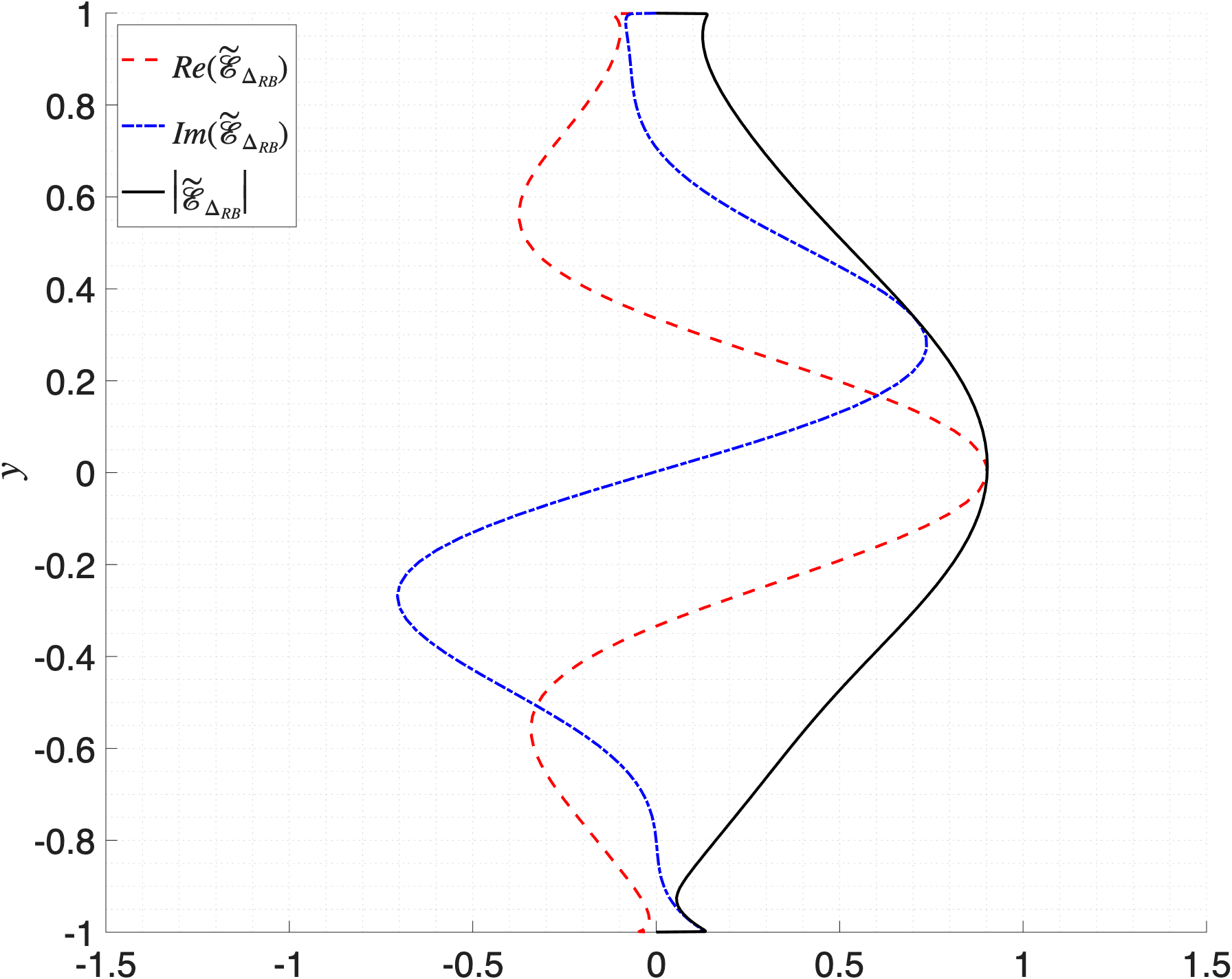}
        \caption{{$\mathcal{E}_{\mathbf{\Delta}_{RB}}$}}
        \label{subfig:e_rb_dif}
    \end{subfigure}
    \hfill
    \begin{subfigure}[b]{0.32\textwidth}
        \centering
        \includegraphics[width=\textwidth]{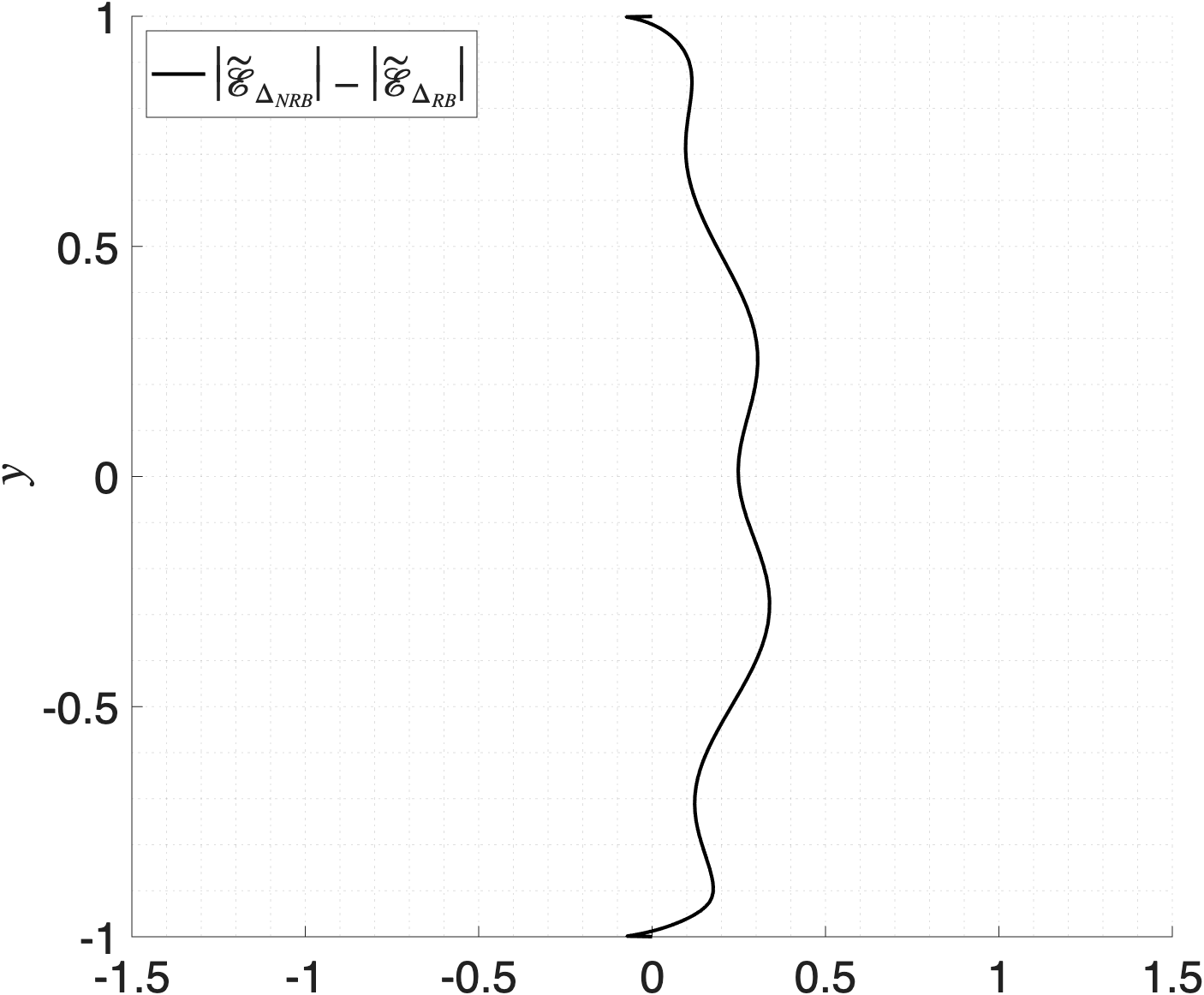} 
\caption{{$|\widetilde{\mathcal{E}}_{\mathbf{\Delta}_{NRB}}| - |\widetilde{\mathcal{E}}_{\mathbf{\Delta}_{RB}}|$}}
        \label{subfig:e_block_dif}
    \end{subfigure}
    
    \vspace{0.5cm}
    
    \begin{subfigure}[b]{0.32\textwidth}
        \centering
        \includegraphics[width=\textwidth]{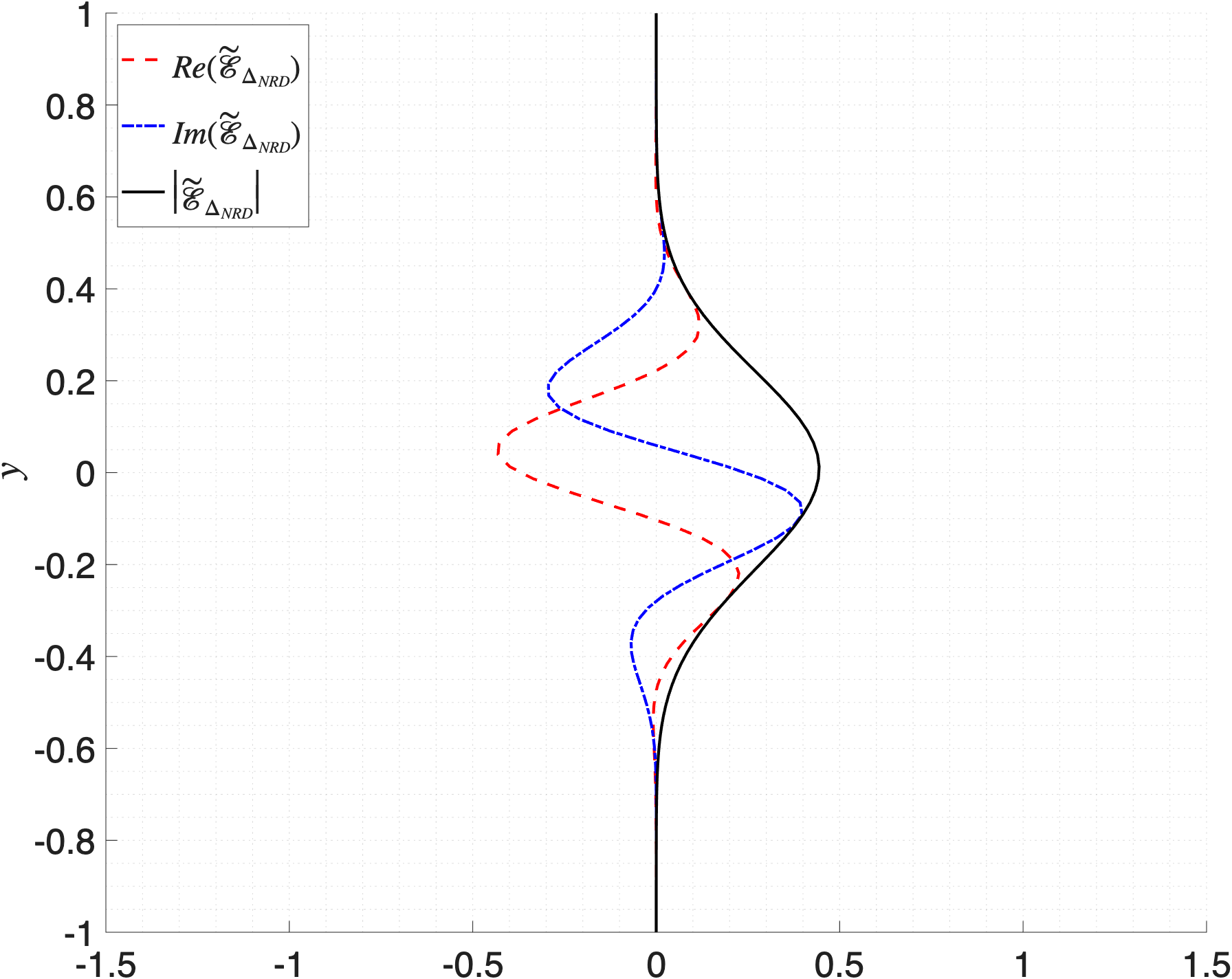}
        \caption{{$\mathcal{E}_{\mathbf{\Delta}_{NRD}}$}}
        \label{subfig:e_deld_dif}
    \end{subfigure}
    \hfill
    \begin{subfigure}[b]{0.32\textwidth}
        \centering
        \includegraphics[width=\textwidth]{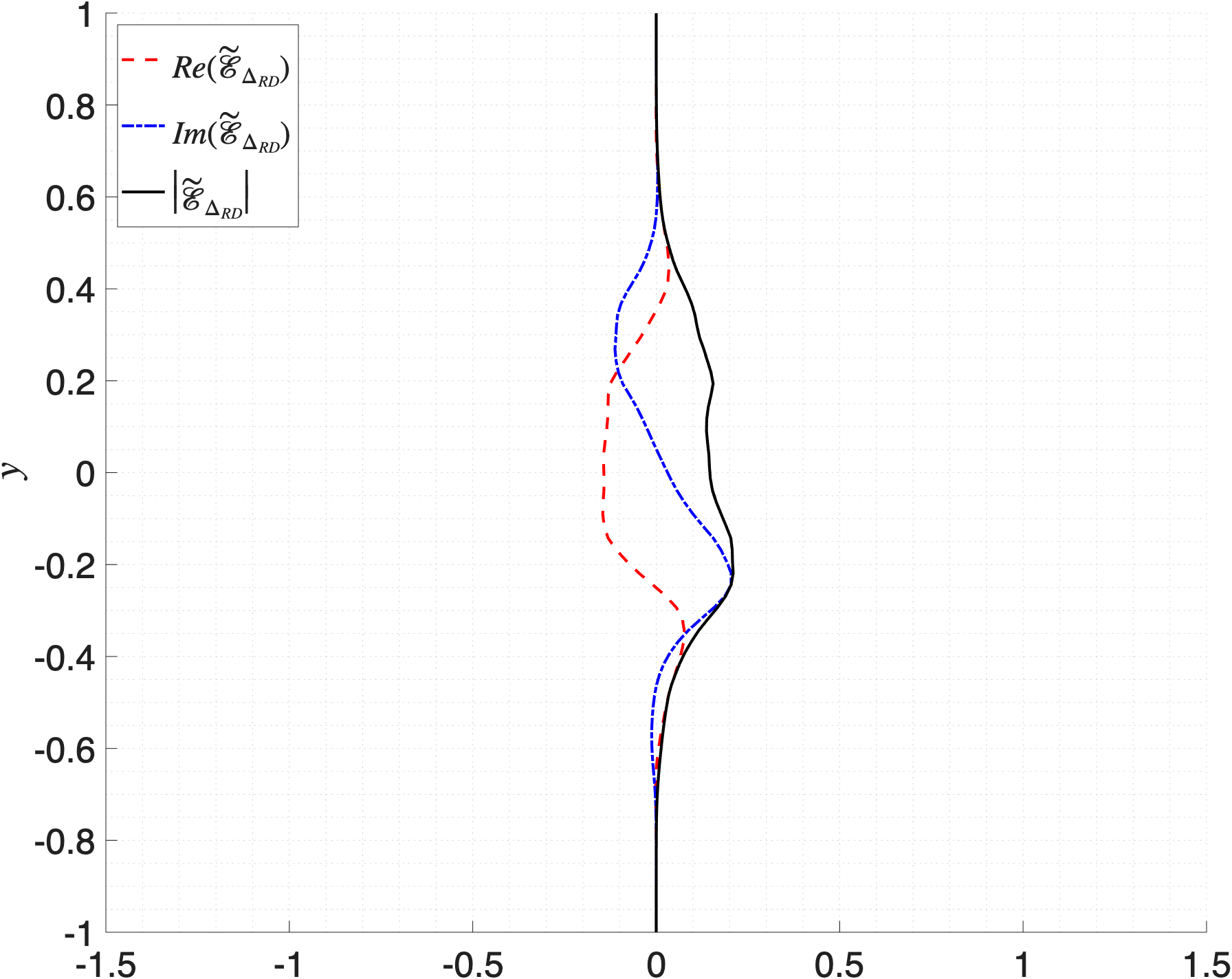}
        \caption{{$\mathcal{E}_{\mathbf{\Delta}_{RD}}$}}
        \label{subfig:e_delu_dif}
    \end{subfigure}
    \hfill
    \begin{subfigure}[b]{0.32\textwidth}
        \centering
        \includegraphics[width=\textwidth]{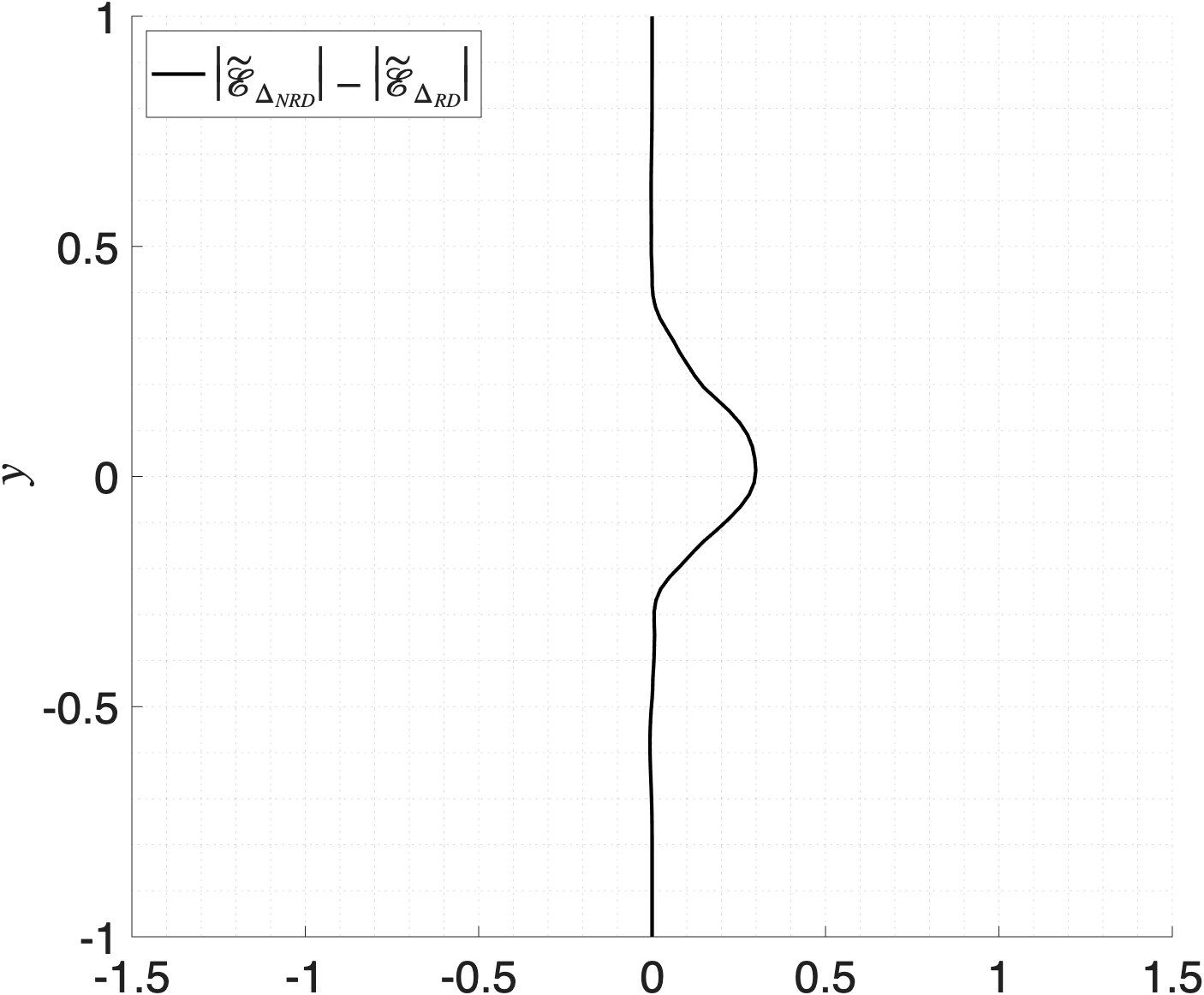} 
        \caption{{$|\widetilde{\mathcal{E}}_{\mathbf{\Delta}_{NRD}}| - |\widetilde{\mathcal{E}}_{\mathbf{\Delta}_{RD}}|$}} 
        \label{subfig:e_diag_dif}
    \end{subfigure}
    
    \caption{$\widetilde{\mathcal{E}}_{\mathbf{\Delta}}$ computed for Couette base flow at $Re=358$, with the wavenumber pair $(k_x,k_z)=(0.196,0.628)$ using 
    the uncertainty structures from~\S\ref{sec:uncertainty} ((a) $\mathbf{\Delta}_{NRB}$, (b) $\mathbf{\Delta}_{RB}$, (d) $\mathbf{\Delta}_{NRD}$, (e) $\mathbf{\Delta}_{RD}$), and the difference in the absolute artificial energy profile between (c) block uncertainties, and (f) diagonal uncertainties.}
    \label{fig:e_dif}
\end{figure}

Fig.~\ref{fig:e_dif} shows $\widetilde{\mathcal{E}}_{\mathbf{\Delta}_{NRB}}$, $\widetilde{\mathcal{E}}_{\mathbf{\Delta}_{RB}}$, $\widetilde{\mathcal{E}}_{\mathbf{\Delta}_{NRD}}$, and $\widetilde{\mathcal{E}}_{\mathbf{\Delta}_{RD}}$, computed for Couette base flow at $Re=358$, using the wavenumber pair $(k_x,k_z)=(0.196,0.628)$, chosen to illustrate the difference between the repeated and non-repeated approaches.

First, the artificial energy production profiles obtained with non-repeated uncertainty structures reach higher values than the corresponding curves for the repeated structure. This observation holds both for the block structures ($\mathbf{\Delta}_{NRB}$ and $\mathbf{\Delta}_{RB}$) and for the diagonal structures ($\mathbf{\Delta}_{NRD}$ and $\mathbf{\Delta}_{RD}$), indicating that enforcing the repeated-entry constraint reduces the artificial energy production. This observation is clearly evident by plotting the differences $|\widetilde{\mathcal{E}}_{\mathbf{\Delta}_{NRB}}| - |\widetilde{\mathcal{E}}_{\mathbf{\Delta}_{RB}}|$ and $|\widetilde{\mathcal{E}}_{\mathbf{\Delta}_{NRD}}| - |\widetilde{\mathcal{E}}_{\mathbf{\Delta}_{RD}}|$, in Fig.~\ref{subfig:e_block_dif} and Fig.~\ref{subfig:e_diag_dif}, respectively. Both plots in Fig.~\ref{subfig:e_block_dif} and Fig.~\ref{subfig:e_diag_dif} illustrate the positive difference caused by enforcing the repeated terms constraint for block and diagonal uncertainties for the entire wall-normal domain, respectively.

Second, there is a significant difference in the values of $\widetilde{\mathcal{E}}_{\mathbf{\Delta}}$ between block and diagonal structures, with diagonal structures yielding significantly lower values. This result indicates that using the methodology proposed here to compute stability thresholds yields lower artificial energy production and better represents the dynamics of the NS system.

In particular, integrating the profiles presented in Fig.~\ref{fig:e_dif} yields the following absolute values (using Eq.~\eqref{eq:P_D}):
\begin{equation}
\begin{split}
    |\widetilde{\mathcal{P}}_{\mathbf{\Delta}_{{NRB}}}|
        &= 1.4018, \qquad
    |\widetilde{\mathcal{P}}_{\mathbf{\Delta}_{{RB}}}|
        = 0.9946,\\
    |\widetilde{\mathcal{P}}_{\mathbf{\Delta}_{{NRD}}}|
        &= 0.23943, \qquad
    |\widetilde{\mathcal{P}}_{\mathbf{\Delta}_{{RD}}}|
        = 0.13511.
\end{split}
\end{equation}
Thus, removing the repeated-block constraint increases the integrated
artificial energy production by approximately $41\%$ for the full-block
structures and by approximately $77\%$ for the diagonal structures. 
The non-repeated uncertainties allow the uncertainty fields associated with the
three momentum equations to vary independently. Consequently, each field can
align separately with the wall-normal regions in which the corresponding
velocity component is strongly amplified. This additional freedom increases
the divergence-like terms in Eq.~\eqref{eq:E_delta}, producing artificial energy that is
not present in the incompressible NSE. The considerably
larger values obtained using the full-block structures also reflect the
additional degrees of freedom permitted by the full uncertainty
blocks, whereas the diagonal structures restrict the uncertainty to diagonal entries only.

\begin{figure}[ht!]
    \centering
    \begin{subfigure}[b]{0.32\textwidth}
        \centering
        \includegraphics[width=\textwidth]{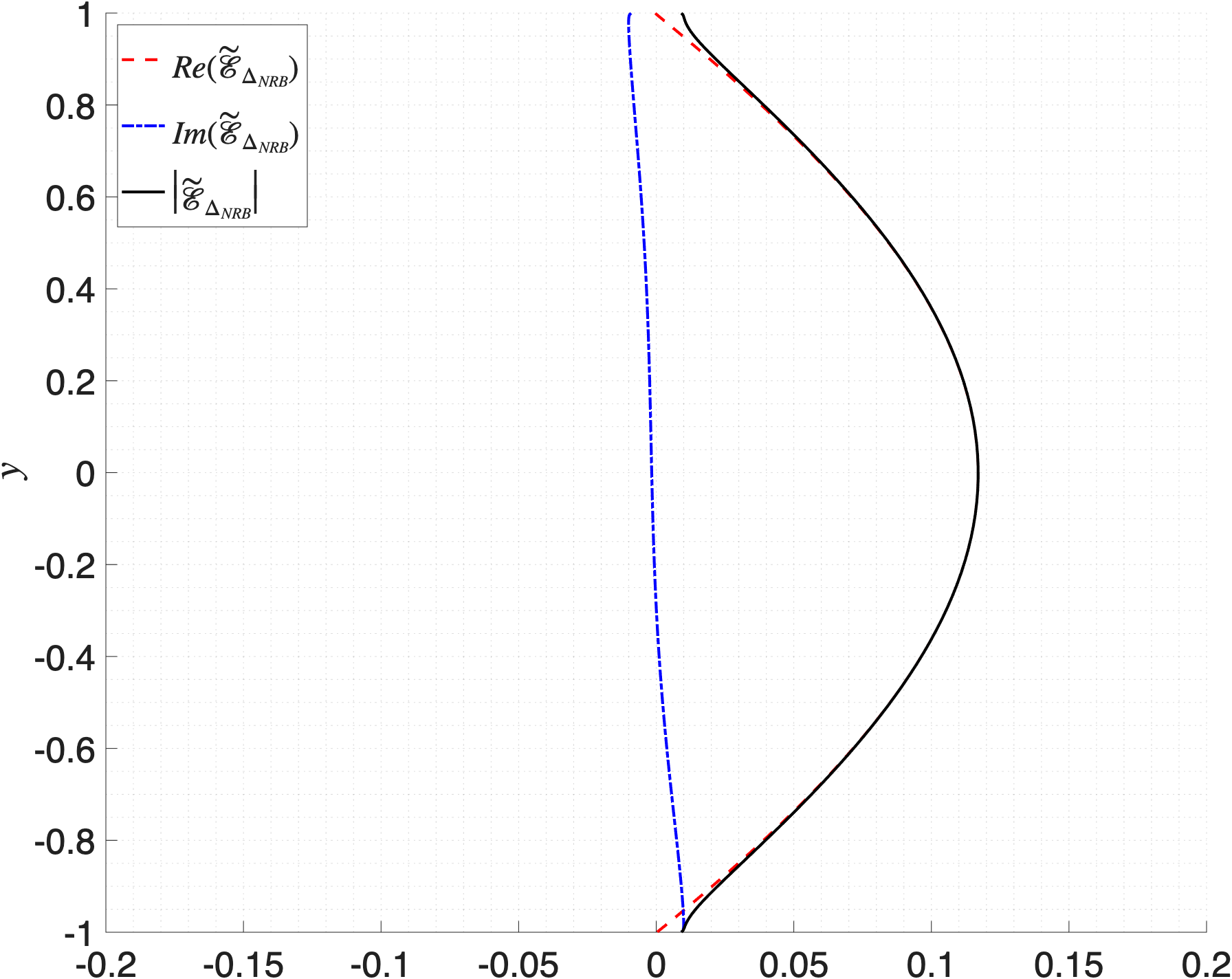}
        \caption{{$\mathcal{E}_{\mathbf{\Delta}_{NRB}}$}}
        \label{subfig:e_NRB_sim}
    \end{subfigure}
    \hfill
    \begin{subfigure}[b]{0.32\textwidth}
        \centering
        \includegraphics[width=\textwidth]{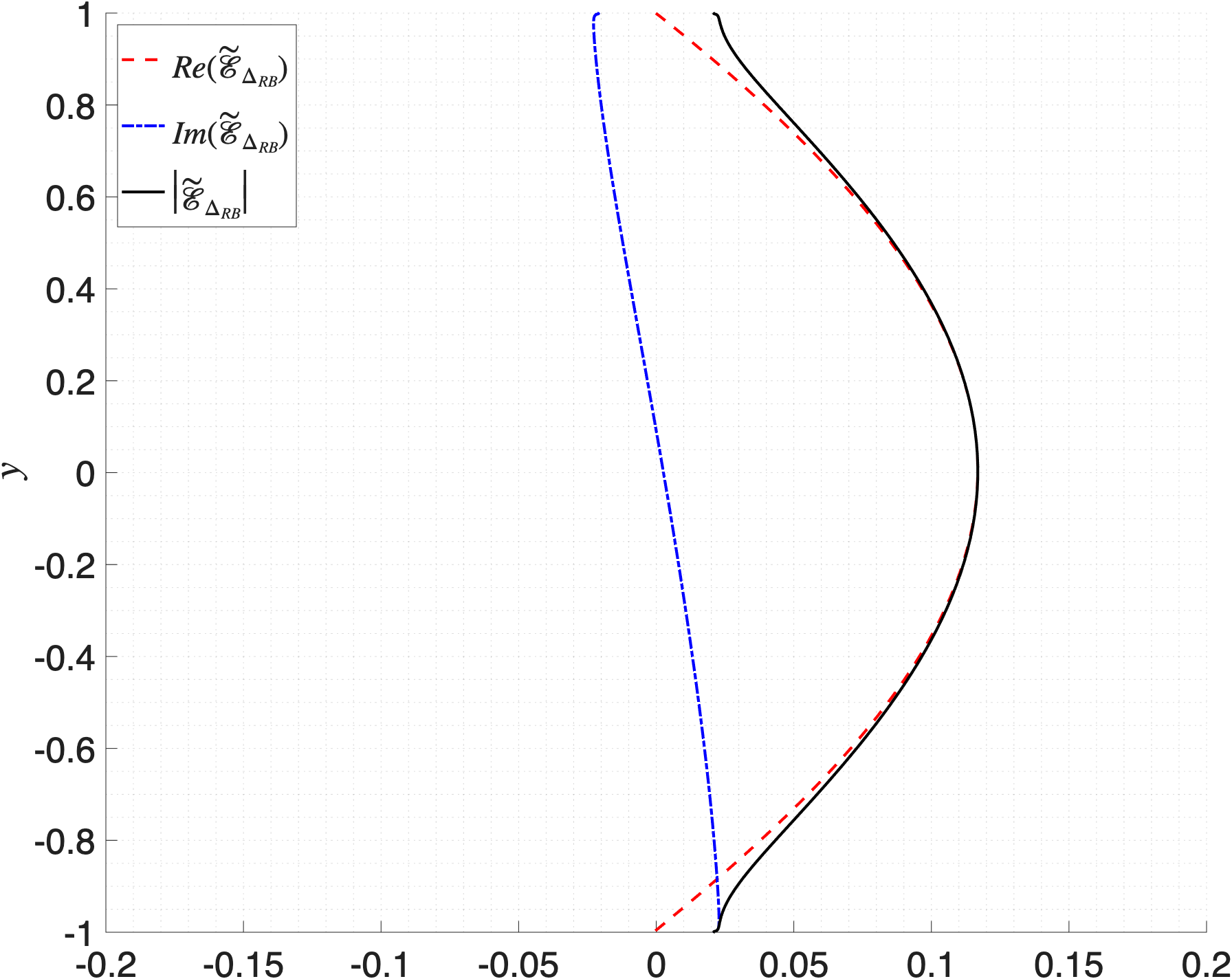}
        \caption{{$\mathcal{E}_{\mathbf{\Delta}_{RB}}$}}
        \label{subfig:e_rb_sim}
    \end{subfigure}
    \hfill
    \begin{subfigure}[b]{0.32\textwidth}
        \centering
        \includegraphics[width=\textwidth]{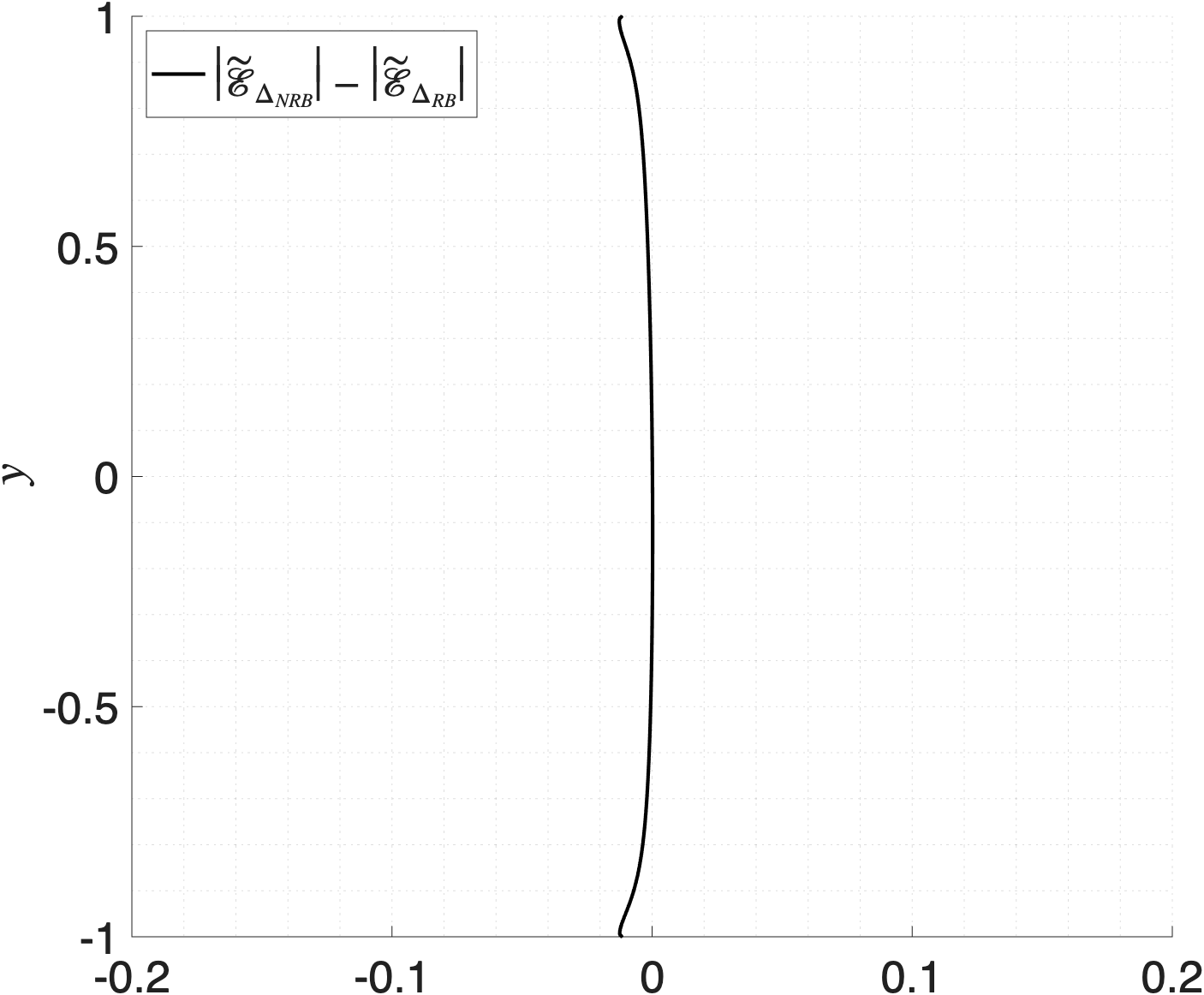} 
        \caption{{$|\widetilde{\mathcal{E}}_{\mathbf{\Delta}_{NRB}}| - |\widetilde{\mathcal{E}}_{\mathbf{\Delta}_{RB}}|$}} 
        \label{subfig:e_block_sim2}
    \end{subfigure}
    
    \vspace{0.5cm}
    
    \begin{subfigure}[b]{0.32\textwidth}
        \centering
        \includegraphics[width=\textwidth]{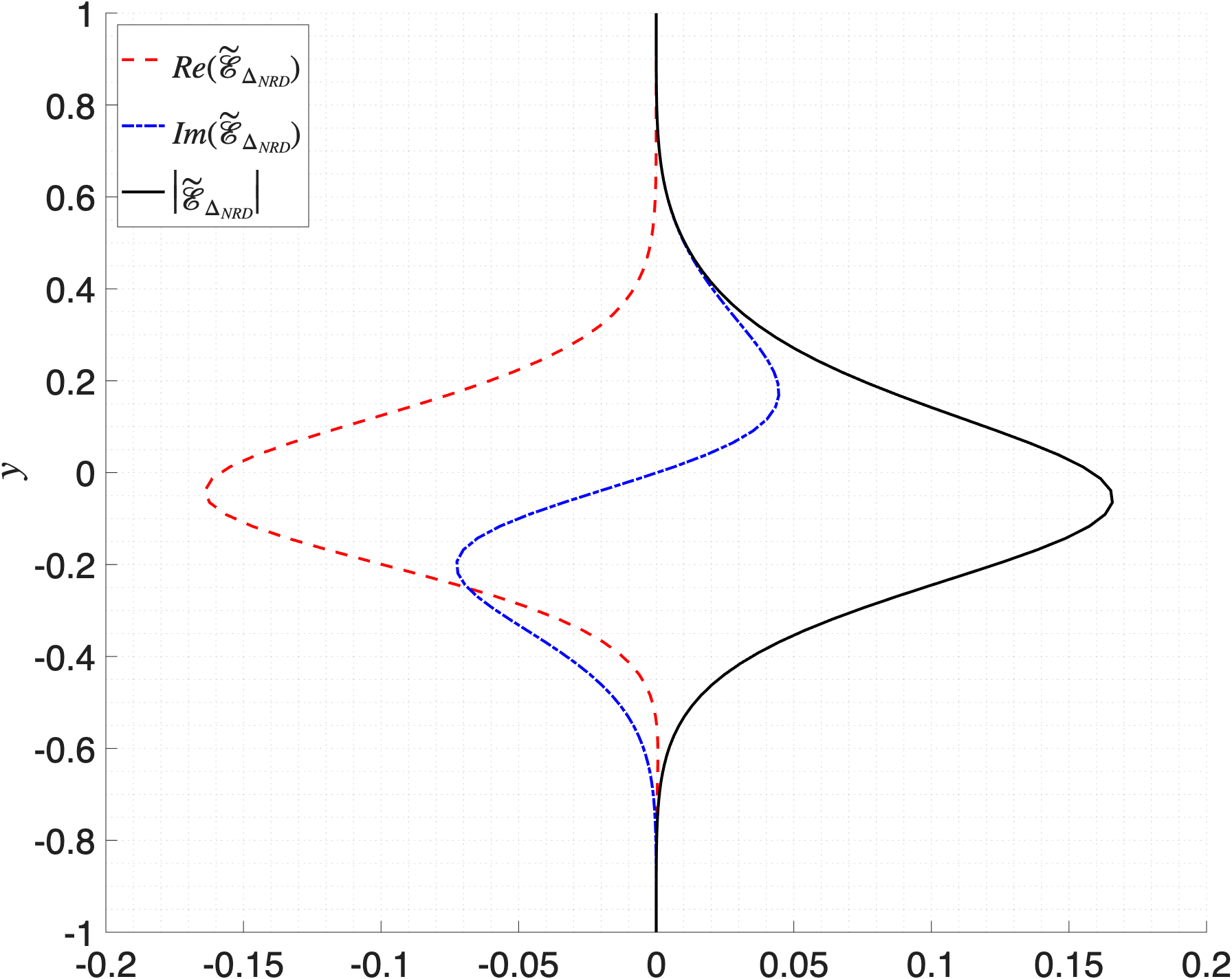}
        \caption{{$\mathcal{E}_{\mathbf{\Delta}_{NRD}}$}}
        \label{subfig:e_deld_sim}
    \end{subfigure}
    \hfill
    \begin{subfigure}[b]{0.32\textwidth}
        \centering
        \includegraphics[width=\textwidth]{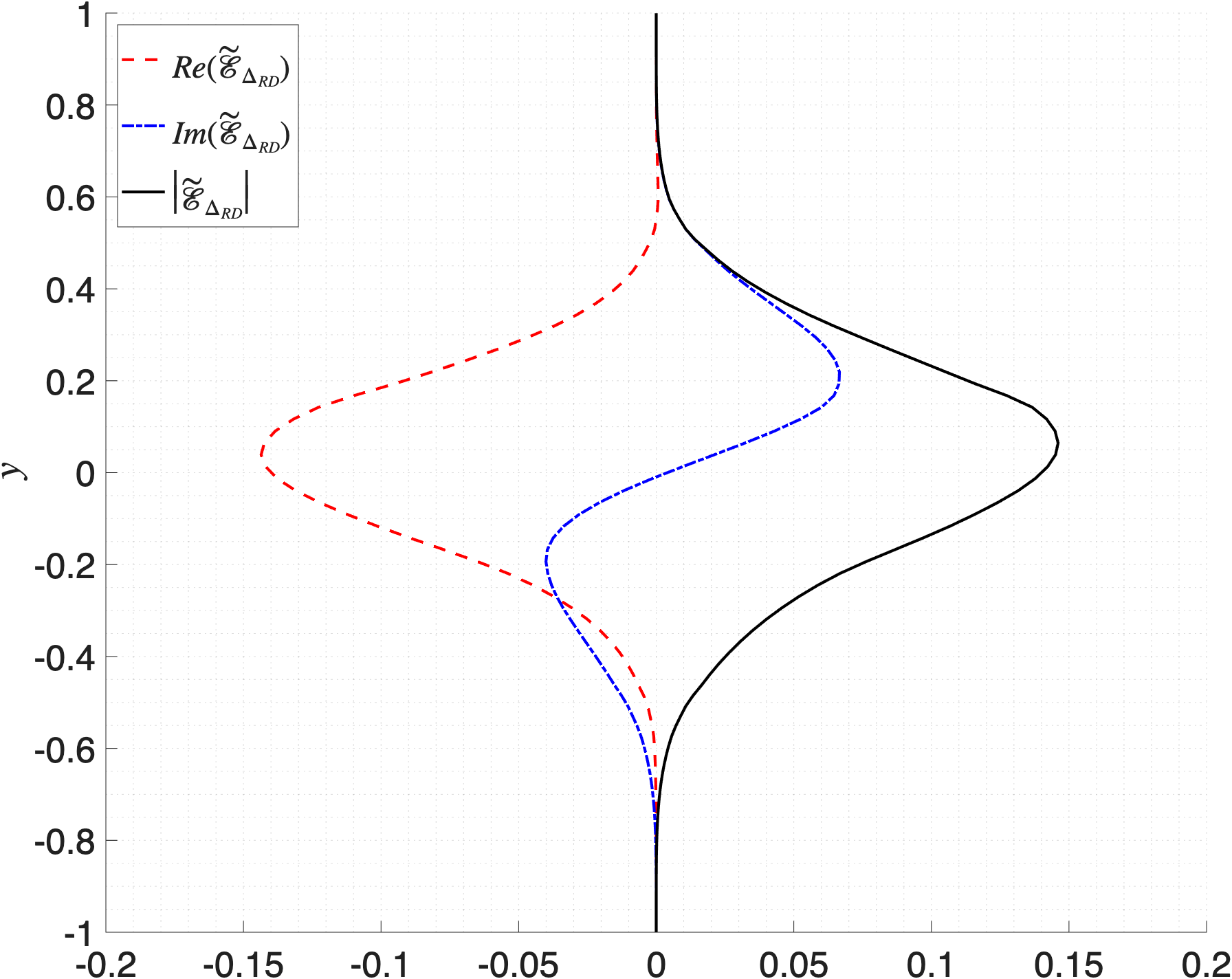}
        \caption{{$\mathcal{E}_{\mathbf{\Delta}_{RD}}$}}
        \label{subfig:e_delu_sim}
    \end{subfigure}
    \hfill
    \begin{subfigure}[b]{0.32\textwidth}
        \centering
        \includegraphics[width=\textwidth]{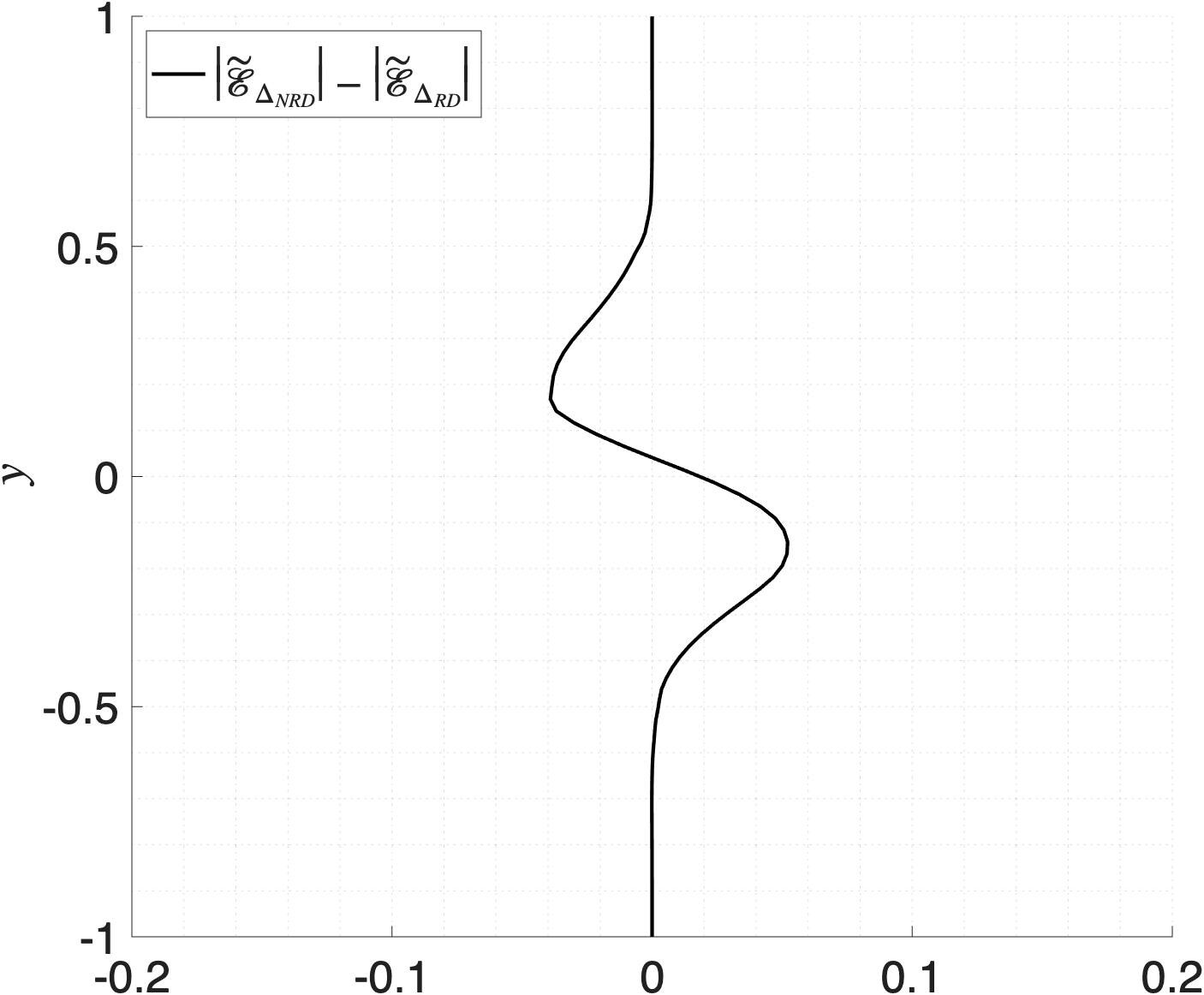}
        \caption{{$|\widetilde{\mathcal{E}}_{\mathbf{\Delta}_{NRD}}| - |\widetilde{\mathcal{E}}_{\mathbf{\Delta}_{RD}}|$}} 
        \label{subfig:e_diag_sim2}
    \end{subfigure}
    
    \caption{$\widetilde{\mathcal{E}}_{\mathbf{\Delta}}$ computed for Couette base flow at $Re=358$, with the wavenumber pair $(k_x,k_z)=(10^{-3},0.5)$ using 
    the uncertainty structures from~\S\ref{sec:uncertainty} ((a) $\mathbf{\Delta}_{NRB}$, (b) $\mathbf{\Delta}_{RB}$, (d) $\mathbf{\Delta}_{NRD}$, (e) $\mathbf{\Delta}_{RD}$), and the difference in the absolute artificial energy profile between (c) block uncertainties, and (f) diagonal uncertainties.}
    \label{fig:e_sim}
\end{figure}

Fig.~\ref{fig:e_sim} shows $\widetilde{\mathcal{E}}_{\mathbf{\Delta}_{NRB}}$, $\widetilde{\mathcal{E}}_{\mathbf{\Delta}_{RB}}$, $\widetilde{\mathcal{E}}_{\mathbf{\Delta}_{NRD}}$, and $\widetilde{\mathcal{E}}_{\mathbf{\Delta}_{RD}}$, computed for Couette base flow for the second wavenumber pair $(k_x,k_z)=(10^{-3},0.5)$ at the same Reynolds number, $Re=358$, chosen to assess how the artificial energy production is affected when the stability thresholds predicted by repeated and non-repeated approaches are similar.
In this case, the profiles generated by each
repeated structure are very similar to those generated by its non-repeated
counterpart. Fig.~\ref{subfig:e_block_sim2} shows that the difference between the repeated block and non-repeated block profiles is negligible; whereas Fig.~\ref{subfig:e_diag_sim2} shows that there is a significant difference between the profiles corresponding to non-repeated diagonal and repeated diagonals. The profile in Fig.~\ref{subfig:e_diag_sim2} contains both positive and negative values, which, as we will show next, cancel out in the integration. While the profiles of $\widetilde{\mathcal{E}}_{\mathbf{\Delta}_{NRD}}$ and $\widetilde{\mathcal{E}}_{\mathbf{\Delta}_{RD}}$ are relatively similar in values, the $\widetilde{\mathcal{E}}_{\mathbf{\Delta}_{RD}}$ profile is shifted to slightly higher $y$ positions than the $\widetilde{\mathcal{E}}_{\mathbf{\Delta}_{NRD}}$ profile, leading to larger values of $|\widetilde{\mathcal{E}}_{\mathbf{\Delta}_{NRD}}| - |\widetilde{\mathcal{E}}_{\mathbf{\Delta}_{RD}}|$, but their overall contribution is canceled out in the integration process.

Integrating the profiles presented in Fig.~\ref{fig:e_sim} yields the following absolute values:
\begin{equation}
\begin{split}
    |\widetilde{\mathcal{P}}_{\mathbf{\Delta}_{NRB}}|
        &= 0.15318, \qquad
    |\widetilde{\mathcal{P}}_{\mathbf{\Delta}_{RB}}|
        = 0.15809,\\
    |\widetilde{\mathcal{P}}_{\mathbf{\Delta}_{NRD}}|
        &= 0.08443, \qquad
    |\widetilde{\mathcal{P}}_{\mathbf{\Delta}_{RD}}|
        = 0.080257.
\end{split}
\end{equation}
The two full-block results differ by only approximately $3\%$, while the two
diagonal results differ by approximately $5\%$. Hence, lifting the repeated
entries constraint provides almost no additional artificial energy-production mechanism for this mode. This suggests that a common uncertainty field is
already sufficiently well aligned with the dominant amplification mechanism,
so that allowing the three uncertainty fields to vary independently offers
little advantage. This result is consistent with Fig.~\ref{fig:prf_couette}, where the repeated
and non-repeated formulations yield similar stability thresholds at this wavenumber pair.

Taken together, Figs.~\ref{fig:e_dif} and~\ref{fig:e_sim} show that the effect of repeated uncertainty
entries is strongly mode dependent. In the oblique mode region, the additional
degrees of freedom introduced by non-repeated uncertainties produce
substantial artificial energy and lead to considerably lower, more
conservative stability thresholds. For the other mode we checked,
the artificial energy production is only weakly affected by the constraint of repeated uncertainty entries, and the corresponding stability thresholds consequently remain
similar. In light of these results, we suggest that the low-threshold
oblique region predicted by the non-repeated formulations is caused, at least in part, by feedback pathways that do not accurately represent the physical advection term and allow artificial energy production that lowers the stability threshold compared to the NS dynamics. Our results also indicate that using the methodology proposed in this work to compute stability thresholds via the structure $\mathbf{\Delta}_{RD}$ yields the lowest artificial energy production out of all of the methods tested.

\subsection{Results for plane Poiseuille flow: artificial energy production}
\label{sec:P_Poiseuille}

\begin{figure}[ht!]
    \centering
    \begin{subfigure}[b]{0.48\textwidth}
        \centering
        \includegraphics[width=\textwidth]{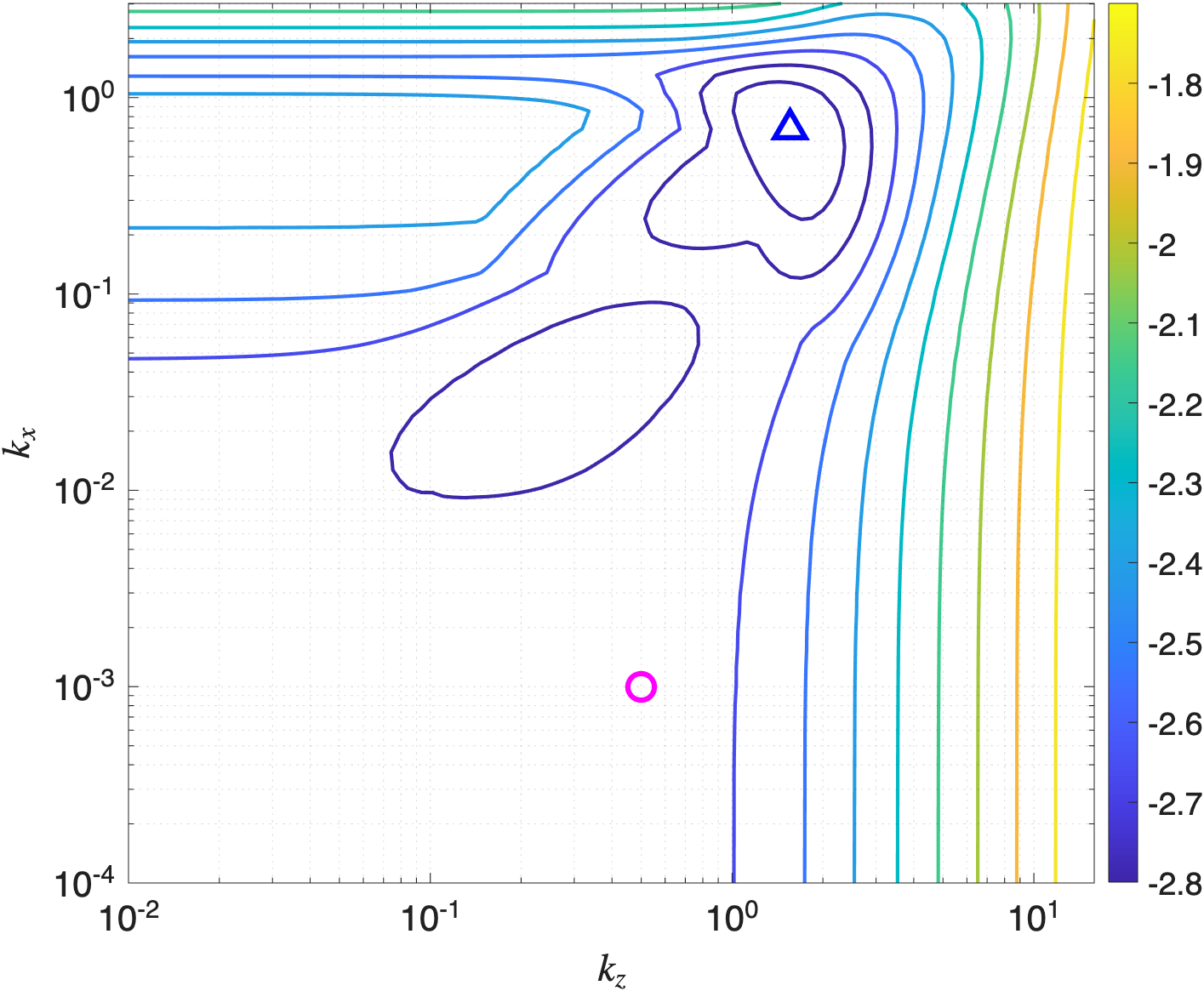}
        \caption{$\norm{\mathscr{H}_{\nabla}}_{\mu_{\boldsymbol{\Delta_{NRB}}}}^{-1}(k_x,k_z)$}
        \label{subfig:poiseuille_NRB}
    \end{subfigure}
    \hfill
    \begin{subfigure}[b]{0.48\textwidth}
        \centering
        \includegraphics[width=\textwidth]{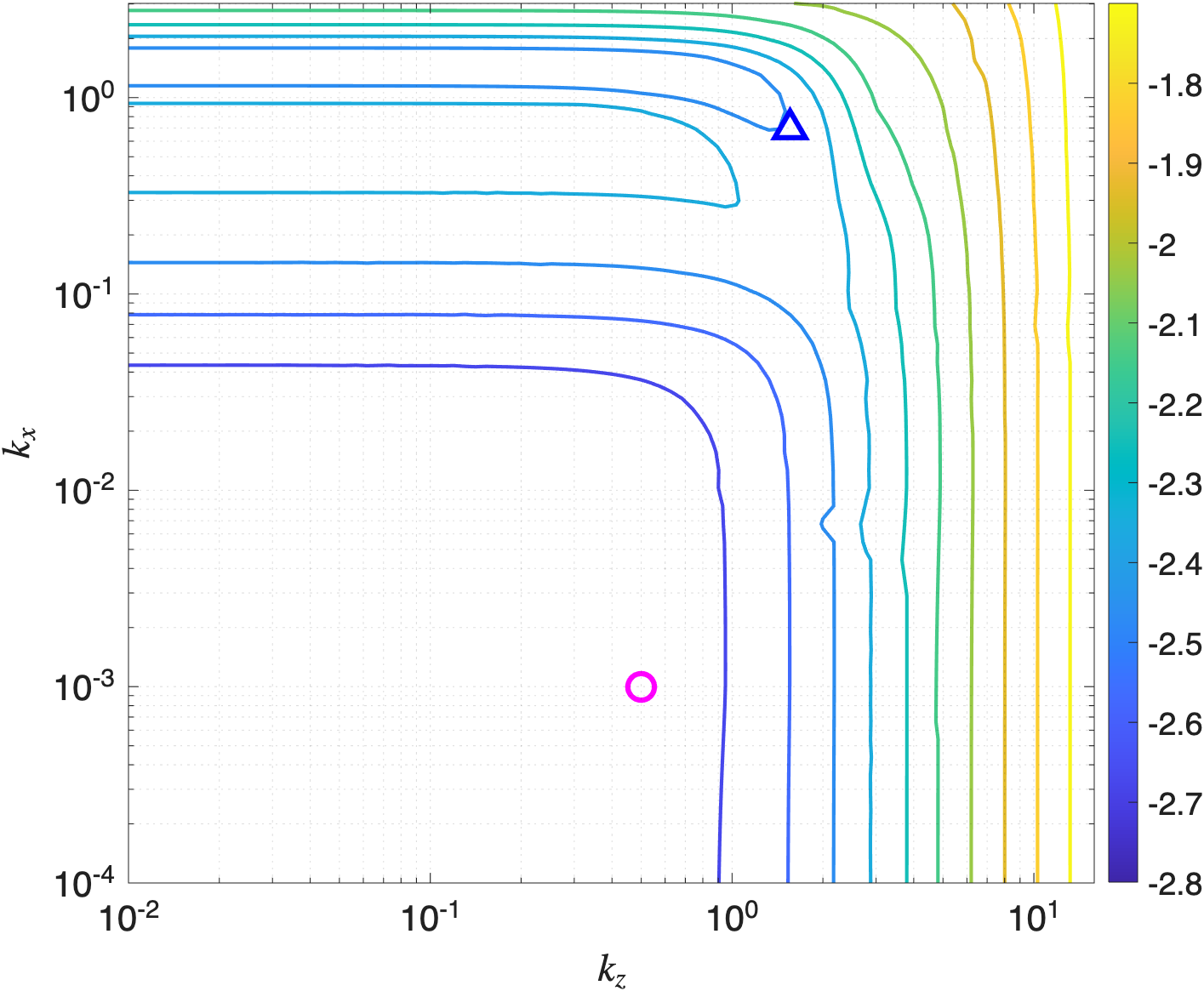}
        \caption{$\norm{\mathscr{H}_{\nabla}}_{\mu_{\boldsymbol{\Delta_{RB}}}}^{-1}(k_x,k_z)$}
        \label{subfig:poiseuille_rb}
    \end{subfigure}
    
    \vspace{0.5cm}
    
    \begin{subfigure}[b]{0.48\textwidth}
        \centering
        \includegraphics[width=\textwidth]{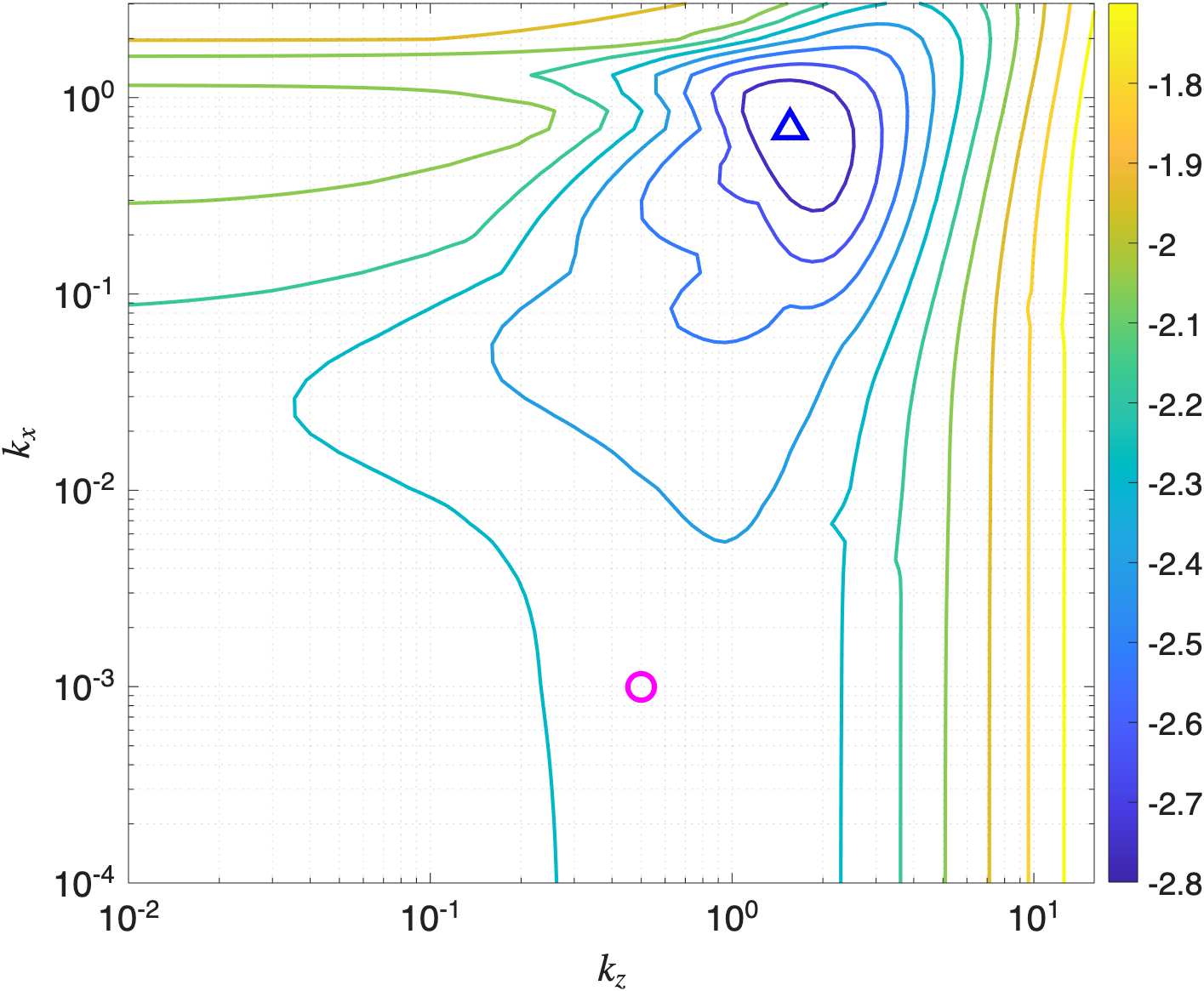}
        \caption{$\sqrt{3}\norm{R\mathscr{H}_{\nabla}L}_{\mu_{\boldsymbol{\Delta_{NRD}}}}^{-1}(k_x,k_z)$}
        \label{subfig:poiseuille_deld}
    \end{subfigure}
    \hfill
    \begin{subfigure}[b]{0.48\textwidth}
        \centering
        \includegraphics[width=\textwidth]{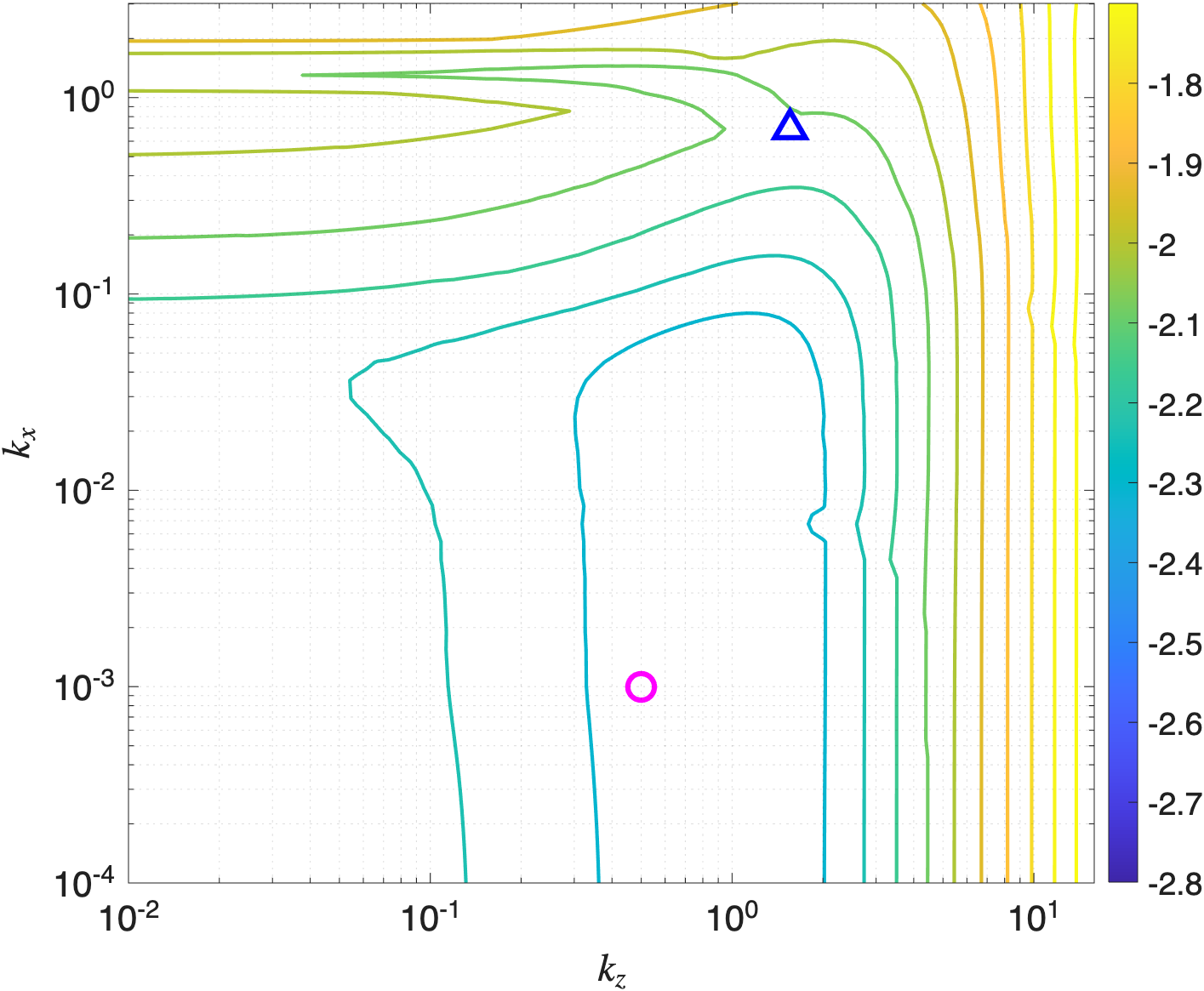}
        \caption{$\sqrt{3}\norm{R\mathscr{H}_{\nabla}L}_{\mu_{\boldsymbol{\Delta_{RD}}}}^{-1}(k_x,k_z)$}
        \label{subfig:poiseuille_delu}
    \end{subfigure}
    
    \caption{Stability analysis of plane Poiseuille flow at $Re=690$ using the uncertainty structures (a) $\mathbf{\Delta}_{NRB}$, (b) $\mathbf{\Delta}_{RB}$, (c) $\mathbf{\Delta}_{NRD}$, (d) $\mathbf{\Delta}_{RD}$. Blue triangle - $k_x = 0.692, k_z = 1.561$, magenta circle - $k_x = 10^{-3}, k_z = 0.5$.}
    \label{fig:prf_poiseuille}
\end{figure}

Next, we consider plane Poiseuille flow, a pressure-driven flow geometry, to show that the effect of non-repeated uncertainty entries causing a region of oblique modes with lower stability thresholds is base-flow-independent. Fig.~\ref{fig:prf_poiseuille} shows contour plots of the four bounds on velocity perturbation magnitude for plane Poiseuille 
flow at $Re=690$. Over these contour plots, we highlight two wavenumber pairs - one where the repeated approaches produce very different results than the non-repeated ones (namely $k_x=0.692, k_z=1.561$), and one where they produce similar results (namely $k_x=10^{-3},k_z=0.5$).

Similarly to Couette flow (Fig.~\ref{fig:prf_couette}), Fig.~\ref{fig:prf_poiseuille} includes a region of oblique modes with lower stability thresholds in the contour plots of $\norm{\mathscr{H}_{\nabla}}_{\mu_{\boldsymbol{\Delta_{NRB}}}}^{-1}$ and $\sqrt{3}\norm{R\mathscr{H}_{\nabla}L}_{\mu_{\boldsymbol{\Delta_{NRD}}}}^{-1}$, the quantities that were computed using uncertainties without repeated entries. This region is not found in the contour plots corresponding to the repeated quantities. In general, all discussion and insights as detailed for Couette flow (Fig.~\ref{fig:prf_couette})  apply to plane Poiseuille flow and, for brevity, will not be repeated here.

\begin{figure}[ht!]
    \centering
    \begin{subfigure}[b]{0.32\textwidth}
        \centering
        \includegraphics[width=\textwidth]{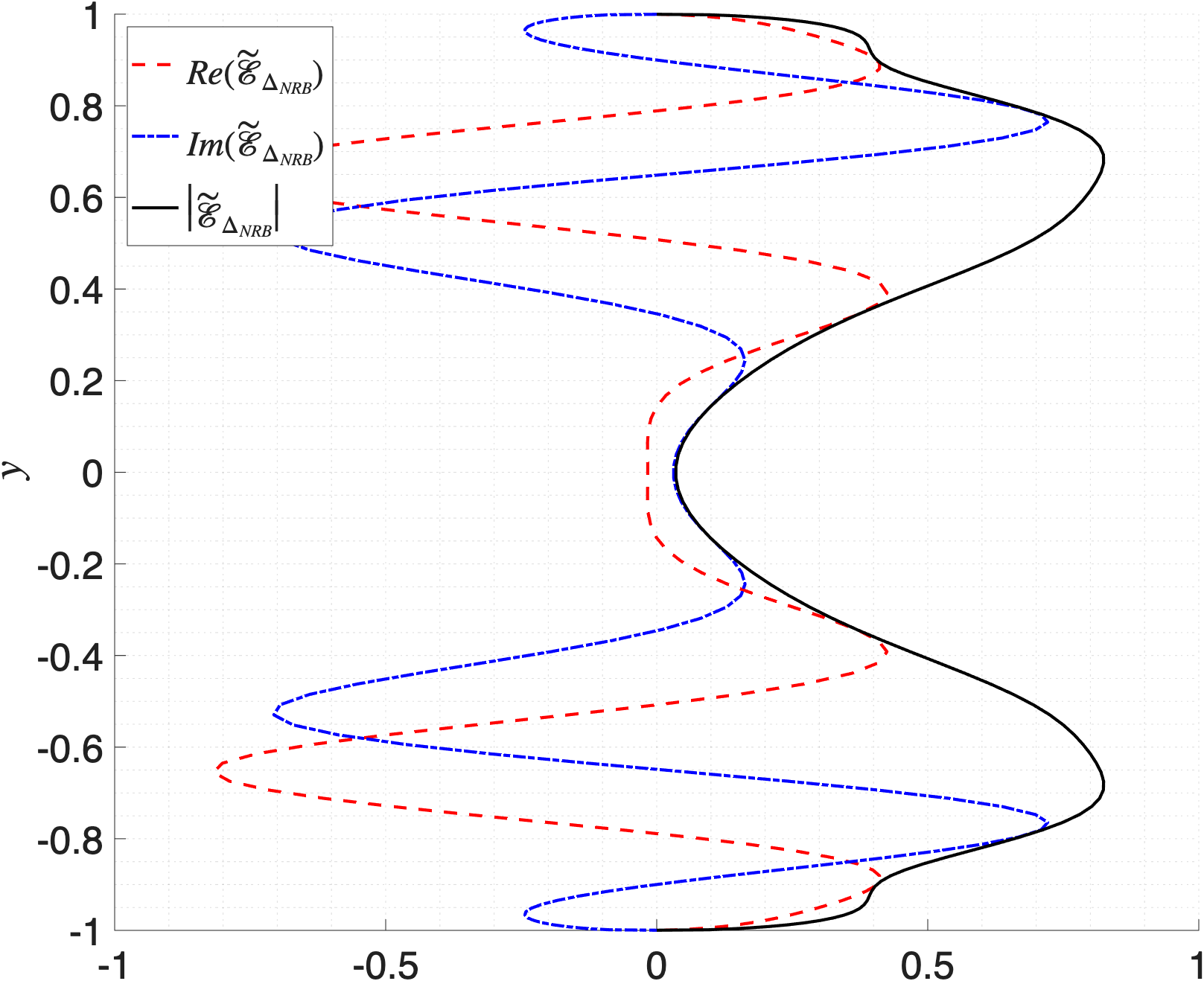}
        \caption{$\mathcal{E}_{\mathbf{\Delta}_{NRB}}$}
        \label{subfig:ep_NRB_dif}
    \end{subfigure}
    \hfill
    \begin{subfigure}[b]{0.32\textwidth}
        \centering
        \includegraphics[width=\textwidth]{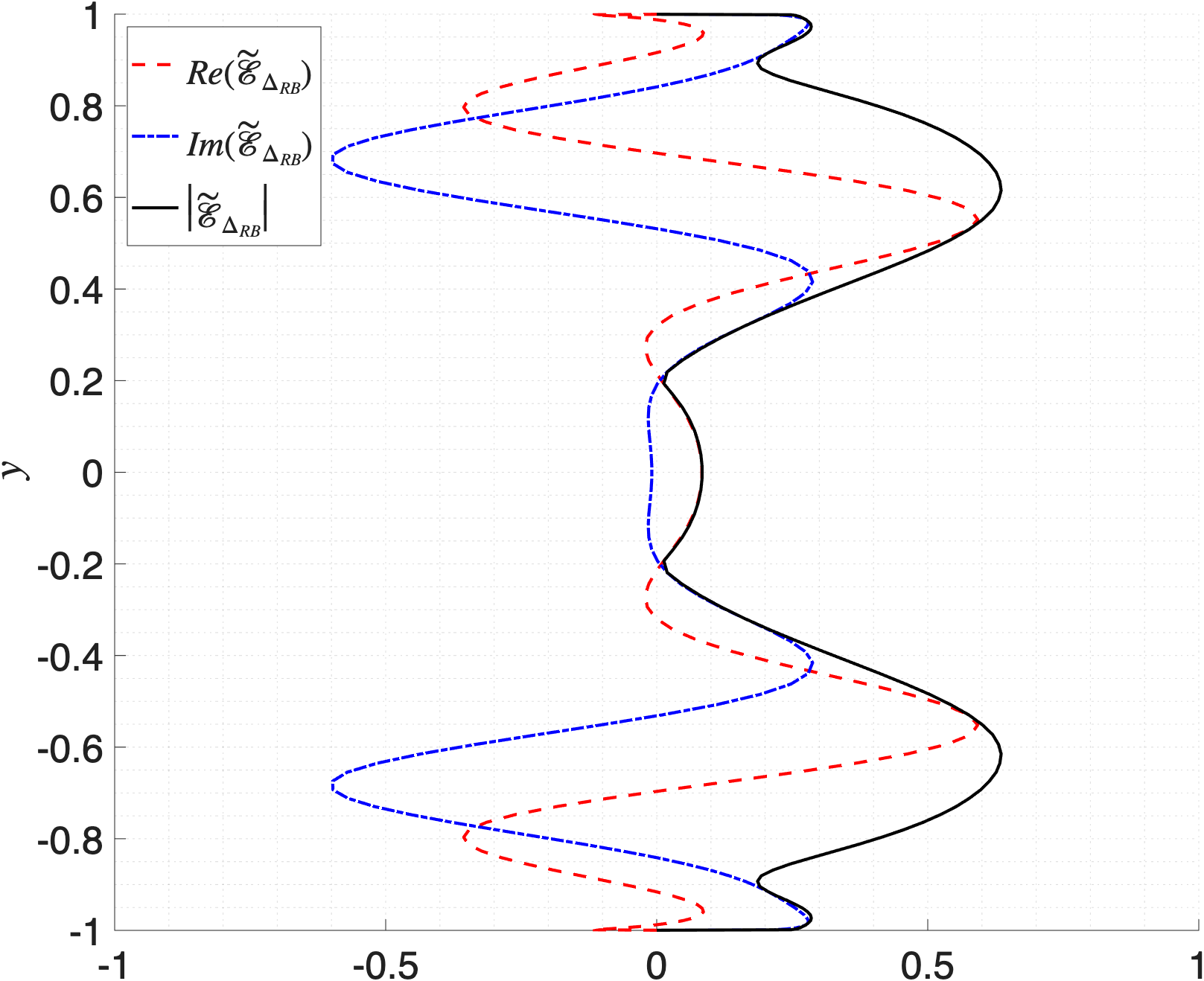}
        \caption{$\mathcal{E}_{\mathbf{\Delta}_{RB}}$}
        \label{subfig:ep_rb_dif}
    \end{subfigure}
    \hfill
    \begin{subfigure}[b]{0.32\textwidth}
        \centering
        \includegraphics[width=\textwidth]{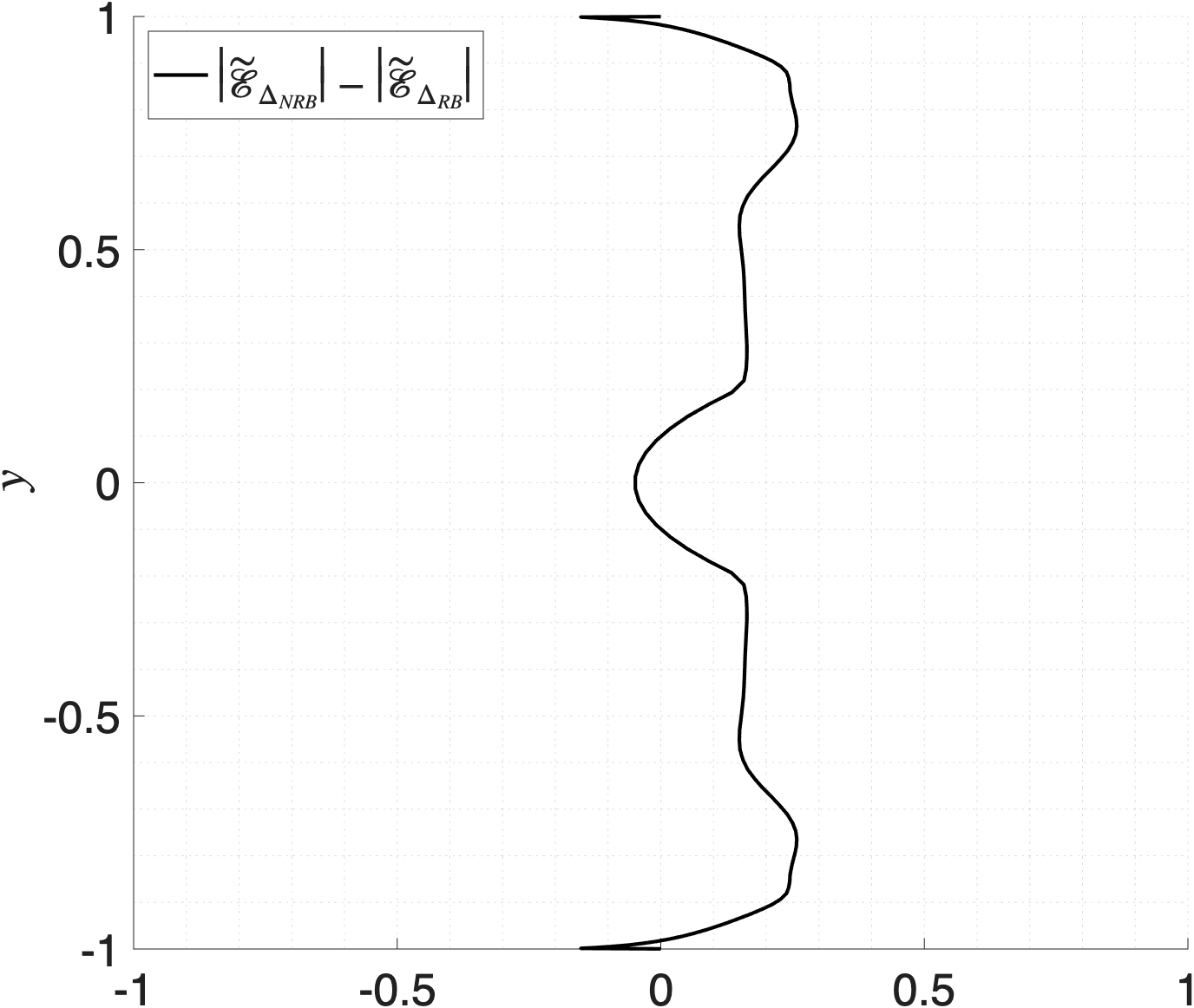} 
        \caption{$|\widetilde{\mathcal{E}}_{\mathbf{\Delta}_{NRB}}| - |\widetilde{\mathcal{E}}_{\mathbf{\Delta}_{RB}}|$} 
        \label{subfig:ep_block_dif}
    \end{subfigure}
    
    \vspace{0.5cm}
    
    \begin{subfigure}[b]{0.32\textwidth}
        \centering
        \includegraphics[width=\textwidth]{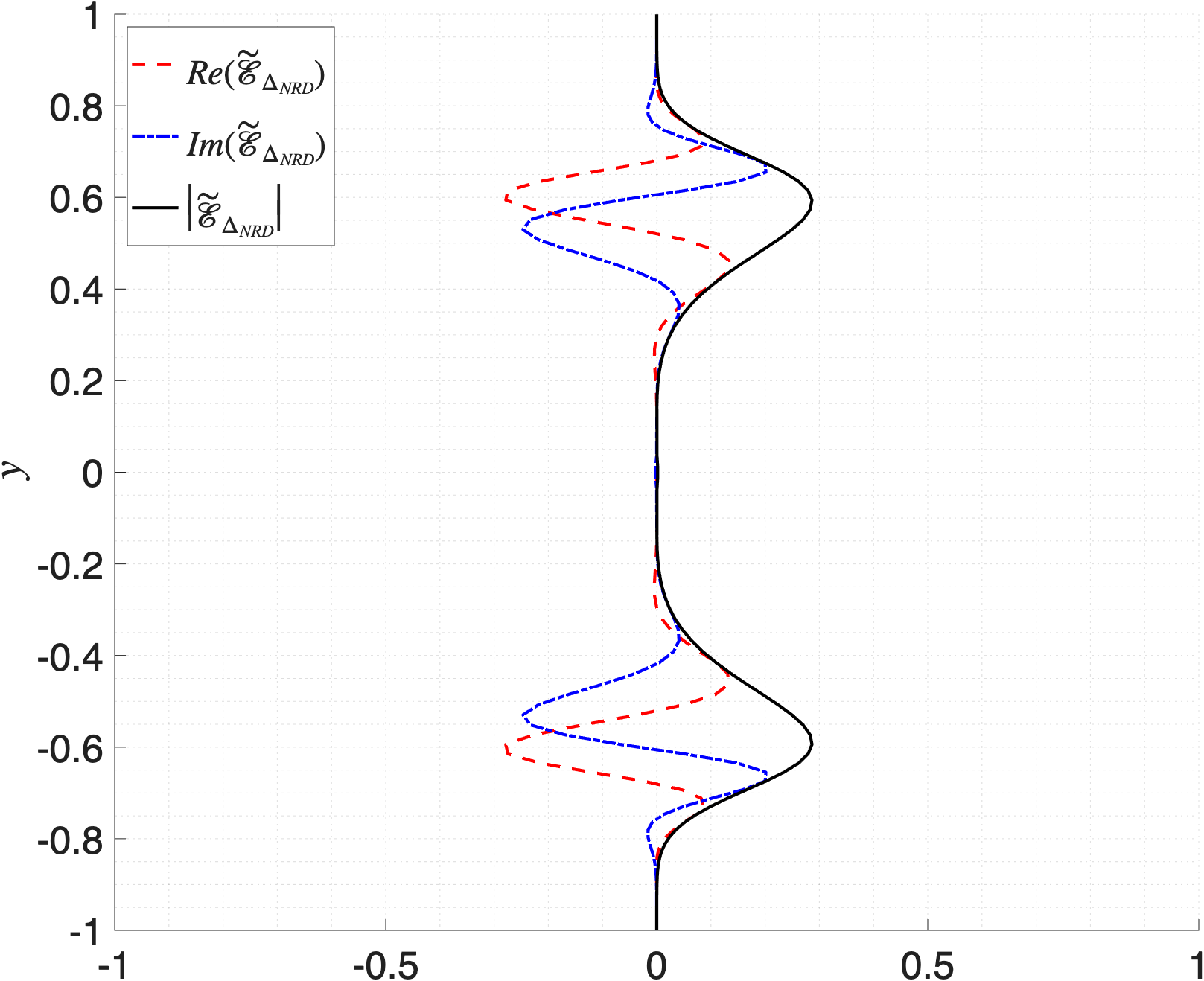}
        \caption{$\mathcal{E}_{\mathbf{\Delta}_{NRD}}$}
        \label{subfig:ep_deld_dif}
    \end{subfigure}
    \hfill
    \begin{subfigure}[b]{0.32\textwidth}
        \centering
        \includegraphics[width=\textwidth]{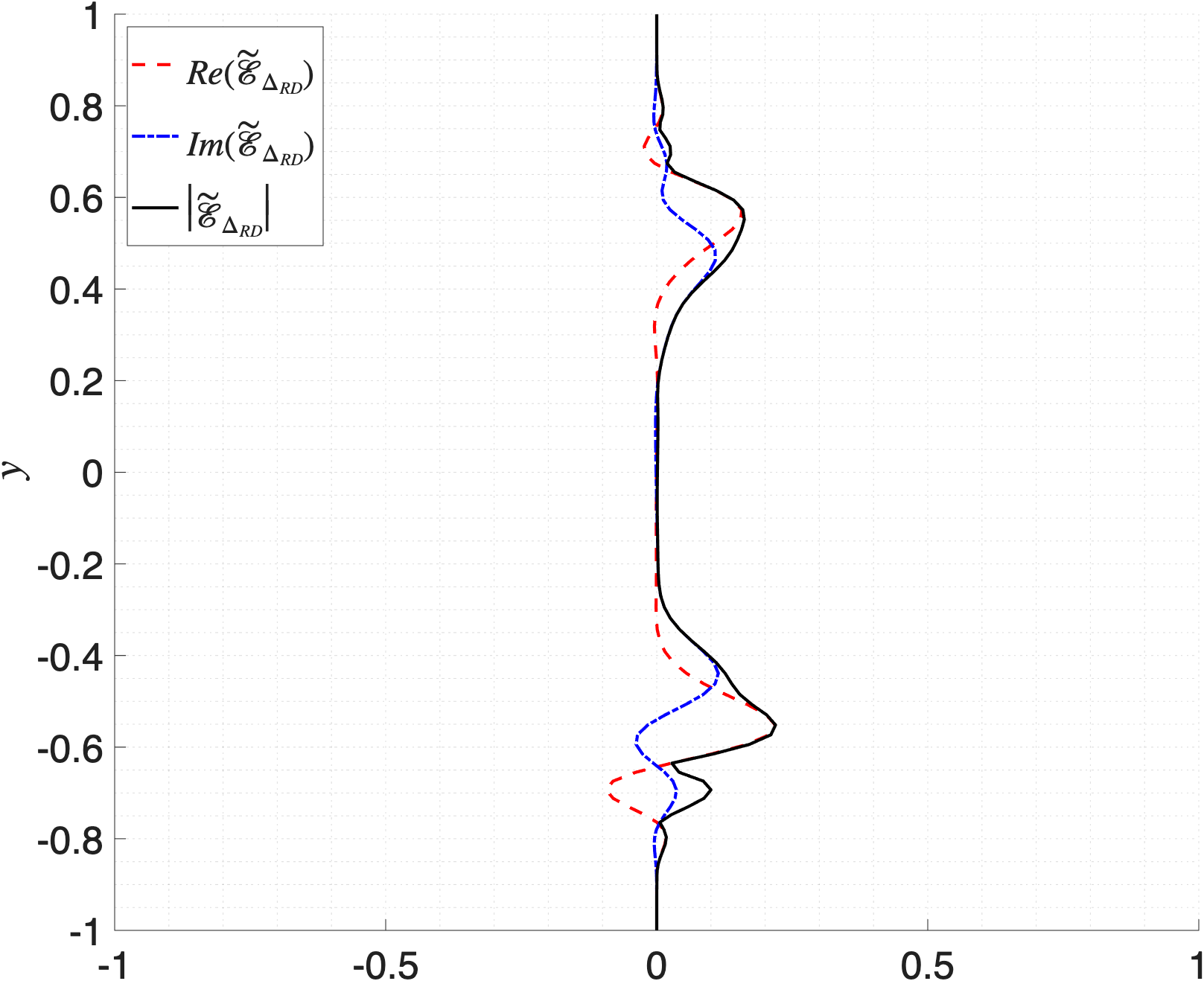}
        \caption{$\mathcal{E}_{\mathbf{\Delta}_{RD}}$}
        \label{subfig:ep_delu_dif}
    \end{subfigure}
    \hfill
    \begin{subfigure}[b]{0.32\textwidth}
        \centering
        \includegraphics[width=\textwidth]{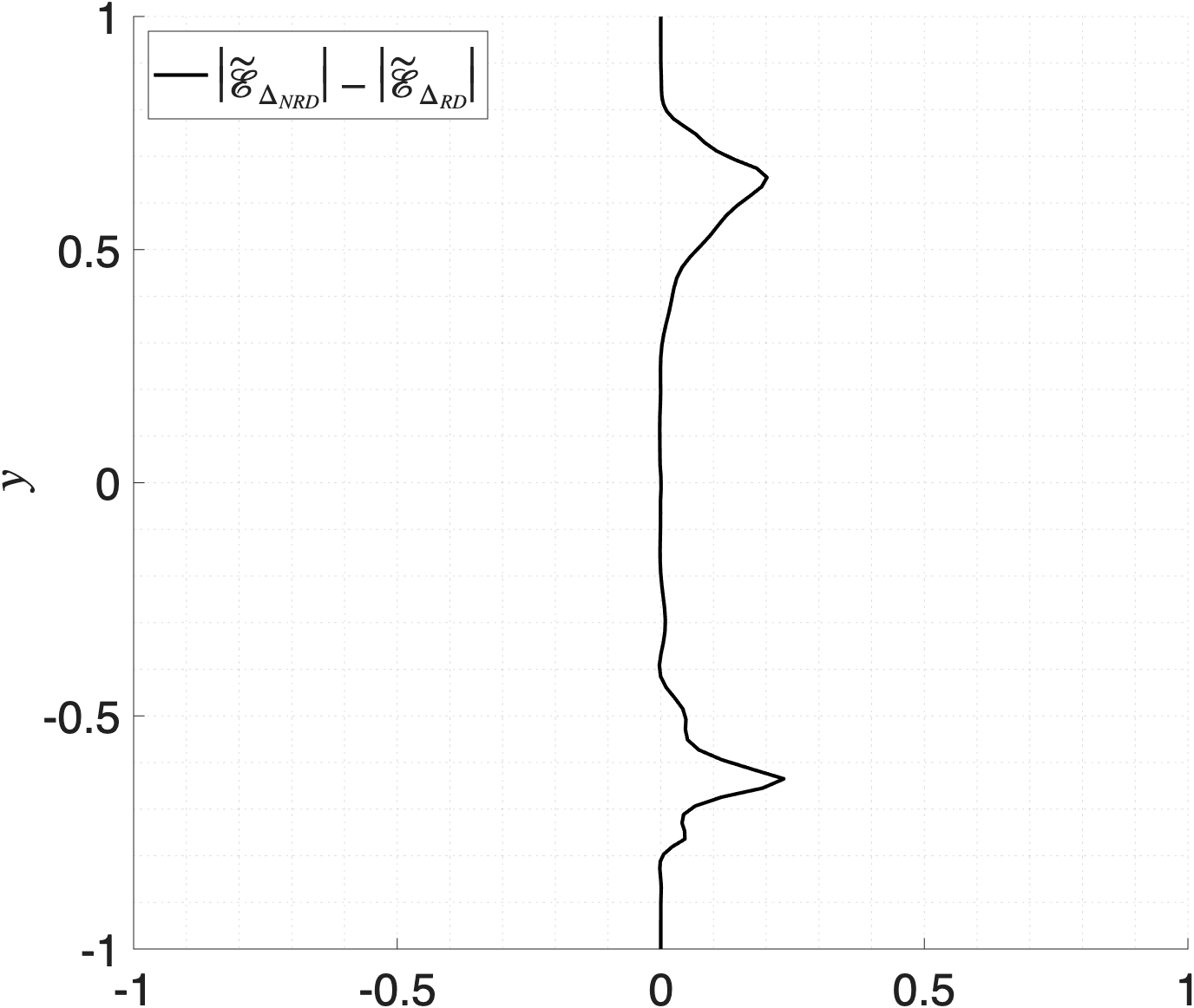} 
        \caption{$|\widetilde{\mathcal{E}}_{\mathbf{\Delta}_{NRD}}| - |\widetilde{\mathcal{E}}_{\mathbf{\Delta}_{RD}}|$} 
        \label{subfig:ep_diag_dif}
    \end{subfigure}
    
    \caption{$\widetilde{\mathcal{E}}_{\mathbf{\Delta}}$ computed for Poiseuille base flow at $Re=690$, with the wavenumber pair $(k_x,k_z)=(0.692,1.561)$ using 
    the uncertainty structures from~\S\ref{sec:uncertainty} ((a) $\mathbf{\Delta}_{NRB}$, (b) $\mathbf{\Delta}_{RB}$, (d) $\mathbf{\Delta}_{NRD}$, (e) $\mathbf{\Delta}_{RD}$), and the difference in the absolute artificial energy profile between (c) block uncertainties, and (f) diagonal uncertainties.}
    \label{fig:ep_dif}
\end{figure}

Fig.~\ref{fig:ep_dif} shows $\widetilde{\mathcal{E}}_{\mathbf{\Delta}_{NRB}}$, $\widetilde{\mathcal{E}}_{\mathbf{\Delta}_{RB}}$, $\widetilde{\mathcal{E}}_{\mathbf{\Delta}_{NRD}}$, and $\widetilde{\mathcal{E}}_{\mathbf{\Delta}_{RD}}$, computed for Poiseuille base flow at $Re=690$, using the wavenumber pair $(k_x,k_z)=(0.692,1.561)$, chosen to illustrate the difference between the repeated and non-repeated approaches for plane Poiseuille flow.
Similarly to Couette flow, the artificial-energy profiles obtained with non-repeated uncertainty structures reach higher values than the corresponding repeated structure curves, both for the block structures ($\mathbf{\Delta}_{NRB}$ and $\mathbf{\Delta}_{RB}$) and for the diagonal structures ($\mathbf{\Delta}_{NRD}$ and $\mathbf{\Delta}_{RD}$). This further strengthens the indication that enforcing the repeated-entries constraint reduces artificial energy production.  This difference can also be seen in Fig.~\ref{subfig:ep_block_dif} and Fig.~\ref{subfig:ep_diag_dif}, which show significant positive values. Fig.~\ref{subfig:ep_block_dif} also contains a small region where the values are negative, but after integration the non-repeated block uncertainty results in larger artificial energy production compared to the repeated block uncertainty, as we will show next.
Once again, there is a significant difference in the values of $\widetilde{\mathcal{E}}_{\mathbf{\Delta}}$ between the block structures and the diagonal structures, with diagonal structures producing significantly reduced values, indicating that using the methodology proposed here results in lower artificial energy production, better representing the dynamics of the NS system.

Integrating the profiles presented in Fig.~\ref{fig:ep_dif} yields the following absolute values:
\begin{equation}
\begin{split}
    |\widetilde{\mathcal{P}}_{\mathbf{\Delta}_{{NRB}}}|
        &= 0.9045, \qquad
    |\widetilde{\mathcal{P}}_{\mathbf{\Delta}_{{RB}}}|
        = 0.6161,\\
    |\widetilde{\mathcal{P}}_{\mathbf{\Delta}_{{NRD}}}|
        &= 0.1591, \qquad
    |\widetilde{\mathcal{P}}_{\mathbf{\Delta}_{{RD}}}|
        = 0.0908.
\end{split}
\end{equation}
Thus, removing the repeated-block constraint increases the integrated
artificial energy production by approximately $47\%$ for the full-block
structures and by approximately $75\%$ for the diagonal structures. We suggest that these differences lead to the lower stability thresholds observed in Fig.~\ref{fig:prf_poiseuille}, as explained in \S~\ref{sec:P_Couette} for Couette flow.

\begin{figure}[ht!]
    \centering
    \begin{subfigure}[b]{0.32\textwidth}
        \centering
        \includegraphics[width=\textwidth]{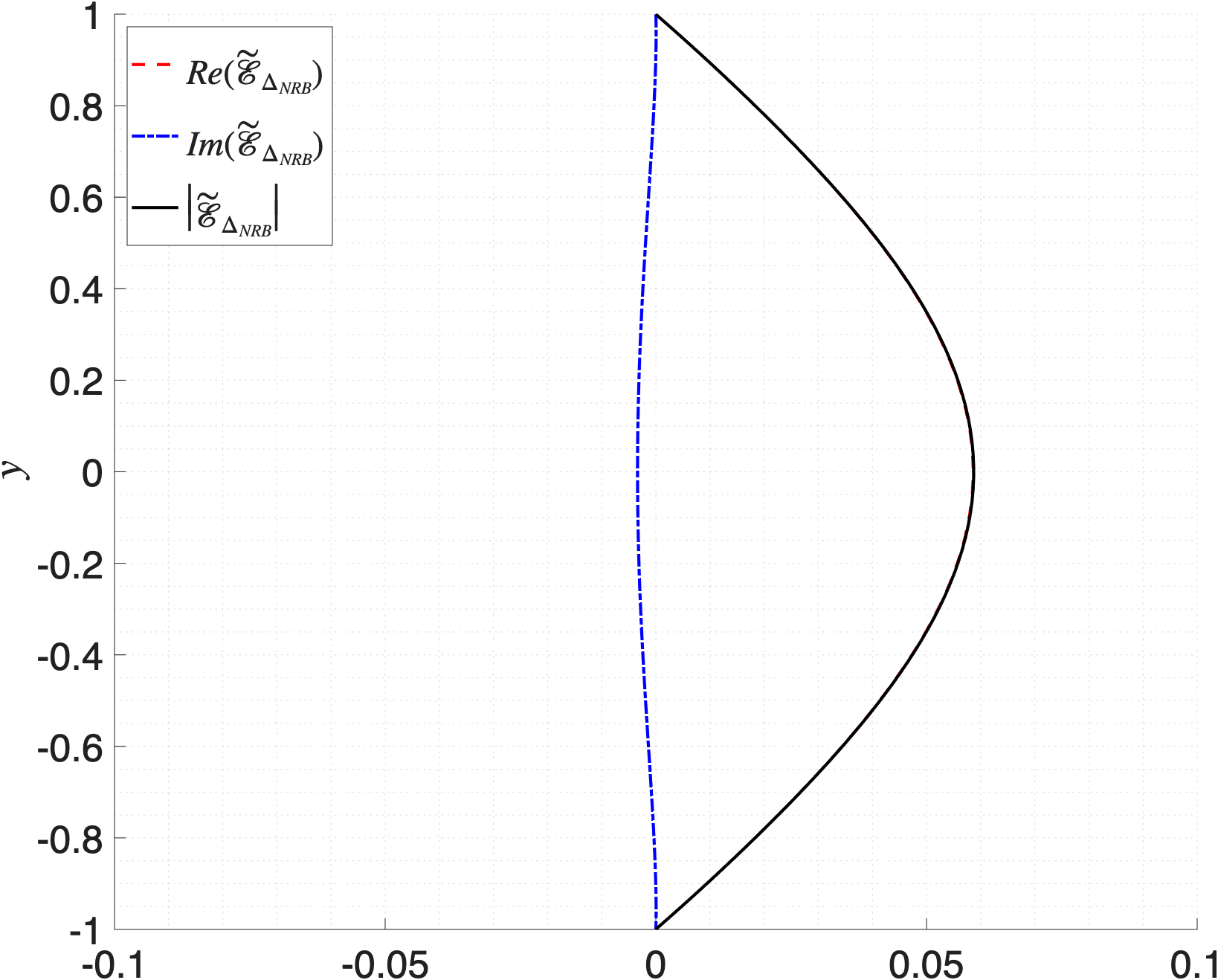}
        \caption{$\mathcal{E}_{\mathbf{\Delta}_{NRB}}$}
        \label{subfig:ep_NRB_sim}
    \end{subfigure}
    \hfill
    \begin{subfigure}[b]{0.32\textwidth}
        \centering
        \includegraphics[width=\textwidth]{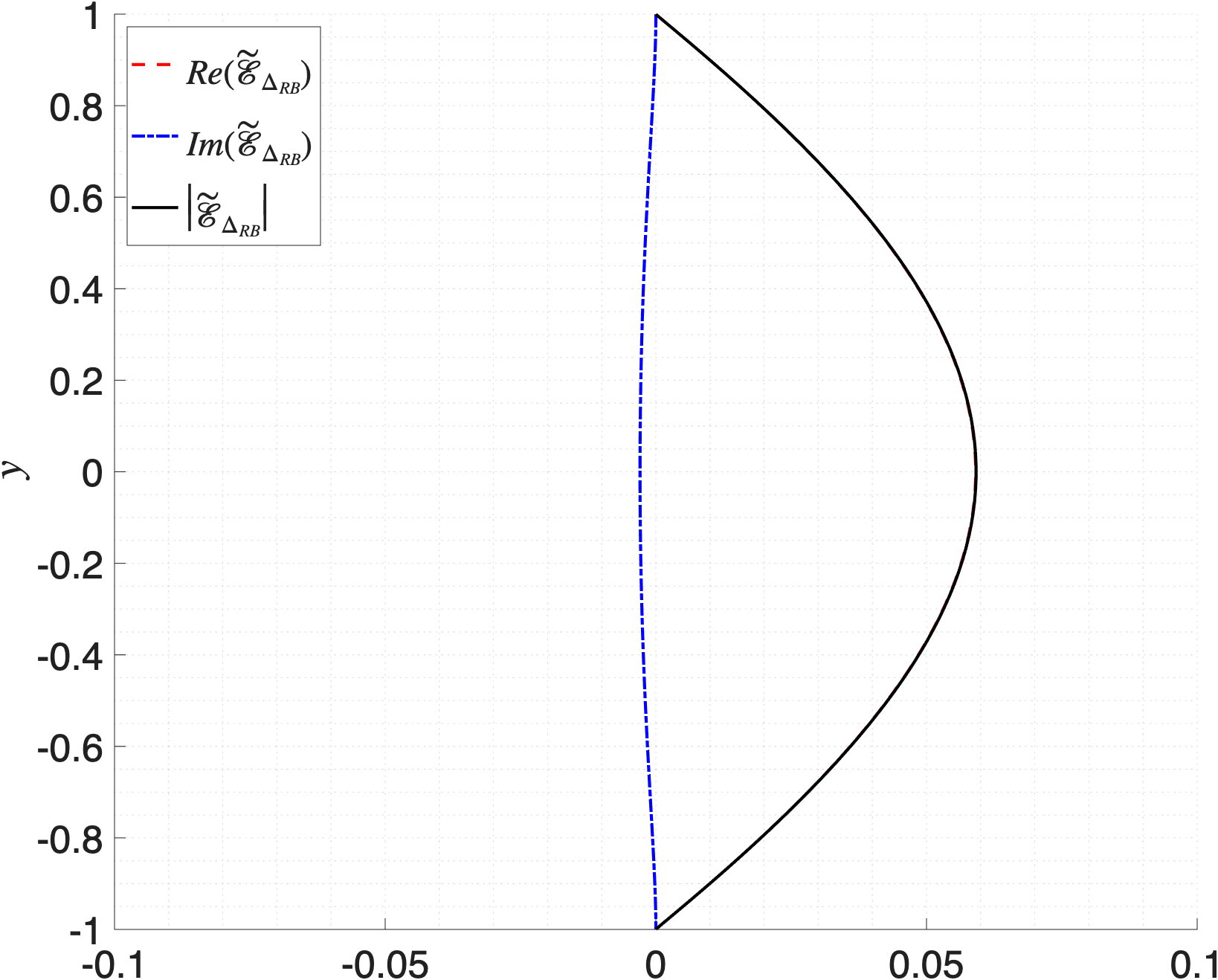}
        \caption{$\mathcal{E}_{\mathbf{\Delta}_{RB}}$}
        \label{subfig:ep_rb_sim}
    \end{subfigure}
    \hfill
    \begin{subfigure}[b]{0.32\textwidth}
        \centering
        \includegraphics[width=\textwidth]{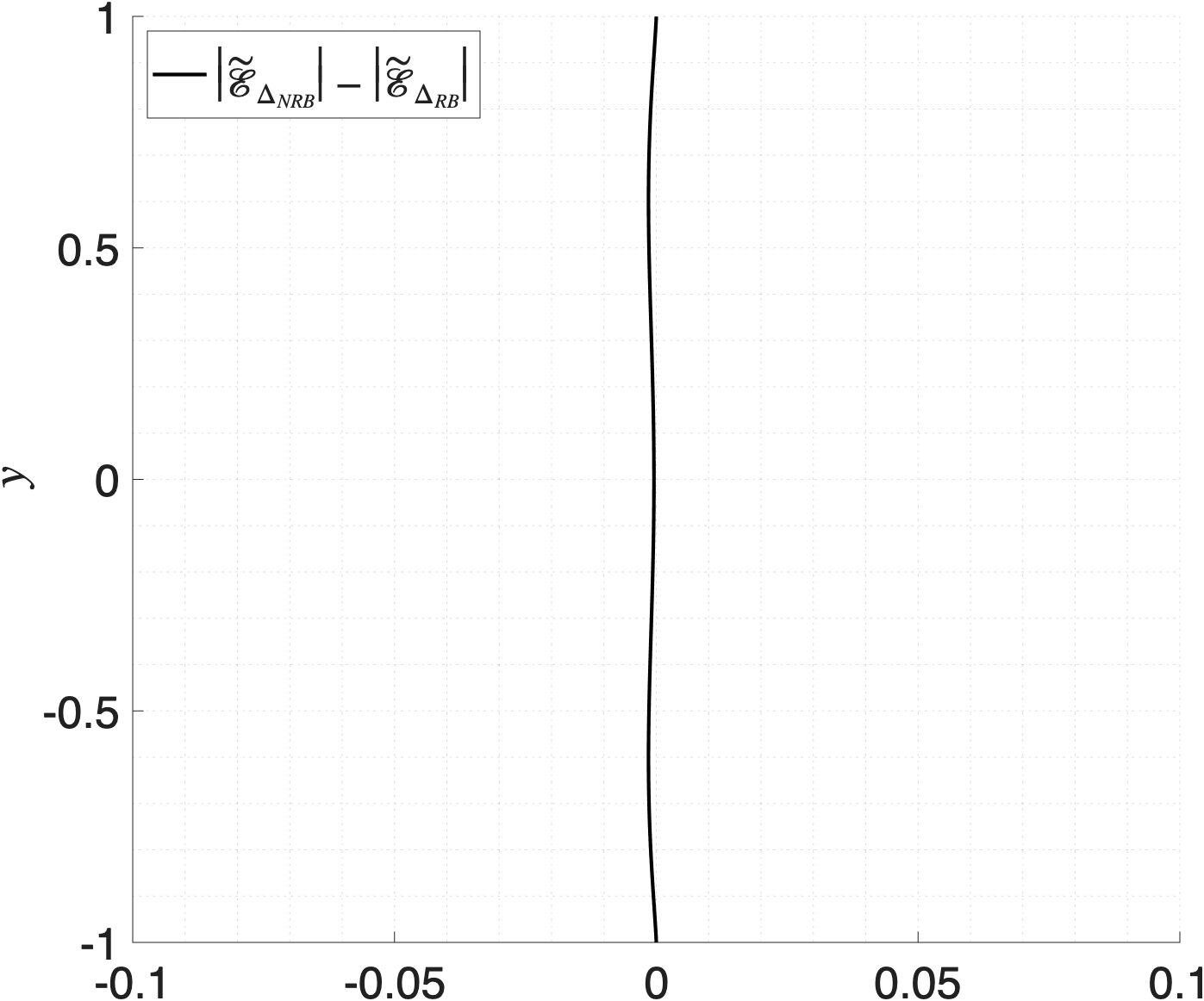} 
        \caption{$|\widetilde{\mathcal{E}}_{\mathbf{\Delta}_{NRB}}| - |\widetilde{\mathcal{E}}_{\mathbf{\Delta}_{RB}}|$} 
        \label{subfig:ep_block_sim}
    \end{subfigure}
    
    \vspace{0.5cm}
    
    \begin{subfigure}[b]{0.32\textwidth}
        \centering
        \includegraphics[width=\textwidth]{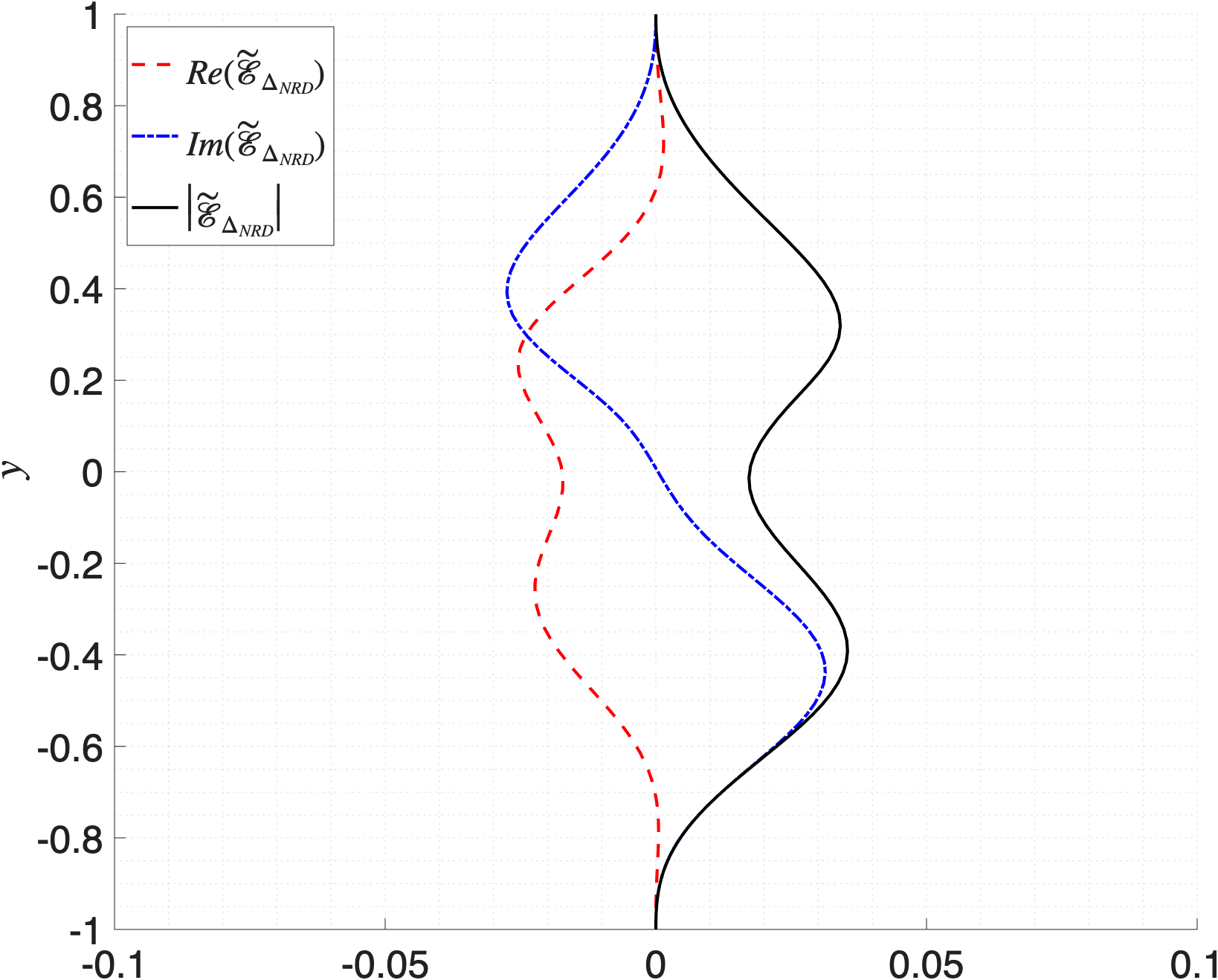}
        \caption{$\mathcal{E}_{\mathbf{\Delta}_{NRD}}$}
        \label{subfig:ep_deld_sim}
    \end{subfigure}
    \hfill
    \begin{subfigure}[b]{0.32\textwidth}
        \centering
        \includegraphics[width=\textwidth]{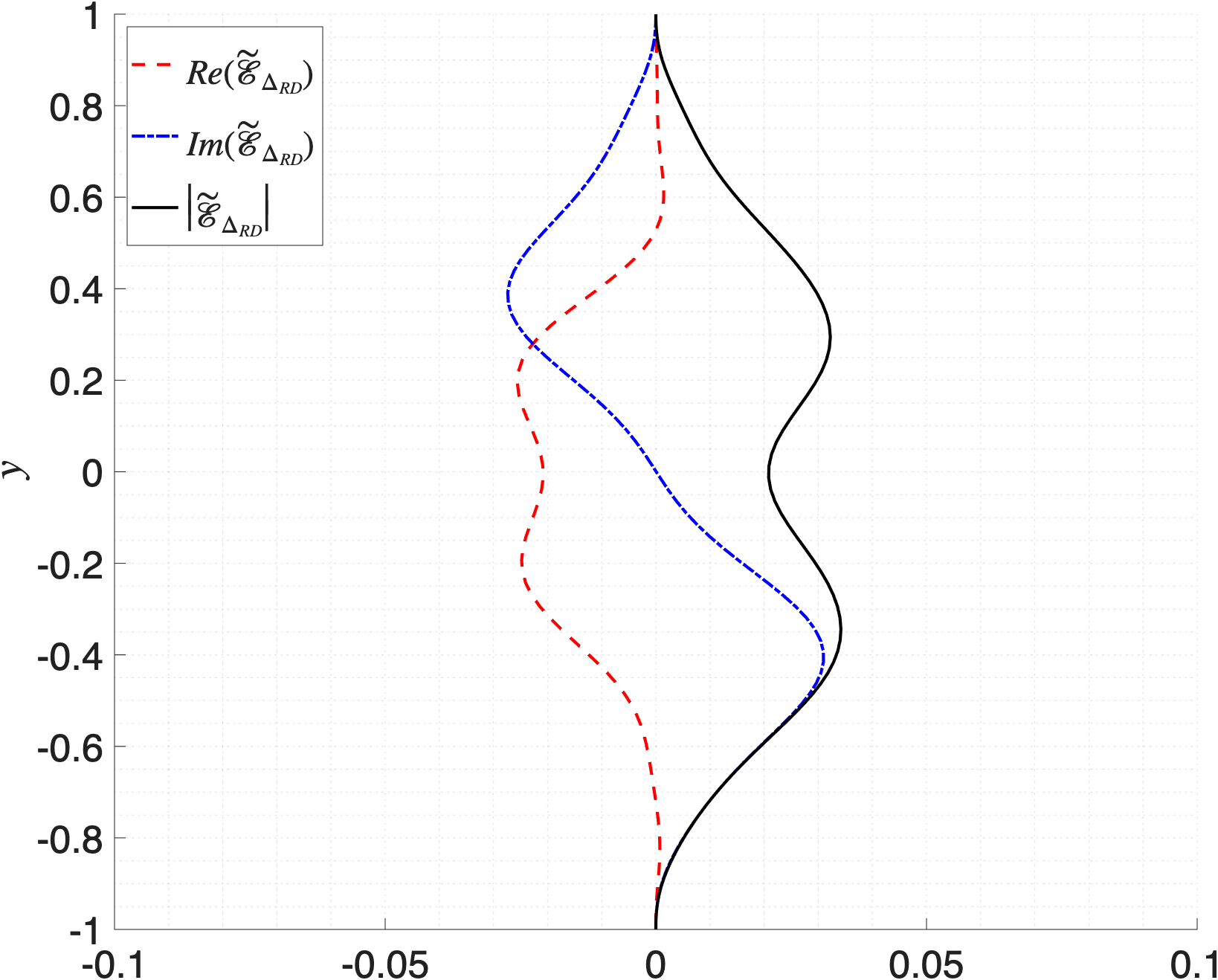}
        \caption{$\mathcal{E}_{\mathbf{\Delta}_{RD}}$}
        \label{subfig:ep_delu_sim}
    \end{subfigure}
    \hfill
    \begin{subfigure}[b]{0.32\textwidth}
        \centering
        \includegraphics[width=\textwidth]{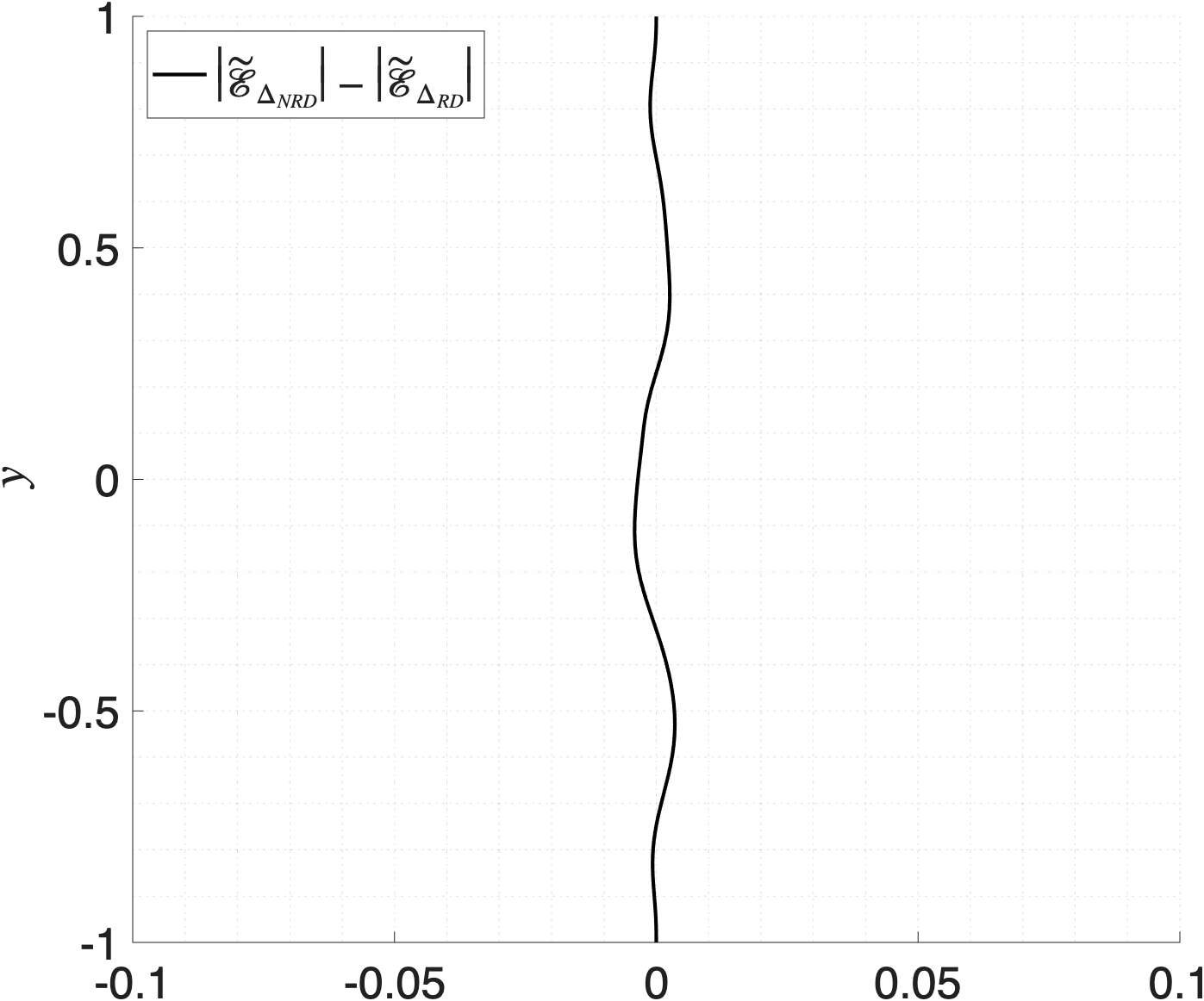} 
        \caption{$|\widetilde{\mathcal{E}}_{\mathbf{\Delta}_{NRD}}| - |\widetilde{\mathcal{E}}_{\mathbf{\Delta}_{RD}}|$} 
        \label{subfig:ep_diag_sim}
    \end{subfigure}
    
    \caption{$\widetilde{\mathcal{E}}_{\mathbf{\Delta}}$ computed for Poiseuille base flow at $Re=690$, with the wavenumber pair $(k_x,k_z)=(10^{-3},0.5)$ using 
    the uncertainty structures from~\S\ref{sec:uncertainty} ((a) $\mathbf{\Delta}_{NRB}$, (b) $\mathbf{\Delta}_{RB}$, (d) $\mathbf{\Delta}_{NRD}$, (e) $\mathbf{\Delta}_{RD}$), and the difference in the absolute artificial energy profile between (c) block uncertainties, and (f) diagonal uncertainties.}
    \label{fig:ep_sim}
\end{figure}

Fig.~\ref{fig:ep_sim} shows $\widetilde{\mathcal{E}}_{\mathbf{\Delta}_{NRB}}$, $\widetilde{\mathcal{E}}_{\mathbf{\Delta}_{RB}}$, $\widetilde{\mathcal{E}}_{\mathbf{\Delta}_{NRD}}$, and $\widetilde{\mathcal{E}}_{\mathbf{\Delta}_{RD}}$, computed for Poiseuille base flow at $Re=690$, using the wavenumber pair $(k_x,k_z)=(10^{-3},0.5)$, chosen to assess how the artificial energy production is affected when the stability thresholds predicted by repeated and non-repeated approaches are similar.

Once again, the profiles generated by each
repeated structure are very similar to those generated by its non-repeated
counterpart,  as can be seen by the very low values in Fig.~\ref{subfig:ep_block_sim} and Fig.~\ref{subfig:ep_diag_sim}. Integrating the profiles presented in Fig.~\ref{fig:ep_sim} yields the following absolute values:
\begin{equation}
\begin{split}
    |\widetilde{\mathcal{P}}_{\mathbf{\Delta}_{NRB}}|
        &= 0.07477, \qquad
    |\widetilde{\mathcal{P}}_{\mathbf{\Delta}_{RB}}|
        = 0.07665,\\
    |\widetilde{\mathcal{P}}_{\mathbf{\Delta}_{NRD}}|
        &= 0.03723, \qquad
    |\widetilde{\mathcal{P}}_{\mathbf{\Delta}_{RD}}|
        = 0.03731.
\end{split}
\end{equation}
The two full-block results differ by only approximately $2.5\%$, while the two
diagonal results differ by approximately $0.2\%$. Herein, lifting the repeated
entries constraint provides almost no additional artificial energy-production mechanism for this mode. This suggests that a common uncertainty field is
already sufficiently well aligned with the dominant amplification mechanism, as was observed for Couette flow. This result is consistent with Fig.~\ref{fig:prf_poiseuille}, where the repeated
and non-repeated formulations yield similar stability thresholds at this
wavenumber pair.

The results from both sections \S~\ref{sec:P_Couette} for Couette and  \S~\ref{sec:P_Poiseuille}  for plane Poiseuille flow provide strong evidence for the claim that the localized region of oblique modes with smaller stability thresholds is caused by a mechanism of fictitious energy production. This fictitious production term arises because the structured uncertainty used in the structured input-output formulation is not divergence-free and appears significantly larger in the region of oblique modes with lower stability thresholds when the repeated-entries constraint is not enforced. Oblique-wave interactions themselves are a well-established physical mechanism in subcritical transition \citep{reddy1998stability}. Experimental observations in plane Poiseuille flow have shown that nonlinear interactions of oblique waves generate streaks and can lead to breakdown \citep{elofsson1998experimental,reddy1998stability}, while nonlinear transition studies of plane Couette flow have identified oblique-wave disturbances as close to minimal transition-triggering perturbations \citep{duguet2010formation}. Therefore, the present results should not be interpreted as indicating that the oblique-wave mechanism itself is fictitious. Rather, they indicate that the additional reduction in stability threshold obtained when the repeated-entry constraint is removed is accompanied by additional artificial energy production absent from the Navier–Stokes energy balance. The comparison with simulation results for an oblique wave transition scenario in Fig.~\ref{fig:modes_Re_Couette}(b) shows that imposing the repeated-entry constraint results in very accurate stability threshold predictions, while lifting this constraint is expected to cause significantly reduced values. This comparison further supports the claim that additional artificial energy production results in smaller, less accurate stability bounds.

\subsection{Reynolds number effect: artificial energy production}
\label{sec:P_Re}

\begin{figure}[ht!]
    \centering
    \begin{subfigure}[b]{0.48\textwidth}
        \centering
        \includegraphics[width=\textwidth]{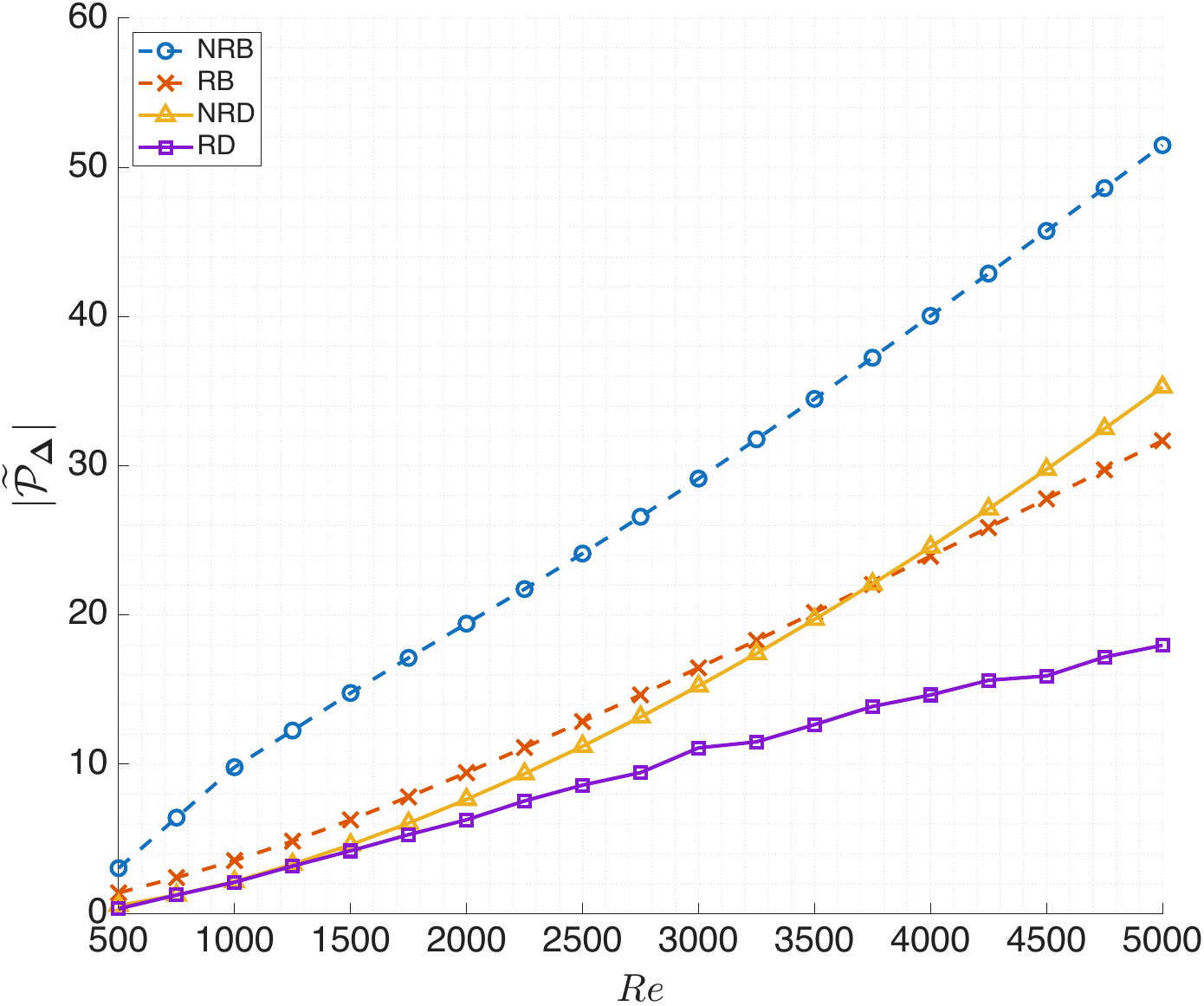}
        \caption{Couette Flow, $(k_x,k_z)=(0.196,0.628)$}
        \label{subfig:P_RE_Couette}
    \end{subfigure}
    \hfill
    \begin{subfigure}[b]{0.48\textwidth}
        \centering
        \includegraphics[width=\textwidth]{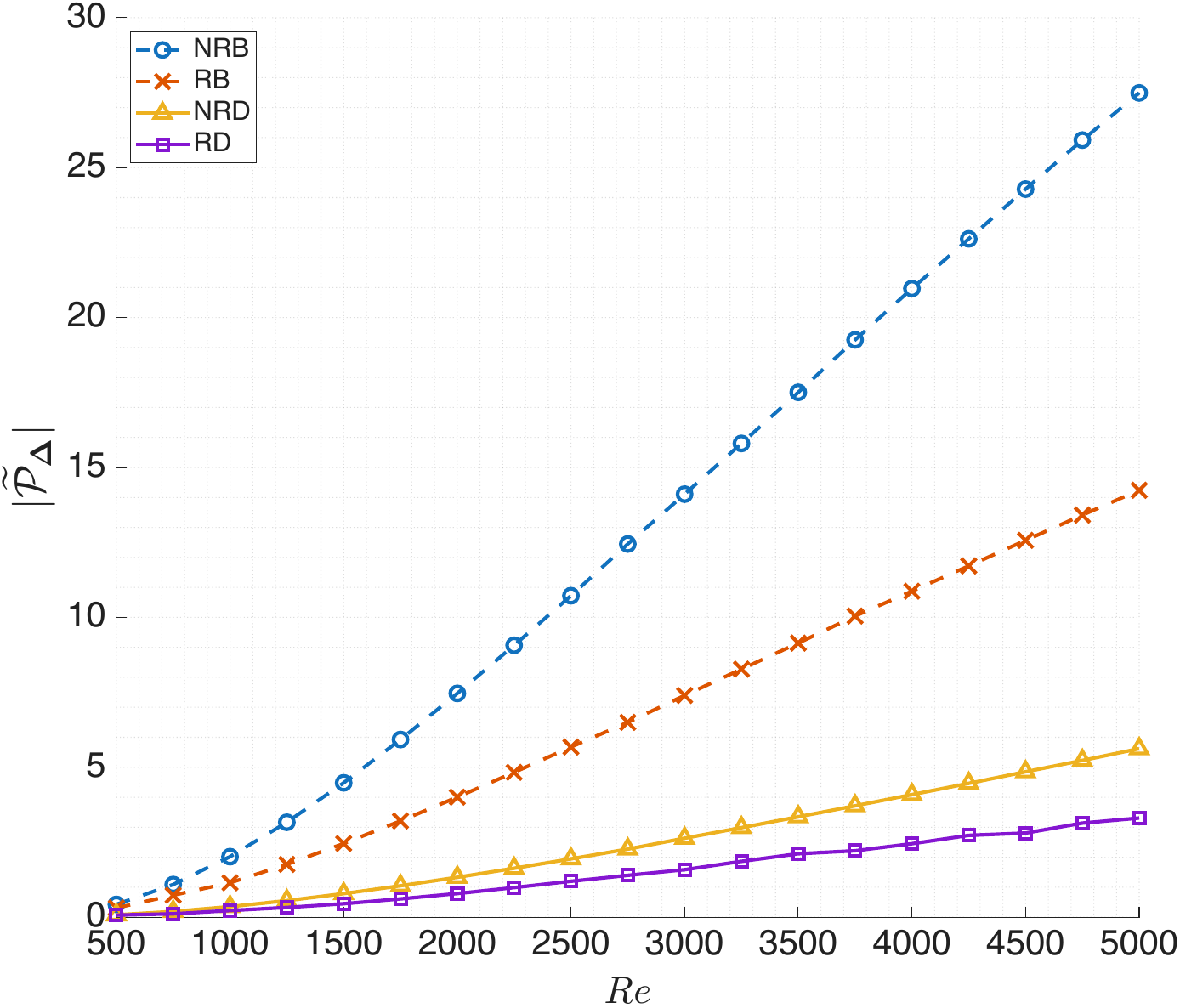}
        \caption{Plane Poiseuille Flow, $(k_x,k_z)=(0.692,1.561)$}
        \label{subfig:P_RE_Poiseiulle}
    \end{subfigure}

    \caption{$|\widetilde{\mathcal{P}}_{\mathbf{\Delta}}|$ computed for (a) Couette base flow with the wavenumber pair $(k_x,k_z)=(0.196,0.628)$, and (b) Poiseuille base flow with the wavenumber pair $(k_x,k_z)=(0.692,1.561)$, computed for $Re\in[250,5000]$.}
    \label{fig:P_RE}
\end{figure}

Lastly, we show that the results presented in this section for a specific Reynolds number also hold for a wide range of Reynolds numbers. We consider the range $Re\in[250,5000]$ for both Couette and plane Poiseuille flows and compute $|\widetilde{\mathcal{P}}_{\mathbf{\Delta}_{NRB}}|$, $|\widetilde{\mathcal{P}}_{\mathbf{\Delta}_{RB}}|$, $|\widetilde{\mathcal{P}}_{\mathbf{\Delta}_{NRD}}|$, and $|\widetilde{\mathcal{P}}_{\mathbf{\Delta}_{RD}}|$ for the modes where the repeated and non-repeated uncertainty structures produce different results. The results of this computation are presented in Fig.~\ref{fig:P_RE}.
These results indicate that for all Reynolds numbers which were tested, the artificial energy production becomes significantly smaller when applying the repeated entries constraint on the uncertainty,  both for the block uncertainties (i.e., $|\widetilde{\mathcal{P}}_{\mathbf{\Delta}_{RB}}|<|\widetilde{\mathcal{P}}_{\mathbf{\Delta}_{NRB}}|$) and for the diagonal uncertainties (i.e., $|\widetilde{\mathcal{P}}_{\mathbf{\Delta}_{RD}}|<|\widetilde{\mathcal{P}}_{\mathbf{\Delta}_{NRD}}|$). While for plane Poiseuille flow the block uncertainties produce significantly larger artificial energy production than the diagonal ones, for Couette flow the separation is less clear and the curves corresponding to $|\widetilde{\mathcal{P}}_{\mathbf{\Delta}_{RB}}|$ and $|\widetilde{\mathcal{P}}_{\mathbf{\Delta}_{NRD}}|$ crossing at $Re=3750$. This observation aligns with our analysis: while we expect diagonal uncertainties to cause less artificial energy production, and thus less conservative stability thresholds, a fair comparison should treat repeated uncertainties separately from the non-repeated ones. For both flows, the uncertainty structure that yields the smallest artificial energy production for all Reynolds numbers is $\boldsymbol{\Delta}_{RD}$, indicating that our approach of using a diagonal uncertainty structure with repeated entries and transforming the interconnected system to obtain an equivalent SSV problem leads to the smallest artificial energy production and to the most accurate stability predictions.

\section{CONCLUSIONS}
\label{sec:conclusions}

This work presents an uncertainty representation derived via linear transformations of the input and output channels that map the feedback loop such that the resulting structured uncertainty has a repeated-diagonal structure. This structure aims to preserve the component-wise pathways of the nonlinear advection term while keeping the structured singular value computation tractable. 
We use the structured singular values computed with this approach to assess the stability of incompressible shear flows under finite-magnitude perturbations, following a methodology similar to that proposed in our previous work utilizing the structured small gain theorem \cite{frank2026stability}.
The newly proposed structure within our stability analysis provides a sufficient condition for flow stability, offering a useful framework for studying flow instability in realistic environments subject to finite-size disturbances. If the disturbance magnitude is below the threshold we determine using the structured small-gain theorem approach, the system will remain stable. If the disturbance magnitude surpasses this threshold, the criterion no longer guarantees stability, and the flow may transition to a turbulent state.

We tested this approach by assessing the stability of Couette and plane Poiseuille flows over a broad range of Reynolds numbers and wavenumbers. The proposed formulation produced less conservative stability thresholds than the repeated full-block approximation while identifying the same dominant instability mechanisms. These thresholds were compared to the results of \cite{reddy1998stability,duguet2010towards,duguet2013minimal,farano2015hairpin,parente2022minimal}, which computed energy thresholds for transition to turbulence using high-fidelity simulations and various optimization methods. This comparison showed strong agreement between thresholds obtained using our new uncertainty structure and the simulation results. The thresholds imposed by the repeated block uncertainty produce significantly less accurate bounds.
We show that the bound based on the desired structured uncertainty $\Delta_\mathbf{u}$ lies within a factor ranging from 1 to $\sqrt{3}$, using our approximated structure  $\Delta_{RD}$ (see Eq.~\eqref{eq:size_bound}).
In particular, among the uncertainty descriptions considered, our repeated-diagonal formulation bound, $\sqrt{3}\norm{R\mathscr{H}{\nabla}L}{\mu_{\boldsymbol{\Delta_{RD}}}}^{-1}$, yielded the least conservative computable thresholds. It also produced the lowest normalized artificial energy contribution and the closest agreement with previous numerical and experimental studies. Therefore, our approach provides more accurate stability thresholds within structured input-output analysis while offering a useful compromise between physical fidelity and computational tractability.

We demonstrated that the new structure captures subcritical bypass transition physics, explaining the flow structures that trigger the initial onset of instability observed in experiments: the bounds show that oblique waves and spanwise-periodic streaks require much smaller perturbations to induce instability at lower Reynolds numbers. As the critical Reynolds number is approached, the predicted thresholds decrease smoothly and approach the LST limit as the Tollmien-Schlichting mode becomes dominant.
The energy analysis shows that artificial energy production increases significantly when the repeated uncertainty-term constraint is lifted for a specific group of oblique flow structures, while other modes are unaffected. This group of oblique modes also shows much lower stability thresholds under non-repeated uncertainty structures. These results provide strong evidence that the localized region of oblique modes with smaller stability thresholds is caused by a mechanism of fictitious energy production. The fact that the stability thresholds produced using $\boldsymbol{\Delta}_{RD}$ (which is a repeated structure) for an oblique transition scenario were found to be very consistent with the results of \cite{reddy1998stability,duguet2010towards,duguet2013minimal} also strengthens this claim. This is not to say that oblique flow structures are not central to the transition process, as oblique-wave interactions are a well-established physical mechanism in subcritical transition \citep{reddy1998stability}. Rather, our results only indicate that the additional reduction in stability threshold obtained when the repeated-entry constraint is removed is accompanied by additional artificial energy production absent from the Navier–Stokes energy balance.

The current framework is well suited to study the onset of instability at early stages of transition, caused by transient growth and bypass transition of finite-size perturbations. For capturing the flow structures during the later stages of the transition process that are associated with secondary instabilities, incorporating higher-order perturbation analyses into the input-output framework should be considered, as proposed in a recent study by \cite{bovzic2026oblique}. 
The uncertainty structures considered within the structured input-output framework proposed to date, including this study, are not explicitly constrained to be divergence-free, so artificial energy production is not eliminated completely. Future work should incorporate the incompressibility constraint directly into the SSV computation process to eliminate all artificial energy production. Furthermore, the stability analysis framework suggested here can be extended in the future to more complex (2D, 3D) and unsteady base flows, utilizing the Lyapunov-based approach proposed recently by  \cite{sahu2026lyapunov}.

\appendix

\section{SSV TRANSFORMATION AND STABILITY THRESHOLD BOUND}
\label{sec:A}

Herein, we show the derivation of Eq.~\eqref{eq:size_bound}. First, we define the following matrices:
\begin{equation}
    P = \begin{bmatrix}
        e_1 & e_4 & e_7 & e_2 & e_5 & e_8 & e_3 & e_6 & e_9
    \end{bmatrix}^T.
\end{equation}
Here, $e_i$ is the $i$-th standard basis vector, defined by the $i$-th entry being equal to $1$ and the rest of the entries being equal to $0$. 
We define the block matrix $K \in \mathbb{R}^{rc\times rc}$ that consists of $r\times c$ blocks of matrices $E_{\ell m}\in\mathbb{R}^{c\times r}$, i.e.,
\begin{equation}
    K_{r,c} = \begin{bmatrix} 
    E_{11} & E_{12} & \cdots & E_{1c} \\ 
    E_{21} & E_{22} & \cdots & E_{2c} \\ 
    \vdots & \vdots & \ddots & \vdots \\ 
    E_{r1} & E_{r2} & \cdots & E_{rc} 
    \end{bmatrix}.
\end{equation}
Each block $E_{\ell m}\in \mathbb{R}^{c\times r}$ is a matrix with elements $e_{i,j}$ satisfying:
\begin{equation}
    e_{i,j} = 
    \begin{cases}
        1, \quad i=m, \, j=\ell \\
        0, \quad \text{else}
    \end{cases}.
\end{equation}
The matrices $L$ and $R$ are defined below:
\begin{equation}
    R = (I_{N_y}\otimes P)K_{N_y,9}, \,\,\,\, L = K_{3,N_y}(I_{N_y}\otimes \begin{bmatrix}
        I_3 & I_3 & I_3
    \end{bmatrix}).
\end{equation}
These matrices were designed to maintain the following relationship:
\begin{equation}
    \boldsymbol{\Delta}_{\mathbf{u}} \equiv \{\mathbf{U}_\Xi: \mathbf{U}_\Xi= L {\mathbf{V}}_\Xi R, \,\, {\mathbf{V}}_\Xi\in \mathbf{\Delta}_{RD} \},
\end{equation}
meaning that for every $\mathbf{U}_\Xi\in \boldsymbol{\Delta}_{\mathbf{u}}$ there exists $\mathbf{V}_\Xi\in \mathbf{\Delta}_{RD}$ such that $\mathbf{U}_\Xi= L {\mathbf{V}}_\Xi R$. Furthermore, these matrices do not affect the nonzero entries of $\mathbf{V}_\Xi$, only reordering them to obtain the diagonal structure $\mathbf{\Delta}_{RD}$. The dimensions of these matrices are defined by the number of discretization points we used, denoted as $N_y$. Herein, we use $\mathbf{U}_\Xi \in \mathbb{C}^{3N_y\times 9N_y}$, $L \in \mathbb{R}^{3N_y \times 9N_y}$, {$R \in \mathbb{R}^{9N_y \times 9N_y}$}, and $\mathbf{V}_\Xi \in \mathbb{C}^{9N_y\times 9N_y}$.

The determinant condition in the SSV definition shown in Eq.~\eqref{eq:SSV} becomes:
\begin{equation}
\label{eq:det1}
    \text{det}[I - \mathscr{H}_{\nabla} {\mathbf{U}}_\Xi] = \text{det}[I - \mathscr{H}_{\nabla} L {\mathbf{V}}_\Xi R].
\end{equation}
By employing the Weinstein–Aronszajn determinant identity (see \cite[][Corollary 2.8.5.]{bernstein2009matrix}), we obtain:
\begin{equation}
\label{eq:det}
    \text{det}[I - \mathscr{H}_{\nabla} L {\mathbf{V}}_\Xi R] = \text{det}[I - R \mathscr{H}_{\nabla} L {\mathbf{V}}_\Xi].
\end{equation}

Next, we relate the singular values of two uncertainty matrices $\mathbf{U}_\Xi\in\boldsymbol{\Delta}_{\mathbf{u}}$ and $\mathbf{V}_\Xi\in \mathbf{\Delta}_{RD}$. It was shown in \cite{frank2026stability} that
\begin{equation}
    \bar{\sigma}(\mathbf{U}_\Xi) = \max_i \sqrt{ |{u_\xi}_i|^2 + |{v_\xi}_i|^2 + |{w_\xi}_i|^2}.
\end{equation}
As in Eq.~\eqref{eq:structexact}, $u_\xi, v_\xi, w_\xi \in \mathbb{C}^{N_y\times 1}$ represent discretized versions of the perturbation velocity components. The notation ${u_{\xi}}_i$ (and similarly for the two other velocity components) denotes the $i$-th entry of ${u_{\xi}}$ in the discretized domain, which contains $N_y$ entries. Since $\mathbf{V}_\Xi$ is a diagonal matrix  of reorganized entries from  $\mathbf{U}_\Xi$, its largest singular value is
\begin{equation}
\label{eq:SV_unc}
    \bar{\sigma}(\mathbf{V}_\Xi) = \max_i \{ |{u_\xi}_i|,|{v_\xi}_i|,|{w_\xi}_i| \}.
\end{equation}
 It is clear that
\begin{equation}
\label{eq:SV_rel}
    \bar{\sigma}(\mathbf{U}_\Xi) \leq \sqrt{3}\bar{\sigma}(\mathbf{V}_\Xi).
\end{equation}
Using the determinant equality in Eq.~\eqref{eq:det} and the singular value relation in Eq.~\eqref{eq:SV_rel} within the minimization
problem in Eq.~\eqref{eq:SSV}, we obtain 
\begin{equation}
\label{eq:mu_relation}
    \frac{1}{\sqrt{3}}\mu_{\mathbf{\Delta}_{RD}}(R\mathscr{H}_\nabla L) \leq \mu_{\boldsymbol{\Delta}_\mathbf{u}}(\mathscr{H}_\nabla).
\end{equation}
Then, based on Eq.~\eqref{eq:H_ssv}, we obtain 
\begin{equation}
\label{eq:H_bound}
     \frac{1}{\sqrt{3}}\norm{R\mathscr{H}_{\nabla}L}_{\mu_{\boldsymbol{\Delta_{RD}}}} \leq \norm{\mathscr{H}_{\nabla}}_{\mu_{\boldsymbol{\Delta_{\mathbf{u}}}}}.
\end{equation}
Notably, Eq.~\eqref{eq:SV_unc} also results in the relationship
\begin{equation}
    \bar{\sigma}(\mathbf{V}_\Xi) \leq \bar{\sigma}(\mathbf{U}_\Xi).
\end{equation}
Following the same logic used in Eq.~\eqref{eq:mu_relation} and Eq.~\eqref{eq:H_bound}, we reach the following relation:
\begin{equation}
    \frac{1}{\sqrt{3}}\norm{R\mathscr{H}_{\nabla}L}_{\mu_{\boldsymbol{\Delta_{RD}}}} \leq \norm{\mathscr{H}_{\nabla}}_{\mu_{\boldsymbol{\Delta_{\mathbf{u}}}}} \leq \norm{R\mathscr{H}_{\nabla}L}_{\mu_{\boldsymbol{\Delta_{RD}}}}.
\end{equation}
Thus, our approach computes stability thresholds that approximate $\norm{\mathscr{H}_{\nabla}}_{\mu_{\boldsymbol{\Delta_{\mathbf{u}}}}}$. In the worst case, this approximation is tight up to a factor of $\sqrt{3}$. We obtain Eq.~\eqref{eq:size_bound} by computing the inverse of Eq.~\eqref{eq:H_bound}. 

\section{ARTIFICIAL ENERGY PRODUCTION DERIVATION}
\label{sec:app_b}
We start by writing the momentum NS equation for velocity perturbations using Einstein's index notation:
\begin{equation}
    \label{eq:NSE_index}
    \frac{\partial u_i}{\partial t}
    + u_j \frac{\partial u_i}{\partial x_j}
    + U_j \frac{\partial u_i}{\partial x_j}
    + u_j \frac{\partial U_i}{\partial x_j}
    =
    -\frac{1}{\rho}\frac{\partial p}{\partial x_i}
    + \nu \frac{\partial^2 u_i}{\partial x_j^2}.
\end{equation}
Here, $u$ represents velocity perturbations, $U$ represents the base flow velocity profile, $\rho$ is the fluid's density, and $\nu$ is the kinematic viscosity.
Following closely the derivation in \cite[chapter 12 section 10]{kundu2024fluid}, we obtain the following form of the energy equation:
\begin{equation}
    \label{eq:energy}
    \frac{d}{dt}\int \frac{u_i^2}{2}dV = -\int u_i u_j \frac{\partial u_i}{\partial x_j} dV - \int u_i U_j \frac{\partial u_i}{\partial x_j} dV - \int u_i u_j \frac{\partial U_i}{\partial x_j} dV - \frac{1}{\rho} \int u_i \frac{\partial p}{\partial x_i} + \nu \int u_i \frac{\partial^2 u_i}{\partial x_j^2} dV,
\end{equation}
where $dV$ is a volume element. By applying the incompressibility assumption to the base flow ($\partial U_j/\partial x_j=0$), Gauss' theorem, and the boundary conditions in Eq.~\eqref{eq:BC}, we reach the following form of the energy equation
\begin{equation}
    \label{eq:energy2}
    \frac{d}{dt}\int \frac{u_i^2}{2}dV = -\int u_i u_j \frac{\partial U_i}{\partial x_j} dV - \nu \int \left(\frac{\partial u_i}{\partial x_j}\right)^2 dV + \frac{1}{2}\int u_i^2 \frac{\partial u_j}{\partial x_j} dV.
\end{equation}

Besides the standard production and dissipation terms, this form of the energy equation includes the term $\frac{1}{2}\int u_i^2 \frac{\partial u_j}{\partial x_j} dV$, which can be written using a non-index convention as $\frac{1}{2}\int \norm{\mathbf{u}}^2 (\nabla\cdot\mathbf{u})) dV$. This term is usually not found in the energy equation (see e.g. \cite{darzin1981hydrodynamic,schmid2002stability}). This term is dropped in most derivations due to the continuity of the velocity perturbation field ($\partial u_j / \partial x_j = 0$), which we did not assume in this derivation. As explained in~\S\ref{sec:math}, we essentially replace $u_j$ in the term $u_j \frac{\partial u_i}{\partial x_j}$ with a structured uncertainty term. This methodology does not enforce a divergence-free condition on the structured uncertainty. Thus, we expect a term similar to $\frac{1}{2}\int u_i^2 \frac{\partial u_j}{\partial x_j} dV$ in the energy balance as a result of our methodology. This term is a fictitious term producing artificial energy, which is not part of the original NSE system and causes a discrepancy between real flow physics and the structured input-output modeling approach based on a fixed structure; thus, our goal is to obtain an uncertainty structure that faithfully represents the non-linear advection term while minimizing this fictitious energy term. In the rest of this section, we follow the derivation leading to Eq.~\eqref{eq:energy2} using our model of a structured uncertainty term, to find the form of the artificial energy production term.

We consider here uncertainty matrices of the form:
\begin{equation}
    \label{eq:U_xi}
    \mathbf{U}_\Xi = 
    \begin{bmatrix}
        \begin{matrix}
            {U_\xi}_{1,1} & {U_\xi}_{1,2} & {U_\xi}_{1,3}
        \end{matrix} & \mathbf{O}_{N_y\times 3N_y} & \mathbf{O}_{N_y\times 3N_y} \\
        \mathbf{O}_{N_y\times 3N_y} & 
        \begin{matrix}
            {U_\xi}_{2,1} & {U_\xi}_{2,2} & {U_\xi}_{2,3}
        \end{matrix}
        & \mathbf{O}_{N_y\times 3N_y} \\
        \mathbf{O}_{N_y\times 3N_y} & \mathbf{O}_{N_y\times 3N_y} & 
        \begin{matrix}
            {U_\xi}_{3,1} & {U_\xi}_{3,2} & {U_\xi}_{3,3}
        \end{matrix}
    \end{bmatrix}
\end{equation}
This is a general form that allows modeling $\boldsymbol{\Delta}_{\mathbf{u}}$, $\boldsymbol{\Delta}_{RB}$, and $\boldsymbol{\Delta}_{NRB}$. As shown in~\S\ref{sec:A}, the transformation we apply makes $\boldsymbol{\Delta}_{RD}$ equivalent to $\boldsymbol{\Delta}_{\mathbf{u}}$. A similar result is obtained for $\boldsymbol{\Delta}_{NRD}$, where the only difference is the lifting of the repeated uncertainty entries constraints.

We reorder the terms in the uncertainty matrix in a third-order uncertainty tensor
$\mathcal{U}_{ijk}$. The components of this tensor are defined as follows:
\begin{equation}
    \begin{cases}
        \mathcal{U}_{1j1} = {U_\xi}_{1,j}, \quad j={1,2,3}
        \\
        \mathcal{U}_{2j2} = {U_\xi}_{2,j}, \quad j={1,2,3}
        \\
        \mathcal{U}_{3j3} = {U_\xi}_{3,j}, \quad j={1,2,3}
        \\
        \mathcal{U}_{ijk} = 0, \quad i\neq k.
    \end{cases}
    \label{eq:convective_uncertainty_tensor}
\end{equation}
It follows directly from Eq.~\eqref{eq:convective_uncertainty_tensor}
that
\begin{equation}
    \mathcal{U}_{ijk}
    =
    \mathcal{U}_{kji}.
    \label{eq:convective_uncertainty_symmetry}
\end{equation}
This tensor is constructed to demonstrate the effect of the structured uncertainty on the flow dynamics. Using this tensor, the perturbation momentum equation can be written as
\begin{equation}
    \frac{\partial u_i}{\partial t}
    =
    -\mathcal{U}_{ijk}
    \frac{\partial u_k}{\partial x_j}
    -U_j\frac{\partial u_i}{\partial x_j}
    -u_j\frac{\partial U_i}{\partial x_j}
    -\frac{1}{\rho}\frac{\partial p}{\partial x_i}
    +\nu\frac{\partial^2 u_i}{\partial x_j^2}.
    \label{eq:uncertain_perturbation_momentum}
\end{equation}

Multiplying Eq.~\eqref{eq:uncertain_perturbation_momentum} by $u_i$
and integrating over the flow volume gives
\begin{align}
    \frac{\mathrm{d}}{\mathrm{d}t}
    \int \frac{u_i^2}{2}\,\mathrm{d}V
    ={}&
    -\int
    u_i\mathcal{U}_{ijk}
    \frac{\partial u_k}{\partial x_j}
    \,\mathrm{d}V
    -\int
    u_iU_j\frac{\partial u_i}{\partial x_j}
    \,\mathrm{d}V
    \\
    &-\int
    u_i u_j\frac{\partial U_i}{\partial x_j}
    \,\mathrm{d}V
    -\frac{1}{\rho}
    \int
    u_i\frac{\partial p}{\partial x_i}
    \,\mathrm{d}V
    +\nu
    \int
    u_i\frac{\partial^2u_i}{\partial x_j^2}
    \,\mathrm{d}V .
    \label{eq:uncertain_energy_initial}
\end{align}
Thus, we can rewrite the energy equation similarly to Eq.~\eqref{eq:energy2} as
\begin{equation}
    \label{eq:energy_uncertainty}
    \frac{d}{dt}\int \frac{u_i^2}{2}dV = -\int u_i u_j \frac{\partial U_i}{\partial x_j} dV - \nu \int \left(\frac{\partial u_i}{\partial x_j}\right)^2 dV + \mathcal{P}_{\Xi},
\end{equation}
where the contribution of the uncertainty-induced nonlinear term is therefore
\begin{equation}
    \mathcal{P}_{\Xi}
    =
    -\int
    u_i\mathcal{U}_{ijk}
    \frac{\partial u_k}{\partial x_j}
    \,\mathrm{d}V .
    \label{eq:uncertain_nonlinear_production_initial}
\end{equation}

We note that the following relationship is always true:
\begin{equation}
    \frac{\partial}{\partial x_j}
    \left(
        u_i\mathcal{U}_{ijk}u_k
    \right)
    =
    \frac{\partial u_i}{\partial x_j} \mathcal{U}_{ijk}u_k + u_i \frac{\partial\mathcal{U}_{ijk}}{\partial x_j}
    u_k + u_i\mathcal{U}_{ijk} \frac{\partial u_k}{\partial x_j}.
    \label{eq:uncertain_product_rule}
\end{equation}
Interchanging the indices $i$ and $k$ and using
Eq.~\eqref{eq:convective_uncertainty_symmetry} yields
\begin{equation}
    \frac{\partial u_i}{\partial x_j}
    \mathcal{U}_{ijk}u_k
    =
    \frac{\partial u_k}{\partial x_j}
    \mathcal{U}_{kji}u_i
    =
    u_i\mathcal{U}_{ijk}
    \frac{\partial u_k}{\partial x_j}.
    \label{eq:uncertain_symmetric_terms}
\end{equation}
Substituting Eq.~\eqref{eq:uncertain_symmetric_terms} into Eq.~\eqref{eq:uncertain_product_rule} gives
\begin{equation}
    \label{eq:uncertain_product_rule2}
    \frac{\partial}{\partial x_j}
    \left(
        u_i\mathcal{U}_{ijk}u_k
    \right)
    =
    2u_i\mathcal{U}_{ijk} \frac{\partial u_k}{\partial x_j} + u_i \frac{\partial\mathcal{U}_{ijk}}{\partial x_j}
    u_k.
\end{equation}

Thus,
\begin{equation}
    u_i\mathcal{U}_{ijk}
    \frac{\partial u_k}{\partial x_j}
    =
    \frac{1}{2}
    \frac{\partial}{\partial x_j}
    \left(
        u_i\mathcal{U}_{ijk}u_k
    \right)
    -
    \frac{1}{2}
    u_i u_k
    \frac{\partial\mathcal{U}_{ijk}}{\partial x_j}.
    \label{eq:uncertain_nonlinear_identity}
\end{equation}

Substitution of Eq.~\eqref{eq:uncertain_nonlinear_identity} into
Eq.~\eqref{eq:uncertain_nonlinear_production_initial} gives
\begin{align}
    \mathcal{P}_{\Xi} = -\frac{1}{2} \int \frac{\partial}{\partial x_j} \left( u_i\mathcal{U}_{ijk}u_k \right) \,\mathrm{d}V
    +
    \frac{1}{2} \int u_i u_k \frac{\partial\mathcal{U}_{ijk}}{\partial x_j}
    \,\mathrm{d}V.
    \label{eq:uncertain_nonlinear_result}
\end{align}
Applying Gauss' theorem to the first term yields
\begin{equation}
    \mathcal{P}_{\Xi}=
    -\frac{1}{2}
    \int
    u_i\mathcal{U}_{ijk}u_k n_j
    \,\mathrm{d}S
    +
    \frac{1}{2}
    \int
    u_i u_k
    \frac{\partial\mathcal{U}_{ijk}}{\partial x_j}
    \,\mathrm{d}V,
    \label{eq:uncertain_nonlinear_gauss}
\end{equation}
where $dS$ is a surface element. The surface integral vanishes when applying the boundary conditions in Eq.~\eqref{eq:BC}. Hence,
\begin{equation}
    \mathcal{P}_{\Xi}
    =
    \frac{1}{2}
    \int
    u_i u_k
    \frac{\partial\mathcal{U}_{ijk}}{\partial x_j}
    \,\mathrm{d}V.
    \label{eq:uncertain_nonlinear_production}
\end{equation}

Using the fact that for $i\neq k$,  $\mathcal{U}_{ijk} = 0$, this term can be expanded as
\begin{equation}
    \mathcal{P}_{\Xi}
    =
    \frac{1}{2}
    \int
    \left[
        u^2\frac{\partial {U_\xi}_{1,j}}{\partial x_j}
        +
        v^2\frac{\partial {U_\xi}_{2,j}}{\partial x_j}
        +
        w^2\frac{\partial {U_\xi}_{3,j}}{\partial x_j}
    \right]
    \,\mathrm{d}V ,
\label{eq:uncertain_nonlinear_production_expanded}
\end{equation}
where $u\equiv u_1$, $v\equiv u_2$, and $w\equiv u_3$.

We define:
\begin{equation}
    \label{eq:uncertainty_vec}
    \begin{cases}
        {U_\xi}_1 = \begin{bmatrix}
        {U_\xi}_{1,1} & {U_\xi}_{1,2} & {U_\xi}_{1,3}
    \end{bmatrix} \\
        {U_\xi}_2 = \begin{bmatrix}
        {U_\xi}_{2,1} & {U_\xi}_{2,2} & {U_\xi}_{2,3}
    \end{bmatrix} \\
        {U_\xi}_3 = \begin{bmatrix}
        {U_\xi}_{3,1} & {U_\xi}_{3,2} & {U_\xi}_{3,3}
    \end{bmatrix}
    \end{cases}
\end{equation}
Using this definition, we obtain 
Eq.~\eqref{eq:P_xi_body} in the main body of the manuscript,
\begin{equation}
    \label{eq:P_xi}
    \mathcal{P}_{\Xi} =
    \frac{1}{2}
    \int
    \left[
        u^2 (\nabla\cdot {U_\xi}_1)
        +
        v^2 (\nabla\cdot {U_\xi}_2)
        +
        w^2 (\nabla\cdot {U_\xi}_3)
    \right]
    \,\mathrm{d}V.
\end{equation}
For repeated uncertainty structures, this term simplifies to
\begin{equation}
    \label{eq:P_xis}
    \mathcal{P}_{\Xi} =
    \frac{1}{2}
    \int
        \norm{\mathbf{u}}^2 (\nabla\cdot {U_\xi}_1)
    \,\mathrm{d}V.
\end{equation}
The term $\mathcal{P}_\Xi$ (either in  \eqref{eq:P_xi} or \eqref{eq:P_xis}) produces artificial energy that is not found in the original NS system and arises because uncertainty matrices are not forced to be divergence-free.

\bibliography{apssamp}

@article{sahu2026lyapunov,
  title={Lyapunov stability analysis of the chaotic flow past two square cylinders},
  author={Sahu, Sidhartha and Papadakis, George},
  journal={Journal of Fluid Mechanics},
  volume={1034},
  pages={A35},
  year={2026},
  publisher={Cambridge University Press}
}

@INPROCEEDINGS{peet2025inflectional,
  author={Peet, Yulia T. and Purra, Varshitha},
  booktitle={2025 IEEE 64th Conference on Decision and Control (CDC)}, 
  title={Inflectional Instability of Linearized Incompressible Euler Equations via Linear Partial Inequality Tests}, 
  year={2025},
  volume={},
  number={},
  pages={8358-8364},
  doi={10.1109/CDC57313.2025.11312828}}

@article{bovzic2026oblique,
  title={From oblique-wave forcing to streak reinforcement: A perturbation-based frequency-response framework},
  author={Bo{\v{z}}i{\'c}, Du{\v{s}}an and Dwivedi, Anubhav and Jovanovi{\'c}, Mihailo R},
  journal={Physical Review Fluids},
  volume={11},
  number={6},
  pages={063901},
  year={2026},
  publisher={APS}
}

@inproceedings{rath2024structured,
  title={A structured input-output approach to evaluating the effects of uniform wall-suction on optimal perturbations in transitional boundary layers},
  author={Rath, Aishwarya and Liu, Chang and Gayme, Dennice F},
  booktitle={2024 IEEE 63rd Conference on Decision and Control (CDC)},
  pages={7714--7719},
  year={2024},
  organization={IEEE}
}

@article{song2026structured,
  title={Structured input--output analysis of compliant wall turbulence},
  author={Song, Jingzhou and Wu, Ting and Wu, Chutian and Liu, Chang and He, Guowei},
  journal={Journal of Fluid Mechanics},
  volume={1035},
  pages={A7},
  year={2026},
  publisher={Cambridge University Press}
}

@INPROCEEDINGS{luiACC23,
  author={Liu, Chang and Shuai, Yu and Rath, Aishwarya and Gayme, Dennice F.},
  booktitle={2023 American Control Conference (ACC)}, 
  title={A structured input-output approach to characterizing optimal perturbations in wall-bounded shear flows}, 
  year={2023},
  volume={},
  number={},
  pages={2319-2325},
  doi={10.23919/ACC55779.2023.10156058}}

@article{farrell1993stochastic,
  title={ {Stochastic} forcing of the linearized {Navier}--{Stokes} equations },
  author={Farrell, B. F. and Ioannou, P. J.},
  journal={Phys. Fluids A},
  volume={5},
  number={11},
  pages={2600--2609},
  year={1993},
}

@article{bamieh2001energy,
  title={ {Energy} amplification in channel flows with stochastic excitation },
  author={Bamieh, B. and Dahleh, M.},
  journal={Phys. Fluids},
  volume={13},
  number={11},
  pages={3258--3269},
  year={2001},
}

@inproceedings{jovanovic2004unstable,
  author    = {Jovanovi{\'c}, M. R. and Bamieh, B.},
  title     = {{Unstable modes versus non-normal modes in supercritical channel flows}},
  booktitle = {Proc. Amer. Control Conf.},
  address   = {Boston, MA},
  pages     = {2245--2250},
  publisher = {IEEE},
  year      = {2004}
}

@article{jovanovic2005componentwise,
  title={ {Componentwise} energy amplification in channel flows },
  author={Jovanovi{\'c}, M. R. and Bamieh, B.},
  journal={J. Fluid Mech.},
  volume={534},
  pages={145--183},
  year={2005},
}

@article{HC10,
  author  = {Hwang, Y. and Cossu, C.},
  title   = {Amplification of coherent streaks in the turbulent {Couette} flow: an input-output analysis at low {Reynolds} number},
  journal = {J. Fluid Mech.},
  year    = {2010},
  volume  = {643},
  pages   = {333--348}
}

@article{mckeon2010critical,
  title={ {A} critical-layer framework for turbulent pipe flow },
  author={McKeon, B. J. and Sharma, A. S.},
  journal={J. Fluid Mech.},
  volume={658},
  pages={336--382},
  year={2010},
}

@article{mckeon2013experimental,
  author  = {McKeon, B. J. and Sharma, A. S. and Jacobi, I.},
  title   = {Experimental manipulation of wall turbulence: a systems approach},
  journal = {Phys. Fluids},
  volume  = {25},
  number  = {3},
  pages   = {031301},
  year    = {2013}
}

@article{mckeon2017engine,
  author  = {McKeon, B. J.},
  title   = {The engine behind (wall) turbulence: perspectives on scale interactions},
  journal = {J. Fluid Mech.},
  volume  = {817},
  pages   = {P1},
  year    = {2017}
}

@article{madhusudanan2019coherent,
  title={ {Coherent} large-scale structures from the linearized {Navier}--{Stokes} equations },
  author={Madhusudanan, A. and Illingworth, S. J. and Marusic, I.},
  journal={J. Fluid Mech.},
  volume={873},
  pages={89--109},
  year={2019},
}

@article{liu2020input,
  title={ {An} input--output based analysis of convective velocity in turbulent channels },
  author={Liu, C. and Gayme, D. F.},
  journal={J. Fluid Mech.},
  volume={888},
  pages={A32},
  year={2020},
}

@article{symon2021energy,
  title={ {Energy} transfer in turbulent channel flows and implications for resolvent modelling },
  author={Symon, S. and Illingworth, S. J. and Marusic, I.},
  journal={J. Fluid Mech.},
  volume={911},
  pages={A3},
  year={2021},
}

@article{liu2022spatial,
  title={ {Spatial} {Input}--{Output} {Analysis} of {Actuated} {Turbulent} {Boundary} {Layers} },
  author={Liu, C. and Gluzman, I. and Lozier, M. and Midya, S. and Gordeyev, S. and Thomas, F. O. and Gayme, D. F.},
  journal={AIAA J.},
  volume={60},
  number={11},
  pages={6313--6327},
  year={2022},
}

@article{GG21,
  author  = {Gluzman, I. and Gayme, D. F.},
  title   = {Input-output framework for actuated boundary layers},
  journal = {Phys. Rev. Fluids},
  year    = {2021},
  volume  = {6},
  number  = {5},
  pages   = {053901}
}

@article{frank2025choosing,
  author  = {Frank-Shapir, O. and Gluzman, I.},
  title   = {Choosing Between Active and Passive Flow Control via Input-Output Analysis: Application to {Couette} Flow},
  journal = {IEEE Control Syst. Lett.},
  year    = {2025},
  volume  = {9},
  pages   = {715--720},
  doi     = {10.1109/LCSYS.2025.3576311}
}

@article{liu2021,
  title={ {Structured} input--output analysis of transitional wall-bounded flows },
  author={Liu, C. and Gayme, D. F.},
  journal={J. Fluid Mech.},
  volume={927},
  pages={A25},
  year={2021},
}

@inproceedings{mushtaq2023,
  title={ {Structured} input-output tools for modal analysis of a transitional channel flow },
  author={Mushtaq, T. and Bhattacharjee, D. and Seiler, P. J. and Hemati, M.},
  booktitle={{AIAA SCITECH 2023 Forum}},
  pages={1805},
  year={2023}
}

@article{shuai2023structured,
  title={ {Structured} input--output analysis of oblique laminar--turbulent patterns in plane {Couette}--{Poiseuille} flow },
  author={Shuai, Y. and Liu, C. and Gayme, D. F.},
  journal={Int. J. Heat Fluid Flow},
  volume={103},
  pages={109207},
  year={2023},
}

@article{mushtaq2024structured,
  title={ {Structured} singular value of a repeated complex full-block uncertainty },
  author={Mushtaq, T. and Bhattacharjee, D. and Seiler, P. and Hemati, M. S.},
  journal={Int. J. Robust Nonlinear Control},
  volume={34},
  number={7},
  pages={4881--4897},
  year={2024},
}

@article{frank2026stability,
  author  = {Frank-Shapir, O. and Gluzman, I.},
  title   = {Stability analysis of transitional flows based on disturbance magnitude},
  journal = {J. Fluid Mech.},
  volume  = {1030},
  pages   = {A8},
  year    = {2026}
}

@article{packard1993,
  title={ {The} complex structured singular value },
  author={Packard, A. and Doyle, J.},
  journal={Automatica},
  volume={29},
  number={1},
  pages={71--109},
  year={1993},
}

@book{Zhou1995-dl,
  author    = {Zhou, K. and Doyle, J. C. and Glover, K.},
  title     = {{Robust and Optimal Control}},
  publisher = {Pearson},
  address   = {Upper Saddle River, NJ},
  year      = {1995}
}

@book{schmid2002stability,
  author    = {Schmid, P. J. and Henningson, D. S.},
  title     = {{Stability and Transition in Shear Flows}},
  series    = {{Applied Mathematical Sciences}},
  volume    = {142},
  publisher = {Springer},
  address   = {New York},
  year      = {2001}
}

@article{cherubini2011minimal,
  title={The minimal seed of turbulent transition in the boundary layer},
  author={Cherubini, S. and De Palma, P. and Robinet, J.-C. and Bottaro, A.},
  journal={J. Fluid Mech.},
  volume={689},
  pages={221--253},
  year={2011},
}

@book{bernstein2009matrix,
  author    = {Bernstein, D. S.},
  title     = {{Matrix Mathematics: Theory, Facts, and Formulas}},
  publisher = {Princeton University Press},
  year      = {2009}
}

@article{vavaliaris2020optimal,
  title={Optimal perturbations and transition energy thresholds in boundary layer shear flows},
  author={Vavaliaris, Chris and Beneitez, Miguel and Henningson, Dan S},
  journal={Phys. Rev. Fluids},
  volume={5},
  number={6},
  pages={062401},
  year={2020},
  publisher={APS}
}

@article{Henningson1996,
   author  = {Henningson, Dan S.},
   title   = {Comment on ``Transition in shear flows. Nonlinear
              normality versus non-normal linearity''},
   journal = {Physics of Fluids},
   volume  = {8},
   number  = {8},
   pages   = {2257--2258},
   year    = {1996},
   doi     = {10.1063/1.868997},
   }

@article{kim1987turbulence,
  title={ {Turbulence} statistics in fully developed channel flow at low {Reynolds} number },
  author={Kim, John and Moin, Parviz and Moser, Robert},
  journal={ J. Fluid Mech. },
  volume={177},
  pages={133--166},
  year={1987},
  publisher={Cambridge University Press}
}

@article{elofsson1998experimental,
  title={ {An} experimental study of oblique transition in plane {Poiseuille} flow },
  author={Elofsson, Per A and Alfredsson, P Henrik},
  journal={ J. Fluid Mech. },
  volume={358},
  pages={177--202},
  year={1998},
  publisher={Cambridge University Press}
}

@article{reddy1998stability,
  title={ {On} stability of streamwise streaks and transition thresholds in plane channel flows },
  author={Reddy, Satish C and Schmid, Peter J and Baggett, Jeffrey S and Henningson, Dan S},
  journal={ J. Fluid Mech. },
  volume={365},
  pages={269--303},
  year={1998},
  publisher={Cambridge University Press}
}

@inproceedings{bhattacharjee2023structured,
  title={ {Structured} input-output analysis of compressible plane {Couette} flow },
  author={Bhattacharjee, Diganta and Mushtaq, Talha and Seiler, Peter J and Hemati, Maziar},
  booktitle={ {AIAA} {SCITECH} 2023 {Forum} },
  pages={1984},
  year={2023}
}

@article{duguet2010formation,
  title={ {Formation} of turbulent patterns near the onset of transition in plane {Couette} flow },
  author={Duguet, Yohann and Schlatter, Philipp and Henningson, Dan S},
  journal={ J. Fluid Mech. },
  volume={650},
  pages={119--129},
  year={2010},
  publisher={Cambridge University Press}
}

@article{farano2015hairpin,
  title={ {Hairpin}-like optimal perturbations in plane {Poiseuille} flow },
  author={Farano, Mirko and Cherubini, Stefania and Robinet, Jean-Christophe and De Palma, Pietro},
  journal={ J. Fluid Mech. },
  volume={775},
  pages={R2},
  year={2015},
  publisher={Cambridge University Press}
}

@article{elofsson2000experimental,
  title={ {An} experimental study of oblique transition in a {Blasius} boundary layer flow },
  author={Elofsson, Per A and Alfredsson, P Henrik},
  journal={ Eur. J. Mech. B/Fluids },
  volume={19},
  number={5},
  pages={615--636},
  year={2000},
  publisher={Elsevier}
}

@article{asai1995boundary,
  title={ {Boundary}-layer transition triggered by hairpin eddies at subcritical {Reynolds} numbers },
  author={Asai, Masahito and Nishioka, Michio},
  journal={ J. Fluid Mech. },
  volume={297},
  pages={101--122},
  year={1995},
  publisher={Cambridge University Press}
}

@misc{darzin1981hydrodynamic,
  title={ {Hydrodynamic} stability },
  author={Darzin, PG and Reid, WH},
  
  year={1981},
  publisher={Cambridge University Press, Cambridge}
}

@article{duguet2013minimal,
  title={ {Minimal} transition thresholds in plane {Couette} flow },
  author={Duguet, Yohann and Monokrousos, Antonios and Brandt, Luca and Henningson, Dan S},
  journal={ Phys. Fluids },
  
  volume={25},
  number={8},
  year={2013},
  publisher={AIP Publishing}
}

@article{duguet2010towards,
  title={ {Towards} minimal perturbations in transitional plane {Couette} flow },
  author={Duguet, Yohann and Brandt, Luca and Larsson, B Robin J},
  journal={ Physical Review E—Statistical, Nonlinear, and Soft Matter Physics },
  volume={82},
  number={2},
  pages={026316},
  year={2010},
  publisher={APS}
}

@article{parente2022minimal,
  title={ {Minimal} energy thresholds for sustained turbulent bands in channel flow },
  author={Parente, Enza and Robinet, J-Ch and De Palma, Paul and Cherubini, Stefania},
  journal={ J. Fluid Mech. },
  volume={942},
  pages={A18},
  year={2022},
  publisher={Cambridge University Press}
}

@article{scherer2001theory,
  title={Theory of robust control},
  author={Scherer, Carsten},
  journal={Delft University of Technology},
  pages={1--160},
  year={2001}
}

@book{kundu2024fluid,
  title={Fluid mechanics},
  author={Kundu, Pijush K and Cohen, Ira M and Dowling, David R and Capecelatro, Jesse},
  year={2024},
  publisher={Elsevier}
}

@book{singh2006computing,
  title={Computing Sensor Location for Nonlinear Systems Under the Influence of Disturbances},
  author={Singh, Abhay K and Hahn, Juergen},
  year={2006},
  publisher={American Institute of Chemical Engineers}
}

@article{liu2022structured,
  title={Structured input--output analysis of stably stratified plane Couette flow},
  author={Liu, Chang and Colm-cille, P Caulfield and Gayme, Dennice F},
  journal={Journal of Fluid Mechanics},
  volume={948},
  pages={A10},
  year={2022},
  publisher={Cambridge University Press}
}

\end{document}